\documentclass[11pt,a4paper]{article}
\usepackage{jheppub}
\usepackage{axodraw2}
\usepackage{amsfonts}
\usepackage{bm,bbm}
\usepackage{xcolor}
\usepackage{cancel}

\newcommand{\la}[1]{\label{#1}}
\newcommand{\be}{\begin{equation}}
\newcommand{\ee}{\end{equation}}
\newcommand{\ba}{\begin{eqnarray}}
\newcommand{\ea}{\end{eqnarray}}
\newcommand{\rmi}[1]{{\mbox{\scriptsize #1}}}
\newcommand{\nr}[1]{\eqref{#1}} 

\newcommand{\nn}{\nonumber \\}

\renewcommand{\vec}[1]{{\bf #1}}

\newcommand{\eq}{eq.~}
\newcommand{\eqs}{eqs.~}
\newcommand{\se}{sec.~}
\newcommand{\ses}{secs.~}
\newcommand{\fig}{fig.~}
\newcommand{\figs}{figs.~}
\newcommand{\app}{appendix~}
\newcommand{\tabl}{table~}
\newcommand{\PT}{\mathbbm{P}^\rmii{T}}

\newcommand{\alphas}{\alpha_{\rm s}}

\newcommand{\F}{\rmii{$F$}}

\newcommand{\mpl}{m_\rmi{pl}}
\newcommand{\Nf}{N_{\rm f}}
\newcommand{\Nc}{N_{\rm c}}

\newcommand{\mE}{m_\rmii{E}}
\newcommand{\miE}{m_\rmiii{E}}

\newcommand{\gammaE}{\gamma_\rmii{E}}
\newcommand{\rmO}{{\mathcal{O}}}

\def\lsi{\raise0.3ex\hbox{$<$\kern-0.75em\raise-1.1ex\hbox{$\sim$}}}
\def\gsi{\raise0.3ex\hbox{$>$\kern-0.75em\raise-1.1ex\hbox{$\sim$}}}

\newcommand{\gsim}{\mathop{\gsi}}

\newcommand{\sign}{\mathop{\mbox{sign}}}
\newcommand{\nF}{n_\rmii{F}} 
\newcommand{\nB}{n_\rmii{B}} 
\newcommand{\rmii}[1]{{\mbox{\tiny\rm{#1}}}}
\newcommand{\rmiii}[1]{{\mbox{\tiny{$\scriptstyle{\rm#1}$}}}}

\newcommand{\im}{\mathop{\mbox{Im}}}
\newcommand{\Tint}[1]{{\hbox{$\sum$}\!\!\!\!\!\!\!\int\,}_{\!\!\!\!\raise-0.9ex\hbox{$\scriptstyle{#1}$}}}
\newcommand{\Tinti}[1]{{{\Sigma}\!\!\!\!\raise0.3ex\hbox{$\int$}_\rmii{${#1}$}}}

\newcommand{\bi}{\begin{itemize}}
\newcommand{\ei}{\end{itemize}}
\newcommand{\hide}[1]{ }
\newcommand{\blind}[1]{\fbox{$ ? $}} 

\newcommand{\deltabar}{\raise-0.02em\hbox{$\bar{}$}\hspace*{-0.8mm}{\delta}}
\newcommand{\A}{\rmii{$A$}} 
\renewcommand{\P}{\mathcal{P}}
\newcommand{\Q}{\mathcal{Q}}

\newcommand{\K}{\mathcal{K}}
\newcommand{\X}{\mathcal{X}}

\newcommand{\ax}{a}
\newcommand{\rhoT}{\varrho^\rmii{T}}
\newcommand{\rhoE}{\widetilde\varrho^{\hspace*{0.5mm}\rmii{E}}}
\newcommand{\rhoTB}{\varrho^\rmii{T,Born}}
\newcommand{\rhoEB}{\widetilde\varrho^{\hspace*{0.5mm}\rmii{E,Born}}}
\newcommand{\US}{\rmi{ultrasoft}}

\newcommand{\IR}{\rmi{soft}}

\newcommand{\UV}{\rmi{hard}}

\newcommand{\now}{\rmi{0}}
\newcommand{\inow}{\rmii{0}}
\newcommand{\fin}{\rmi{fin}} 
\newcommand{\ini}{\rmi{ini}} 
\newcommand{\ifin}{\rmii{fin}} 
\newcommand{\iini}{\rmii{ini}} 

\def\TAsc(#1,#2)(#3,#4,#5)%
{\SetWidth{2.0}\CArc(#1,#2)(#3,#4,#5)\SetWidth{1.0}}
\def\Lwidth{3}

\def\TAgl(#1,#2)(#3,#4,#5){\SetWidth{2.0}\PhotonArc(#1,#2)(#3,#4,#5){\Lwidth}%
{6.283 #3 mul 360 div #4 #5 sub #4 #5 sub mul sqrt mul Tdensity mul}%
\SetWidth{1.0}}
\def\TLgl(#1,#2)(#3,#4){\SetWidth{2.0}\Photon(#1,#2)(#3,#4){\Lwidth}
{#1 #3 sub #1 #3 sub mul #2 #4 sub #2 #4 sub mul add sqrt Tdensity mul}%
\SetWidth{1.0}}

\def\Lwidth{1.3}

\newcommand{\picg}[1]{\;\parbox[c]{62pt}{\begin{picture}(160,80)(0,-27)
\SetWidth{1.0}\SetScale{0.45} #1 \end{picture}}\;}

  \def\GraphggCtwo{\picg{
  \SetWidth{1.5}
    \Photon(50,70)(0,60){3}{6}
  \DashLine(50,70)(100,100){6}
  \GCirc(50,70){5}{0}
  \Photon(50,10)(100,20){3}{6}
  \Photon(100,20)(150,10){3}{6}
  \Photon(100,20)(100,60){3}{5}
  \Photon(50,70)(100,60){3}{6}
  \Photon(100,60)(150,70){3}{6}
 }}

 \def\GraphggC{\picg{
  \SetWidth{1.5}
  \Photon(0,10)(50,20){3}{6}
  \Photon(50,20)(100,10){3}{6}
  \Photon(50,20)(50,60){3}{5}
  \Photon(0,70)(50,60){3}{6}
  \Photon(50,60)(100,70){3}{6}
  \Photon(100,70)(155,60){3}{7}
  \DashLine(100,70)(150,100){6}
  \GCirc(100,70){5}{0}
 }}

\makeatletter \@addtoreset{equation}{section} \makeatother
\renewcommand{\theequation}{\arabic{section}.\arabic{equation}}
\makeatletter
\renewcommand\section{\@startsection {section}{1}{\z@}%
                                   {-5.5ex \@plus -1ex \@minus -.2ex}
                                   {2.3ex \@plus.2ex}%
                                   {\normalfont\large\bfseries}}
\renewcommand\subsection{\@startsection{subsection}{2}{\z@}%
                                     {-3.25ex\@plus -1ex \@minus -.2ex}%
                                     {1.5ex \@plus .2ex}%
                                     {\normalfont\normalsize\bfseries}}
\renewcommand\thesection {\@arabic\c@section}
\renewcommand\thesubsection   {\thesection.\@arabic\c@subsection}
\renewcommand{\@seccntformat}[1]{%
\csname the#1\endcsname.\hspace{1.0em}}
\makeatother

\newcommand{\hiddenappsubsection}[2][]{%
  \begingroup
    \renewcommand{\addcontentsline}[3]{} 
    \refstepcounter{subsection}
    \vspace{3.25ex plus1ex minus.2ex}
    \noindent
    \textbf{\thesubsection.\; #2}\par
    \vspace{1.5ex plus .2ex}
    \label{#1}
  \endgroup
}

\begin{document}


\begin{flushright}
April 2026
\end{flushright}

\arxivnumber{2601.08221}

\vspace*{-0.9cm}


\title{\boldmath
  Energy and momentum dependence of  \\[1mm]
  the soft-axion interaction rate 
}


\author[a]
{Killian Bouzoud,} 
\author[a]
{Jacopo Ghiglieri,} 
\author[b]
{M.~Laine,} 
\author[c]
{G.S.S.~Sakoda} 


\affiliation[a]{
SUBATECH, Nantes Universit\'e, IMT Atlantique, IN2P3/CNRS, \\
4 rue Alfred Kastler, La Chantrerie BP 20722, 44307 Nantes, France}


\affiliation[b]{
AEC, 
Institute for Theoretical Physics, 
University of Bern, \\ 
Sidlerstrasse 5, CH-3012 Bern, Switzerland}


\affiliation[c]{
Instituto de F\'isica, Universidade de S\~ao Paulo, \\
S\~ao Paulo, SP 05508-090, Brazil}



\emailAdd{kbouzoud@subatech.in2p3.fr}
\emailAdd{jacopo.ghiglieri@subatech.in2p3.fr}
\emailAdd{laine@itp.unibe.ch}
\emailAdd{gustavo.sakoda@usp.br}


 

 
\abstract{
Axions coupled to thermal non-Abelian gauge fields may have 
cosmological significance. As the heat bath defines a frame, 
its influence depends separately on energy and momentum. 
A light-like momentum ($k \approx \omega$) is relevant for 
the axion contribution to the effective number of light
neutrinos, $\Delta N^{ }_\rmii{eff}$, whereas 
a vanishing momentum ($k=0$) plays a role for warm natural inflation 
or ultralight dark matter, 
and has been employed in lattice estimates 
(both classical and quantum-statistical) of the strong
sphaleron rate. 
Focussing on  
soft energies ($\alphas^{ }T \ll \omega \ll \pi T$), we carry out 
an HTL computation to show how the domains $k=0$
and $k \approx \omega$ interpolate to each other. We then 
compare with lattice data at $k=0$, 
and connect our analysis to NLO computations
at $k \approx \omega \ge \pi T$. 
Assembling the current best input, 
we re-investigate light QCD axion decoupling dynamics
at $T \ge 200$~MeV, 
showing that efficient interactions in the ultrasoft domain
increase $\Delta N^{ }_\rmii{eff}$ from $\sim 0.03$
to $\sim 0.04$ at $f^{ }_\ax = 4\times 10^8_{ }$~GeV.
}


 
\maketitle
\flushbottom


%
\section{Introduction}
\la{se:intro}

An axion-like field, $\varphi$, 
describes a Beyond the Standard Model
particle that couples to the Standard Model 
through dimension-five pseudoscalar operators. 
It was introduced in the context of the strong
CP problem~\cite{cp1,cp2,cp3}, and was soon conceived 
to have potential phenomenological significance. 
In particular, it could play a role 
in cosmology, as a dark matter 
candidate~\cite{dm1,dm2,dm3}, 
or as a main player~\cite{ai} or 
participant~\cite{tw} in inflationary dynamics.
There are on-going experimental 
searches for axion-like particles,
both terrestrial and in astrophysical environments. 
Given that no signal has been found, axions could either be very
heavy, or very weakly coupled. Many of the current efforts  
concentrate on the latter option, with physical 
zero-temperature axion masses assumed to be in the eV range or below
(cf., e.g., ref.~\cite{experiment}).
That said, heavy QCD axions  
continue also to be investigated (cf., e.g., ref.~\cite{heavy_ax}). 

Quite a few gauge-invariant pseudoscalar operators can be
constructed out of the Standard Model fields. Their
coefficients at some high scale depend on the UV completion of the 
theory, but when the scale is lowered with 
the help of the renormalization group, QCD induces mixings between 
the coefficients, whereby in general all operators are present.   
For our purposes it is sufficient to focus 
on the operator that couples directly to gluons, 
\be
 \mathcal{L} 
 \; \supset \;
 \frac{ 1 }{2} \partial^\mu\varphi\, \partial_\mu\varphi
 - V^{\vphantom{|}}_0(\varphi)  
 - \frac{\varphi \,  \chi}{f^{ }_a}
 \;, 
 \quad
 \chi \;\equiv\;
 \frac{ 
 \alphas{ } \,
 \epsilon^{\mu\nu\rho\sigma}_{ }
 F^{c}_{\mu\nu} F^{c}_{\rho\sigma}
 }{16\pi}
 \;, \la{L}
\ee
where the metric convention ($+$$-$$-$$-$) is assumed, 
$
 F^c_{\mu\nu} 
$ 
is the field strength tensor, 
$\alphas^{ } \equiv g^2_{ }/(4\pi)$ is the QCD coupling, 
$ c \in \{ 1, ..., \Nc^2 - 1 \}$ is a colour index, 
$ \Nc^{ } \equiv 3$, 
and $f^{ }_\ax$ is the axion decay constant. 
The value of $f^{ }_\ax$ reflects 
the scale of the UV completion of the theory.  
In the dark matter context, 
the bare potential, $V^{\vphantom{|}}_0$, is often assumed
to vanish, though it could be non-zero as a reflection of UV physics. 
The (temperature-dependent) axion mass, $m^{ }_\ax$,
gets a contribution both from $V^{\vphantom{|}}_0$, and 
from the dynamical effects induced by the operator $\chi$.

When we think about the cosmological role of axions, 
the gluons that constitute the $\chi$ operator are part of
a thermal ensemble, at a temperature $T$. 
The kinetic 
equilibration rate of the plasma is $\Gamma \sim \alphas^2 T$~\cite{equil}, 
whereas the axion interaction rate is 
$\Upsilon \sim \alphas^3 T^3_{ }/ f_\ax^2$
(the precise functional form and value are
discussed in this work).
In a reheated universe, where $H\sim T^2_{ }/\mpl^{ }$ 
is the Hubble rate, we normally have $H \ll \Gamma$.  
Then the QCD
plasma can be considered thermalized at the time scales 
that are relevant for axion dynamics. On the other hand, 
in the inflationary or reheating 
context, when the axion energy density can be the 
dominant component in~$H$, 
the validity of the assumption $H \ll \Gamma$
is less clear~\cite{therm}, particularly for helicity 
which evolves at a slower rate $\sim \alphas^3 T$~\cite{hook}.
Nevertheless, we treat the QCD plasma as fully thermalized
in the present work. We stay at temperatures $T \ge 200$~MeV, 
so that the plasma is deconfined, noting that interesting
work has also been carried out for $T \ll 200$~MeV 
(cf.,\ e.g.,\ refs.~\cite{chpt0,chpt1,chpt2} and references therein).

The concrete question that we address concerns  
the axion interaction rate, $\Upsilon$
(also called the equilibration or thermalization rate). 
It plays a role both for axion particles, which could contribute
to dark radiation or dark matter, and for an axion condensate, 
which could act as an inflaton or as ultralight dark matter. 
For axion particles, the interaction rate appears 
in the kinetic equation for their phase space
distribution, which we denote by $f^{ }_\varphi$. 
To leading non-trivial order in $1/f_\ax^2$, the evolution
of $f^{ }_\varphi$ is governed by~\cite{sangel}\footnote{%
 Here we assume translational invariance, but in some cases
 spatial inhomogeneities could play a role. For instance, 
 if the axion originates as a pseudo-Goldstone 
 mode from spontaneous symmetry breaking in a reheated universe,
 a network of cosmic strings is generated, which can also 
 emit axions (cf.,\ e.g.,\ refs.~\cite{strings,strings2} 
 and references
 therein). In some constructions the angular direction 
 manifests a $\mathbbm{Z}$($N$) symmetry, and then domain walls exist
 as well (cf.,\ e.g.,\ ref.~\cite{zn} and references
 therein). 
 } 
\ba
 \bigl(\, 
 \partial^{ }_t - H k \partial^{ }_k 
 \,\bigr)
 \, f^{ }_\varphi(t,k)
 & = &  - \Upsilon(\omega,k)
     \, \bigl[\,
       f^{ }_\varphi(t,k) - \nB^{ }(\omega) 
     \, \bigr]
   + 
   \rmO\biggl( \frac{1}{f_\ax^4} \biggr)
  \;, \la{kinetic}
  \\[3mm]
  \omega & \equiv & \sqrt{k^2_{ } + m_\ax^2}
  \;, \la{omega}
\ea
where $k$ is a physical momentum and 
$\nB^{ }$ is the Bose distribution. 
We refer to $\omega$ alternately as ``frequency'' or ``energy''.
Depending on the initial conditions, the dynamics 
described by \eq\nr{kinetic}
is referred to with different names: 
if $f^{ }_\varphi \approx \nB^{ }$, 
we talk about {\em equilibration}; 
if $f^{ }_\varphi \ll \nB^{ }$, 
about {\em particle production}; 
if $f^{ }_\varphi \gg \nB^{ }$, 
about {\em damping}. 

A separate evolution equation is satisfied by 
an axion condensate, $\bar\varphi$. It is defined by writing
$\varphi(t,\vec{x})
 \equiv \bar\varphi(t) +  \delta\varphi(t,\vec{x})$, 
where $\langle \delta\varphi \rangle = 0$. 
After a Fourier transform,  
the momentum $k$ now refers to the 
spatial variation of $\delta\varphi$,
whereas for the condensate, $k=0$. 
In natural inflation, it is $\bar\varphi$ which
drives exponential expansion, and in scenarios of 
ultralight dark matter, the oscillations of $\bar\varphi$
could be responsible for a period of matter-dominated expansion. 

When considering $k=0$, 
there is a subtlety, 
related to the fact that the operator $\chi$ from 
\eq\nr{L} also appears in the chiral anomaly equation. 
We clarify the circumstances under which 
this plays a role
in appendix~\ref{se:fermions}. With this
reservation, and going to the plasma rest frame, 
the evolution equation 
for $\bar\varphi$ reads~\cite{reheat1,mms,warm,db}
\be
 \ddot{\bar \varphi} + 
 \bigl(\, 3H + \Upsilon_\rmi{sph} \bigr) \,\dot{ \bar \varphi}
 + \partial^{ }_\varphi V(\bar{\varphi}) 
 \; 
 \approx 
 \; 
 0 
 \;, \la{eom_varphi}
\ee
where $\Upsilon_\rmi{sph} \equiv \Upsilon(0^+_{ },0)$ 
is defined in \eq\nr{Ups_sph}, 
and the effective potential, $V$, 
incorporates both $V^{\vphantom{|}}_0$
and a dynamical effect from $\chi$
(cf.\ \app\ref{se:fermions}). 
It has been suggested that 
with a non-vanishing $V^{\vphantom{|}}_0$, 
the QCD value of $\Upsilon_\rmi{sph}$ 
permits for a successful (warm) inflation~\cite{sm1,sm2,sm3}. 
Apart from \eq\nr{kinetic}, \eq\nr{eom_varphi} therefore
offers for a second context calling for 
an estimate of~$\Upsilon$. We note that if 
$\Upsilon_\rmi{sph} \ll 3 H$, the dynamics of $\bar\varphi$
is not much affected by $\Upsilon_\rmi{sph}$, 
however by energy conservation, 
$\Upsilon_\rmi{sph}$ still dictates 
how efficiently the gauge plasma heats up. 

Our work is organized as follows. 
We start by setting up the basic notation and summarizing
the kinematic domains considered (cf.\ \se\ref{se:kinematics}).
Subsequently we describe the details of our 
Hard Thermal Loop (HTL) computation (cf.\ \se\ref{se:htl}). 
Analytic results can be obtained if we go to large frequencies, 
$\omega \gg \mE^{ }$ 
(cf.\ \se\ref{se:asy}). 
Another interesting limit is to go to small frequencies,
$\omega \ll \mE^{ }$, 
where the problem becomes non-perturbative, 
and has been addressed with two different classes of lattice
methods (cf.\ \se\ref{se:lattice}). Assembling together 
the input from the preceding sections and from 
\app\ref{se:hard}, 
we update the estimate for $\Delta N^{ }_\rmi{eff}$ originating from
cosmologically stable light QCD axions
(cf.\ \se\ref{se:assembled}). 
After summarizing 
our main findings (cf.\ \se\ref{se:concl}), 
we elaborate on the role of light chiral fermions 
in $\Upsilon_\rmi{sph}$ (cf.\ appendix~\ref{se:fermions});
on why the HTL computation fails at $\omega \ll \mE^{ }$
(cf.\ \app\ref{se:IR}); 
and on how our work connects 
to previous computations at $\omega \ge \pi T$
(cf.\ \app\ref{se:hard}).

%
\section{Basic definitions and kinematic domains}
\la{se:kinematics}

In this work we consider the axion interaction rate, $\Upsilon(\omega,k)$, 
up to leading non-trivial order in an expansion 
in $1/f_\ax^2$. However, the
setup applies to any order in the QCD coupling, $\alphas^{ }$. 
Therefore resummations, NLO computations, or even lattice simulations, can 
be discussed. When restricting to order $1/f_\ax^2$, the formalism
happens to coincide with that of the linear response theory, 
however the $1/f_\ax^2$ expansion is more general~\cite{sangel}. 

Promoting $\chi$ from \eq\nr{L} to a quantum-mechanical operator, 
$\hat\chi$, its {\em retarded correlator} is defined as 
\be
 G_\chi^\rmii{R}(\K)
 \; \equiv \;
 \int_{\X} 
 e^{i(\omega t - \vec{k}\cdot\vec{x}) }_{ } 
 \, 
 \bigl\langle\,
 i 
 \bigl[\,
 \hat\chi(\X),\hat\chi(0) 
 \,\bigr] \theta(t)
 \,\bigr\rangle
 \;, \quad
 \X
 \; \equiv \; (t,\vec{x})
 \;, \quad
 \K 
 \; \equiv \; (\omega,\vec{k})
 \;. \la{G_R}
\ee
Viewing $\omega$ as a complexified variable, and approaching
the real axis from above, 
the imaginary part of $G_\chi^\rmii{R}$ 
yields the {\em spectral function}, 
\be
 \rho^{ }_\chi (\K) 
 \; \equiv \; 
 \im G^\rmii{R}_\chi(\omega+ i\hspace*{0.3mm} 0^+_{ },\vec{k})
 \; = \; 
 \int_{\X} 
 e^{i(\omega t - \vec{k}\cdot\vec{x}) }_{ } 
 \, 
 \biggl\langle\,
 \frac{1}{2}
 \bigl[\,
 \hat\chi(\X),\hat\chi(0) 
 \,\bigr] 
 \,\biggr\rangle
 \;. \la{rho}
\ee
The spectral function has the dimension GeV$^4_{ }$. The axion
interaction rate, of dimension GeV, 
is obtained by dividing it with the factor
$f_\ax^2$ from \eq\nr{L} and with the frequency, 
\be
 \Upsilon(\omega,k) 
 \; \equiv \; 
 \frac{1}{f_\ax^2}
 \frac{\rho^{ }_\chi(\K)}{\omega}
 \;, \quad
 k \; \equiv \; |\vec{k}|
 \;. \la{Ups_def}
\ee
The function $\rho^{ }_\chi(\K)$ is a well-defined
property of QCD even without axions. 

\begin{figure}[t]

\hspace*{-0.1cm}
\centerline{%
 \epsfysize=7.3cm\epsfbox{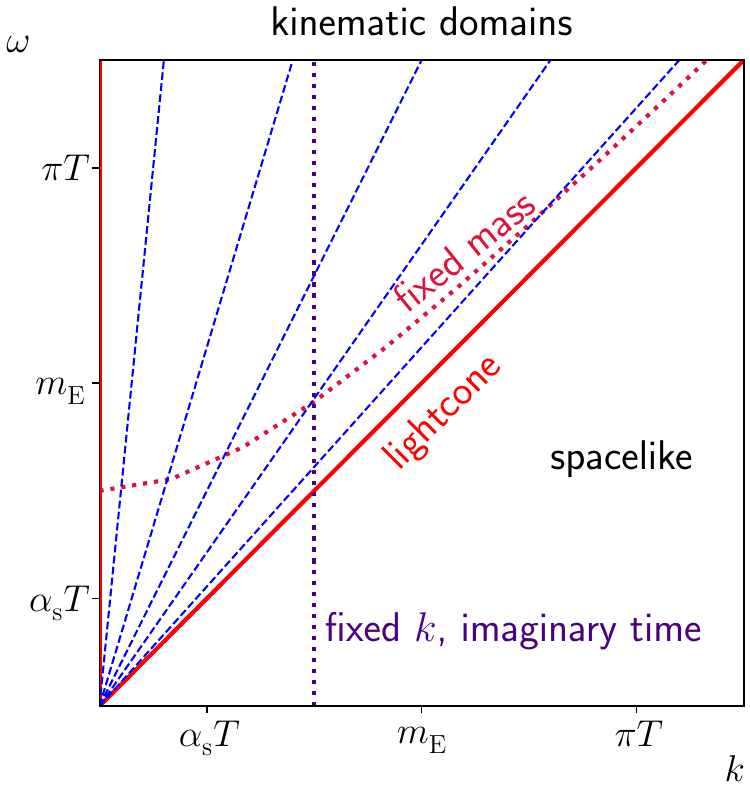}%
}

\caption[a]{\small
  Illustration of the kinematic domains discussed in this work. 
  On-shell axions have a fixed mass,
  $m_\ax^2 = \omega^2_{ } - k^2_{ }$,
  shown with a dotted crimson line. If their mass is 
  small, $m^{ }_\ax \ll \pi T$, they are almost lightlike
  for typical thermal momenta, $k\sim \pi T$. In contrast, 
  when we consider how an axion condensate interacts with
  a thermal plasma, we restrict ourselves to the axis $k=0$. 
  All existing lattice simulations also focus on $k=0$.
  The dotted indigo line shows the curve probed by would-be 
  imaginary-time lattice measurements at $k > 0$. 
  The dashed blue rays span the domain
  studied in the present work,
  and will be used
  for illustrating our results
  in \fig\ref{fig:chi_htl}. 
  }

\la{fig:kinematics}
\end{figure}

The kinematic domains relevant for the current work are illustrated
in \fig\ref{fig:kinematics}.
Quite a bit is already known about $\rho^{ }_\chi(\K)$ along
the lines $k=0$ and $k=\omega$.
At $k=0$, an NLO computation for $\omega \ge \pi T$
was presented in ref.~\cite{Bulk_wdep}. In this domain, NLO 
corrections are of relative magnitude $\rmO(\alphas)$.
In the same work, a HTL computation was presented for $\omega\sim \mE^{ }$, 
where 
$
 \mE \sim \sqrt{\alphas^{ }\pi\vphantom{|}}\hspace*{0.3mm} T
$ 
denotes a Debye mass 
(it was not declared as a full HTL computation, though
in hindsight it is one). A classical-statistical lattice 
determination for $\omega \le \mE^{ }$ was
worked out in ref.~\cite{clgt} (the value at $\omega\to 0$ had
been extensively investigated in ref.~\cite{mt}). 
There are also estimates of 
the limit $\omega\to 0$ from
quantum-statistical 
lattice simulations~\cite{eucl1,eucl2,eucl3}, though they have their
own systematic uncertainties, to which we return in \se\ref{se:lattice}. 

For $k = \omega$, there is an even longer history of computations
(cf., e.g., refs.~\cite{trad_1,trad_2,trad_3,salvio,trad_4,bg,mainz}). 
A full leading-order analysis at $k = \omega \ge \pi T$,
and an estimate of its uncertainties, was presented
in ref.~\cite{bg}, together with a critical 
review of previous work. Ref.~\cite{bg} also outlined an NLO 
computation for the same domain, with the NLO corrections being
of relative magnitude $\rmO(\alphas^{1/2})$, though it was not declared 
as a full NLO result 
(we discuss this in more detail in appendix~\ref{se:hard}).
A HTL computation for the domain $\omega\sim \mE^{ }$
was worked out in ref.~\cite{mainz},  who also suggested 
a procedure to extrapolate the result to lower frequencies, 
though only for Abelian gauge fields. 

The main technical novelty of the present study is to interpolate between
the $k = 0$ and $k = \omega$ domains, while assuming that 
$\omega \sim \mE^{ }$. In practical terms, the interpolation can 
be conveniently illustrated by considering rays of constant
$k/\omega$, as indicated with the dashed blue lines
in \fig\ref{fig:kinematics}. We show in the next section that
such an interpolation can be determined with the help of a full
HTL computation. The motivation for presenting this 
interpolation is manifold: to permit for the study of axions
with a non-zero mass (cf.\ the dotted crimson line of a fixed
mass in \fig\ref{fig:kinematics}); 
to present a case for extending lattice simulations
to the $k = \omega$ axis
(either classical-statistical or quantum-statistical);
to provide perturbative support for the analytic continuation
that the latter require (cf.\ the dotted indigo line of a fixed $k$ and 
imaginary time in \fig\ref{fig:kinematics}); and, 
most importantly, 
to assemble an educated guess for a full $\Upsilon(k,k)$, 
incorporating $k=0$ lattice information to the 
extent possible (cf.\ \se\ref{se:assembled}). 

%
\section{Main steps of the HTL computation}
\la{se:htl}

When we address momenta and energies in the domain 
$k\sim\omega\sim \mE^{ } \ll \pi T$, perturbation theory needs
to be resummed, in order to account for large effects from the
hard scale, $\pi T$~\cite{ht1,ht2,ht3,ht4}. In this section 
we present the main steps of the resummed computation. 
Perturbative HTL computations can be carried out both in the 
imaginary-time and in the real-time formalism of thermal field theory, 
with identical results if the same observable is considered. Here we
use the language of the former, referring to ref.~\cite{sch} for 
a general exposition of the latter. We suggest that a reader not
interested in technical details might skip directly to \se\ref{se:assembled}
(p.~\pageref{se:assembled}).

%
\subsection{Feynman rules}
\la{ss:feynman}

When we embark on an HTL computation, the first issue is how the 
operator $\chi$ from \eq\nr{L} is represented in terms of the HTL 
degrees of freedom. 
In the jargon of the field, this is the question of what
is the HTL correction to the axion--gauge vertex. It has been 
shown in appendix~C of ref.~\cite{Bulk_wdep} that there is none. 
Therefore, we can write
\be
 \chi^{ }_\rmii{HTL} 
 \; 
 \underset{\rmii{\cite{Bulk_wdep}}}{\overset{\rmii{\nr{L}}}{=}}
 \; 
 c^{ }_\chi \epsilon^{\mu\nu\rho\sigma}_{ }
 F^c_{\mu\nu} F^c_{\rho\sigma}
 \;, \quad
 c^{ }_\chi 
 \; \equiv \; 
 \frac{\alphas^{ }}{16\pi}
 \;, \la{c_chi}
\ee
where $F^c_{\mu\nu}$ is now expressed in terms of the HTL-resummed
gluon fields. As a technical remark, in a loop computation involving
UV divergences, the fields and couplings appearing in \eq\nr{c_chi} 
are the bare ones~\cite{Bulk_wdep}, but here we can replace them with 
the renormalized values, since we remain at relative accuracy 
$\rmO(\alphas^0)$ or $\rmO(\alphas^{1/2})$.

If we go to Euclidean spacetime, either for 
perturbative computations in the imaginary-time formalism, 
or for 4d lattice simulations, the Levi-Civita symbol
in \eq\nr{c_chi} implies
that one of the Lorentz indices is temporal. Therefore,
Wick rotation inserts an imaginary unit to $\chi$, 
and a minus sign to its 2-point correlators. In the following, 
we display the overall sign as if we remained in Minkowskian spacetime. 
We denote imaginary-time four-momenta with $P \equiv (p^{ }_n,\vec{p})$,
and Minkowskian four-momenta with $\P \equiv (p^0_{ },\vec{p})$, with 
the Wick rotation corresponding to 
$
 p^{ }_n \to -i [p^0_{ } + i 0^+_{ }]
$, 
and with $p \equiv |\vec{p}|$.
The appearance of $P$ or $\P$ as a subscript indicates implicitely
whether Wick rotation has been carried out. 
To streamline the notation, we normally employ $p^{ }_0$
rather than $p^0_{ }$ as the Minkowskian energy variable;
in our choice of metric, their values coincide.

The HTL-resummed gluon propagator can be written as 
\be
 \Delta^{-1}_{P;\mu\nu}
 \; = \; 
 \frac{ \PT_{\mu\nu} }{P^2_{ } + \Pi^\rmii{T}_P } 
 + 
 \frac{(\delta^{ }_{\mu\nu} - \PT_{\mu\nu}) p^2_{ } }{P^2_{ }
  (p^2_{ } + \widetilde{\Pi}^\rmii{E}_P)}
 + 
 \frac{P^{ }_\mu P^{ }_\nu}{P^4_{ }}
 \biggl( \xi - \frac{p^2_{ }}{ p^2_{ } + \widetilde{\Pi}^\rmii{E}_P} \biggr)
 \;, \la{Delta} 
\ee
where $\xi$ is a gauge parameter, 
$
 \PT_{\mu\nu} 
 \equiv 
 \delta^{ }_{\mu i}\delta^{ }_{\nu j}
 (\delta^{ }_{ij} - p^{ }_i p^{ }_j/p^2_{ })
$,
and (after analytic continuation)
\ba
 \Pi^\rmii{T}_{\P}
 & = & 
 \frac{\mE^2}{2} \,
 \biggl\{
  \frac{p_0^2}{p^2_{ }} + 
  \frac{p^{ }_0}{2p}
  \biggl( 1 - \frac{p_0^2}{p^2_{ }} \biggr)
  \ln \biggl[ \frac{p^{ }_0 + p + i 0^+_{ }}
                   {p^{ }_0 - p + i 0^+_{ }} \biggr] 
 \biggr\}
 \;, \la{PiT} 
 \\[3mm]
 \widetilde \Pi^\rmii{E}_{\P}
 & = & 
 \mE^2 \,
 \biggl\{ 
   1 - 
  \frac{p^{ }_0}{2p}
  \ln \biggl[ \frac{p^{ }_0 + p + i 0^+_{ }}
                   {p^{ }_0 - p + i 0^+_{ }} \biggr] 
 \biggr\}
 \;. \la{PiE}
\ea
At leading order, the Debye mass squared reads
$
 \mE^2 = 4 \pi \alphas T^2_{ } (\Nc/3 + \Nf/6)
$, 
where $\Nf$ is the number of light quark flavours. 
Some key properties of the self-energies are 
$
 \Pi^\rmii{T}_{(0,\vec{p})} = 0
$,
$
 \widetilde \Pi^\rmii{E}_{(0,\vec{p})} = \mE^2
$, 
and 
$
 \Pi^\rmii{T}_{(p,\vec{p})} = \mE^2/2
$.
It is a helpful crosscheck of practical computations
that the last term of \eq\nr{Delta}, 
proportional to $P^{ }_\mu P^{ }_\nu$, 
must cancel exactly from physical observables. 

%
\subsection{Contractions, Matsubara sums, and angular integrals}
\la{ss:matsubara}

When we insert \eq\nr{Delta} into the 2-point correlator 
of $\chi^{ }_\rmii{HTL}$, 
the first step is to contract the Lorentz indices. This is not 
entirely trivial, due to the appearance of two Levi-Civita symbols. 
After some work, the Euclidean 2-point correlator, of which 
\eq\nr{G_R} is an analytic continuation, can be written as 
\ba
 && \hspace*{-1.0cm}
 G^\rmii{E}_\chi(K)
 \; = \; 
 \;-\,32 (\Nc^2 - 1)\, c_\chi^2\; \Tint{P,Q} \! \deltabar(P+Q-K) 
 \la{G_E}
 \\[3mm]
 & \times & 
 \biggl\{\; 
 \bigl[\, 
   p^2_{ } q^2_{ }
  -
  (\vec{p}\cdot\vec{q})^2_{ }
 \,\bigr]
 \biggl[\, 
   \frac{1}{(P^2_{ } + \Pi^\rmii{T}_P)(q^2_{ } + \widetilde\Pi^\rmii{E}_Q)} 
 + 
   \frac{1}{(Q^2_{ } + \Pi^\rmii{T}_Q)(p^2_{ } + \widetilde\Pi^\rmii{E}_P)} 
 \,\biggr]
 \nn[3mm]
 &  & \; + \,  
 \biggl[\,
 \biggl( \frac{p_n^2}{p^2_{ }} + \frac{q_n^2}{q^2_{ }} \biggr) 
 \bigl[\,
   p^2_{ } q^2_{ }
 + 
  (\vec{p}\cdot\vec{q})^2_{ } 
 \,\bigr]
 - 4 p^{ }_n q^{ }_n \hspace*{0.3mm} \vec{p}\cdot\vec{q}
 \,\biggr]
 \frac{1}{(P^2_{ } + \Pi^\rmii{T}_P)(Q^2_{ } + \Pi^\rmii{T}_Q)}
 \;\biggr\}  
 \;. \hspace*{5mm} \nonumber
\ea
Here $\Tinti{P}$ denotes a Matsubara sum-integral, and 
$\;\deltabar$ is normalized such that $\Tinti{P}\;\deltabar(P) = 1$.

The next step is to carry out the Matsubara sums. For this we write
the propagators in a spectral representation,  
\be
 \frac{1}{P^2_{ } + \Pi^\rmii{T}_P} 
 \; = \; 
 \int_{-\infty}^{\infty} \! \frac{{\rm d} p^{ }_0}{\pi}
 \frac{\rhoT_\P}{p^{ }_0 - i p^{ }_n}
 \;, \quad
 \rhoT_\P 
 \; \equiv \; 
 \im \biggl[\, \frac{1}{P^2_{ } + \Pi^\rmii{T}_P} 
 \,\biggr]^{ }_{p^{ }_n\to -i (p^{ }_0 + i 0^+_{ })}
 \;. \la{rho_T}
\ee
Noting that 
$
 \im \{ \ln[(p^{ }_0 + p + i 0^+_{ })/(p^{ }_0 - p + i 0^+_{ })] \}
 = 
 - \pi
$
for $|p^{ }_0| < p$, we find that 
\be
 \Pi^\rmii{T}_\P 
 \; 
 \overset{\rmii{\nr{PiT}}}{
 \underset{|p^{ }_0|\; < \;p \hspace*{2mm}}{\equiv}}
 \underbrace{
 \frac{\mE^2}{2} \,
 \biggl\{
  \frac{p_0^2}{p^2_{ }} + 
  \frac{p^{ }_0}{2p}
  \biggl( 1 - \frac{p_0^2}{p^2_{ }} \biggr)
  \ln \biggl| \frac{p^{ }_0 + p }
                   {p^{ }_0 - p } \biggr| 
 \biggr\}
 }_{\equiv\; \Sigma^\rmii{T}_\P }
 \; - \;
 \underbrace{
 \frac{i \pi \mE^2\, p^{ }_0}{4 p} 
   \biggl( 1 - \frac{p_0^2}{p^2_{ }} \biggr)
 }_{ \equiv\; i \hspace*{0.3mm} \Gamma^\rmii{T}_\P }
 \;. \la{Gamma_T} 
\ee
This implies that $\rhoT_\P > 0$ for $p^{ }_0 > 0$.
In the E channel, 
we insert a minus sign, 
\be
 \frac{-\,1}{p^2_{ } + \widetilde\Pi^\rmii{E}_P} 
 \; = \; 
 \int_{-\infty}^{\infty} \! \frac{{\rm d} p^{ }_0}{\pi}
 \frac{\rhoE_\P}{p^{ }_0 - i p^{ }_n}
 \;, \quad
 \rhoE_\P 
 \; \equiv \; 
 \im \biggl[\, \frac{-\,1}{p^2_{ } + \widetilde\Pi^\rmii{E}_P} 
 \,\biggr]^{ }_{p^{ }_n\to -i (p^{ }_0 + i 0^+_{ })}
 \;. \la{rho_E}
\ee
The self-energy becomes
\be
 \widetilde \Pi^\rmii{E}_\P 
 \; 
 \overset{\rmii{\nr{PiE}}}{
 \underset{|p^{ }_0|\; < \;p \hspace*{2mm}}{\equiv}}
 \; 
 \underbrace{
  \mE^2 \,
 \biggl\{ 
   1 - 
  \frac{p^{ }_0}{2p}
  \ln \biggl| \frac{p^{ }_0 + p}
                   {p^{ }_0 - p} \biggr| 
 \biggr\}
 }_{ \equiv \;  \widetilde\Sigma^\rmii{E}_\P }
 \; + \;
 \underbrace{ 
 \frac{i \pi \mE^2\, p^{ }_0}{2 p} 
 }_{ 
 \equiv \;
 i \hspace*{0.3mm} \widetilde\Gamma^\rmii{E}_\P
 }
 \;. \la{Gamma_E} 
\ee
The opposite sign of the imaginary part guarantees that 
$\rhoE_\P > 0$ for $p^{ }_0 > 0$.

The Matsubara sums can now be carried out, with the help of  
\be
 T \sum_{p^{ }_n}
 T \sum_{q^{ }_n}
 \frac{\delta^{ }_{p^{ }_n + q^{ }_n - k^{ }_n}}{T}
 \frac{1}{(p^{ }_0 - i p^{ }_n)(q^{ }_0 - i q^{ }_n)}
 \; = \; 
 \frac{1 + \nB^{ }(p^{ }_0) + \nB^{ }(q^{ }_0)}
      { p^{ }_0 + q^{ }_0 - i k^{ }_n}
 \;. \la{m_sum}
\ee
The spectral function is obtained from 
$
 \rho^\rmii{HTL}_\chi (\omega) = 
 \im G^\rmii{E}_\chi (k^{ }_n\to -i [\omega + i 0^+_{ }])
$, 
which turns the denominator of \eq\nr{m_sum} into a Dirac-$\delta$.
If we have Matsubara frequencies in the numerator, 
they get effectively converted as 
$p^{ }_n\to -i \hspace*{0.3mm}p^{ }_0$ and 
$q^{ }_n\to -i \hspace*{0.3mm}q^{ }_0$, though verifying this 
rigorously takes some effort. All in all, \eq\nr{G_E} turns into
\ba
 \rho^\rmii{HTL}_\chi(\K)
 & = & 
 32 (\Nc^2 - 1)\, c_\chi^2 \! 
 \int_{\vec{p},\vec{q}} \!\!\!
 (2\pi)^3_{ } \delta^{(3)}_{ }(\vec{k-p-q})
 \int_{-\infty}^{\infty}
 \! \frac{{\rm d} p^{ }_0}{\pi}
 \, \bigl[\, 1 + \nB^{ }(p^{ }_0) + \nB^{ }(q^{ }_0)
    \,\bigr]^{ }_{q^{ }_0 \,=\, \omega - p^{ }_0}
 \hspace*{5mm}
 \nn[3mm]
 & \times & 
 \biggl\{\; 
 \bigl[\, 
   p^2_{ } q^2_{ }
  -
  (\vec{p}\cdot\vec{q})^2_{ }
 \,\bigr]
 \bigl[\, 
  \rhoT_\P \, \rhoE_\Q 
 +  
  \rhoT_\Q \, \rhoE_\P 
 \,\bigr]
 \nn[3mm]
 &  & \; + \,  
 \biggl[\,
 \biggl( \frac{p_0^2}{p^2_{ }} + \frac{q_0^2}{q^2_{ }} \biggr) 
 \bigl[\,
   p^2_{ } q^2_{ }
 + 
  (\vec{p}\cdot\vec{q})^2_{ } 
 \,\bigr]
 - 4 p^{ }_0 q^{ }_0 \hspace*{0.3mm} \vec{p}\cdot\vec{q}
 \,\biggr] \, 
   \rhoT_\P \, \rhoT_\Q
 \;\biggr\}  
 \;.  
 \la{rho_K_pre}
\ea

Finally, we can carry out an angular integral, with the help of 
\ba
 && \hspace*{-2.5cm}
    \frac{1}{(2\pi)^3_{ }}
    \int \! {\rm d}^3_{ }\vec{p} 
    \int \! {\rm d}^3_{ }\vec{q} 
    \; \delta^{(3)}_{ }(\vec{k-p-q})
    \, f(p) \, g(q) \, h(\vec{p}\cdot\vec{q})
 \nn[3mm]
 & = & 
   \frac{1}{(2\pi)^3_{ }}
    \int \! {\rm d}^3_{ }\vec{p} 
    \, f(p) \, g(|\vec{k-p}|) \, h(\vec{p}\cdot\vec{k} - p^2_{ })
 \nn[3mm]
 & \overset{\vec{p}\cdot\vec{k} \; = \; p\hspace*{0.3mm} k z}{=} & 
   \frac{1}{(2\pi)^2_{ }}
    \int_0^\infty \! {\rm d}p \, p^2_{ } 
    \, f(p) 
    \int_{-1}^{+1} \! {\rm d}z  
    \, g(\sqrt{p^2_{ } + k^2_{ } - 2 p k z}) 
    \, h(p k z - p^2_{ })
 \nn[3mm]
 & \overset{q \; \equiv \; \sqrt{p^2_{ } + k^2_{ } - 2 p k z} }{=} & 
   \frac{1}{(2\pi)^2_{ } k }
    \int_0^\infty \! {\rm d}p \, p 
    \int_{|p-k|}^{p+k} \! {\rm d}q \, q  
    \, f(p) 
    \, g(q) 
    \, h\biggl( \frac{k^2_{ } - p^2_{ } - q^2_{ } }{2} \biggr)
 \;. \la{angular}
\ea
Inserting \eq\nr{angular} in \eq\nr{rho_K_pre}, we find
\ba
 \rho^\rmii{HTL}_\chi(\K)
 & 
 \overset{\rmii{\nr{rho_K_pre}}}{\underset{\rmii{\nr{angular}}}{=}} 
 & 
 \frac{ 2 (\Nc^2 - 1)\, c_\chi^2 }{\pi^3_{ }k} \! 
    \int_0^\infty \! {\rm d}p \, p 
    \int_{|p-k|}^{p+k} \! {\rm d}q \, q  
 \int_{-\infty}^{\infty}
 \! {\rm d} p^{ }_0 
 \, \bigl[\, 1 + \nB^{ }(p^{ }_0) + \nB^{ }(q^{ }_0)
    \,\bigr]^{ }_{q^{ }_0 \,=\, \omega - p^{ }_0}
 \hspace*{5mm}
 \nn[3mm]
 & \times & 
 \biggl\{\; 
 \bigl[\, k^2_{ } - (p-q)^2_{ } \,\bigr]
 \bigl[\, (p+q)^2_{ } - k^2_{ } \,\bigr]
 \bigl[\, 
  \rhoT_\P \, \rhoE_\Q 
 +  
  \rhoT_\Q \, \rhoE_\P 
 \,\bigr]
 \la{rho_K}
 \\[3mm]
 &  & \; + \,  
 \biggl[\,
 \biggl( \frac{p_0^2}{p^2_{ }} + \frac{q_0^2}{q^2_{ }} \biggr) 
 \bigl[\,
  \bigl( p^2_{ } + q^2_{ } - k^2_{ } \bigr)^2_{ } 
 + 
  4 p^2_{ } q^2_{ }
 \,\bigr]
 + 8 p^{ }_0 q^{ }_0 
   \bigl( p^2_{ } + q^2_{ } - k^2_{ } \bigr) 
 \,\biggr] \, 
   \rhoT_\P \, \rhoT_\Q
 \;\biggr\}  
 \;.  \nonumber 
\ea
We remark that even though the computations
described in this subsection 
are non-trivial, the final result from \eq\nr{rho_K} is
well-known in the literature (cf., e.g., ref.~\cite{salvio}). 
We have nevertheless presented the intermediate steps, because to our
knowledge they have not been spelled out before.

\begin{figure}[t]

\hspace*{-0.1cm}
\centerline{%
 \epsfysize=7.3cm\epsfbox{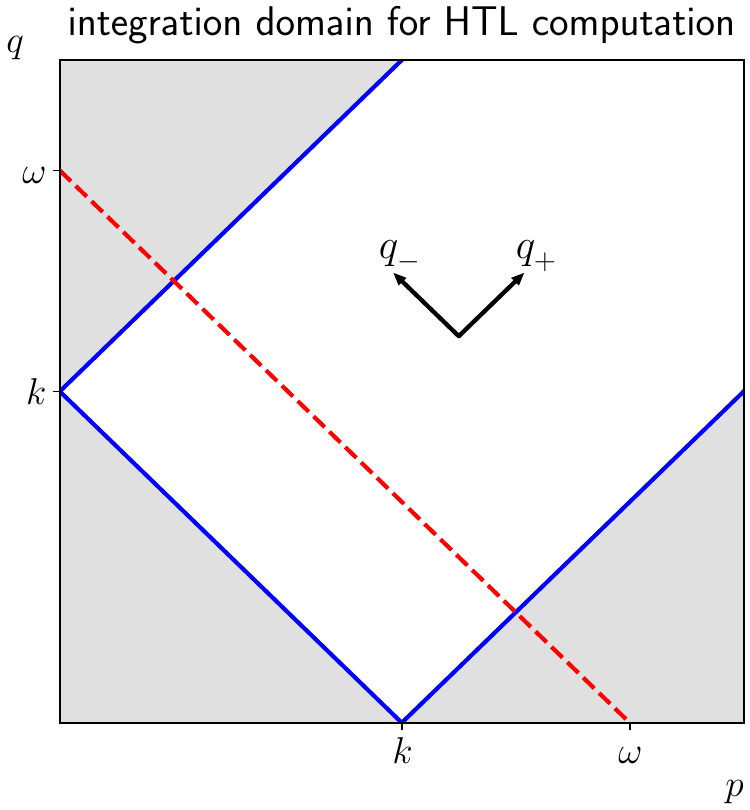}%
}

\caption[a]{\small
  Illustration of the integration domain
  pertinent to \eqs\nr{rho_K} and \nr{new_domain}. 
  The red dashed line corresponds to 
  $q^{ }_+ = \omega/2$,  
  separating different
  types of physical processes
  (cf.\ table~\ref{table:domains} on p.~\pageref{table:domains}).
  }

\la{fig:domains}
\end{figure}

For future reference, we note that 
it is helpful to introduce the variables
\be
 q^{ }_+ \; \equiv \; \frac{q+p}{2}
 \;, \quad
 q^{ }_- \; \equiv \; \frac{q-p}{2}
 \;, \la{qpm}
\ee
and write the integral as 
\be
 \int_0^\infty \! {\rm d}p \, 
 \int_{|p-k|}^{p+k} \! {\rm d} q 
 \; = \; 
 2
 \int_{-k/2}^{+k/2} \! {\rm d}q^{ }_-
 \int_{k/2}^{\infty} \! {\rm d}q^{ }_+ 
 \bigg|^{p \, = \, q^{ }_+ \,-\, q^{ }_-}_{q \, = \, q^{ }_+ \,+\, q^{ }_- }
 \;, \la{new_domain}
\ee
making the integration boundaries constant. 
This domain is illustrated in \fig\ref{fig:domains}.
As for $p^{ }_0$, the symmetry 
$p^{ }_0 \to \omega - p^{ }_0$ implies that we could 
reflect the integrand around $p^{ }_0 = \omega/2$, however  
in the following we keep the original full 
$p^{ }_0$-domain. 

%
\subsection{Cut-cut contribution}
\la{ss:cc}

We divide the integration domain of \eq\nr{rho_K} into three parts, 
depending on the nature of the spectral functions 
$\rhoT_{ }$ and $\rhoE_{ }$. 
If $|p^{ }_0| < p$, the finite widths from \eqs\nr{Gamma_T} and
\nr{Gamma_E} imply that the spectral functions are smooth. 
The smooth shapes are conventionally referred to as 
{\em cuts} (physically, they reflect the phenomenon 
of {\em Landau damping}). Picking up the cut parts from both
spectral functions then yields the cut-cut contribution. 

To determine the cut-cut contribution, we need to identify
which part of the integration domain satisfies simultaneously
$|p^{ }_0| < p$
 and 
$|q^{ }_0| < q$, 
with the constraint $q^{ }_0 = \omega - p^{ }_0$. Given 
that the domains of $p$ and $q$ are interconnected
(cf.\ \fig\ref{fig:domains}), 
this represents a non-trivial geometrical exercise. In terms
of the variables from \eq\nr{qpm}, the result is as shown 
in table~\ref{table:domains}. 
Within the given domain, \eq\nr{rho_K} can be integrated
numerically. In \eq\nr{IRlimit}, we also show how its 
limiting value at $\omega\ll \mE^{ }$ can be obtained analytically. 

%
\begin{table}[t]


{\fontsize{9pt}{11pt}\selectfont
$$
\begin{array}{|c|c|cc|c|} 
  \hline %
  & & & & \\[-3mm]
  \mbox{nature of process} &
  q^{ }_- & 
  \multicolumn{2}{|c|}{ q^{ }_+ } & 
  p^{ }_0
 \\[1mm]
  \hline %
  & & & & \\[-3mm]
  \mbox{cut-cut} & 
   (-k/2,+k/2)  & 
     & (\omega/2,\infty)  & 
   (\omega-q,p)
   \\[1mm]
  \mbox{pole-pole}\;p^{ }_0 > 0\,,\,q^{ }_0 > 0 & 
   (-k/2,+k/2)  & 
   (k/2,\omega/2) &  & 
   (p,\omega-q)
   \\[1mm]
  \mbox{pole-pole}\;p^{ }_0 < 0\,,\,q^{ }_0 > 0 & 
   (-k/2,+k/2)  
   & 
   \multicolumn{2}{c|}{
   (k/2,\infty)
   } 
   & 
   (-\infty,-p)
   \\[1mm]
  \mbox{pole-pole}\;p^{ }_0 > 0\,,\,q^{ }_0 < 0 & 
   (-k/2,+k/2)  
   & 
   \multicolumn{2}{c|}{
   (k/2,\infty)
   }
   & 
   (\omega + q,\infty)
   \\[1mm]
  \mbox{pole~in}\;p^{ }_0\,,\;
  \mbox{cut~in}\;q^{ }_0 & 
   (-k/2,+k/2)  & 
   (k/2,\omega/2) &  & 
   (\omega-q,\omega+q)
   \\[1mm]
  & 
  & 
    & (\omega/2,\infty)  & 
   (p,\omega+q)
   \\[1mm]
  \mbox{cut~in}\;p^{ }_0\,,\;
  \mbox{pole~in}\;q^{ }_0 & 
   (-k/2,+k/2)  & 
   (k/2,\omega/2) & & 
   (-p,p)
   \\[1mm]
  & 
  & 
    & (\omega/2,\infty)  & 
   (-p,\omega-q)
   \\[1mm]
  \hline
\end{array}
$$
}

\vspace*{-3mm}

\caption[a]{\small
 The integration domains for the separate physical processes
 discussed in 
 \ses\ref{ss:cc}--\ref{ss:pc}. 
 We make use of the variables in \eq\nr{qpm}, and assume
 $\omega \ge k \ge 0$.
 }

 \la{table:domains}
\end{table}
%

%
\subsection{Pole-pole contribution}
\la{ss:pp}

If 
$|p^{ }_0| > p$
 and 
$|q^{ }_0| > q$, 
the thermal widths, 
$\Gamma^\rmii{T}_\P$ from \eq\nr{Gamma_T} 
and 
$\widetilde\Gamma^\rmii{E}_\P$ from \eq\nr{Gamma_E}, 
are absent. Then then spectral functions take the forms
\ba
 \rhoT_\P 
 & 
 \underset{|p^{ }_0|\, > \, p}{
 \overset{\rmii{\nr{rho_T}}}{=}}
 & 
 \pi\, \sign(p^{ }_0) \, 
 \delta\bigl(\,
       p_0^2 - p^2_{ } - \Sigma^\rmii{T}_\P
 \,\bigr)
 \;, \la{rhoT_pole}
 \\[3mm]
 \rhoE_\P 
 & 
 \underset{|p^{ }_0|\, > \, p}{
 \overset{\rmii{\nr{rho_E}}}{=}}
 & 
 \pi\, \sign(p^{ }_0) \, 
 \delta\bigl(\,
       p^2_{ } + \widetilde\Sigma^\rmii{E}_\P
 \,\bigr)
 \;. \la{rhoE_pole}
\ea
We can integrate over the Dirac-$\delta$'s analytically, however
it needs to be decided which of the three integrals are 
eliminated this way, and in which order. 

Before proceeding, let us determine the integration domain 
in which the pole-pole contribution can get realized. In the 
$(p^{ }_0,q^{ }_0)$-plane, there are 4 disjoint regions
where this could happen. We start with the 
first quadrant, with $p^{ }_0 > p > 0$ and $q^{ }_0 > q > 0$, 
returning to the other quadrants at the end of this section. 
The corresponding ranges are shown in table~\ref{table:domains}.

We then first integrate over $p^{ }_0$. Denoting by 
$p^\rmii{T}_0(p)$ and $p^\rmii{E}_0(p)$ the positive poles
following from \eqs\nr{rhoT_pole} and \nr{rhoE_pole}, respectively, 
and determining the corresponding Jacobians, we get
\ba
 \int_p^{\omega - q} \! {\rm d}p^{ }_0 \, \phi(p^{ }_0)\, \rhoT_\P
 &
 \overset{\rmii{\nr{rhoT_pole}}}{=}
 & 
 \theta( \omega - q - p^{ }_0 )
 \, \phi( p^{ }_0 )
 \, \frac{\pi\, p^{ }_0(p_0^2 - p^2_{ })}
         {\mE^2\, p_0^2 - (p_0^2 - p^2_{ })^2_{ }}
 \bigg|^{ }_{p^{ }_0 \;=\; p^\rmiii{T}_0(p)}
 \;, \la{poleintT_p0} 
 \\[3mm]
 \int_p^{\omega - q} \! {\rm d}p^{ }_0 \, \phi(p^{ }_0)\, \rhoE_\P
 &
 \overset{\rmii{\nr{rhoE_pole}}}{=}
 & 
 \theta( \omega - q - p^{ }_0 )
 \, \phi( p^{ }_0 )
 \, \frac{\pi\, p^{ }_0 (p_0^2 - p^2_{ })}
         {p^2_{ } [ \mE^2  - (p_0^2 - p^2_{ }) ] }
 \bigg|^{ }_{p^{ }_0 \;=\; p^\rmiii{E}_0(p)}
 \;. \la{poleintE_p0}
\ea 

It requires more effort to integrate over the other Dirac-$\delta$.
It is helpful to leave $q^{ }_-$ as the outer integral, 
and instead integrate over $q^{ }_+$, with the domain specified
in table~\ref{table:domains}. While doing so, we have to keep in mind 
that 
$p = q^{ }_+ - q^{ }_-$, 
$q = q^{ }_+ + q^{ }_-$,
$p^{ }_0 = p^\rmii{T}_0(q^{ }_+ - q^{ }_-)$, and
$q^{ }_0 = \omega - p^\rmii{T}_0(q^{ }_+ - q^{ }_-)$, 
are all functions of $q^{ }_+$, and need to be included in the 
computation of the Jacobian. 
We also know that 
$
 q_0^2 - q^2_{ } - \Sigma^\rmii{T}_\Q = 0
$
for all $q^{ }_-$. Differentiating this 
with respect to $q^{ }_-$ yields a useful relation between
partial derivatives of $ \Sigma^\rmii{T}_\Q $, permitting 
for a simplification of the Jacobian. All in all, this yields
\ba
 \int_{{k}/{2}}^{{\omega}/{2}}
 \! {\rm d}q^{ }_+ 
 \, \chi(p,q) 
 \, \rhoT_\Q 
 & = & 
 \theta(\;\exists\;\mbox{pole})
 \, \chi(p,q)
 \, \frac{\pi\, \sign(q^{ }_0) \,  q (q_0^2 - q^2_{ })}
         {2[ \mE^2\, q_0^2 - 3 (q_0^2 - q^2_{ })^2_{ } ]}
 \bigg|^{ }_{q^\rmiii{T}_0(q) \,=\, \omega - p^\rmiii{T/E}_0(p) }
 \;, \hspace*{8mm} \la{poleintT_qp}
 \\[3mm]
 \int_{{k}/{2}}^{{\omega}/{2}}
 \! {\rm d}q^{ }_+ 
 \, \chi(p,q) 
 \, \rhoE_\Q 
 & = & 
 \theta(\;\exists\;\mbox{pole})
 \, \chi(p,q)
 \, \frac{\pi\, \sign(q^{ }_0)
 \, (q_0^2 - q^2_{ }) }{2 q \, [ \mE^2  - 3 (q_0^2 - q^2_{ }) ]}
 \bigg|^{ }_{q^\rmiii{E}_0(q) \,=\, \omega - p^\rmiii{T}_0(p) }
 \;. \hspace*{8mm} \la{poleintE_qp}  
\ea
A simpler avenue leading to the same result 
is to determine the Jacobian first in $p$ and $q$, 
and to go over to $q^{ }_\pm$ only afterwards. 

The remaining challenge is to determine under which conditions
the poles are found. We first recall that 
$p^\rmii{T}_0(p), p^\rmii{E}_0(p) \ge \mE / \sqrt{3}$
(cf.\ \eqs\nr{p0T} and \nr{p0E}). Given that 
\eqs\nr{poleintT_qp} and \nr{poleintE_qp} require that the poles
sum to $\omega$, a pole-pole contribution with 
$p^{ }_0 > 0$ and $q^{ }_0 > 0$ can only exist
for $\omega \ge 2 \mE / \sqrt{3} $. However, for $k > 0$, 
the constraint is stronger, and we need a numerical procedure
for deciding whether it is fullfilled. 

We start by noting that if the constraints following from 
\eqs\nr{poleintT_qp} and \nr{poleintE_qp} are satisfied, which means
that \eq\nr{def_F} has a zero, then it can be shown that 
$
 p^{ }_0
 \; = \; 
 p^\rmii{T/E}_0(p) 
$ 
is necessarily smaller than 
$
 \omega - q 
$.
This means that the $\theta$-constraints in 
\eqs\nr{poleintT_p0} and \nr{poleintE_p0}
are automatically true, 
i.e.\ that the $p^{ }_0$-pole lies
within the integration range, and we do not need
to worry about this further. 

The main task is therefore to resolve the constraint from 
\eqs\nr{poleintT_qp} and \nr{poleintE_qp}. 
This is equivalent to asking whether the function
\be
 \mathcal{F}(q^{ }_-, q^{ }_+) 
 \; \equiv \; 
 p^\rmii{T/E}_0(\overbrace{q^{ }_+ - q^{ }_-}^{p^{ }}) + 
 q^\rmii{T/E}_0(\overbrace{q^{ }_+ + q^{ }_-}^{q^{ }}) - \omega
 \la{def_F}
\ee
crosses zero, as we vary $q^{ }_+$ between the boundary values, 
$k/2$ and $\omega/2$. If not, we set the integrand to zero 
at this value of $q^{ }_-$. If yes, we evaluate the integrand
with the help of \eqs\nr{poleintT_p0}--\nr{poleintE_qp}. 
Subsequently, we integrate over $q^{ }_-$ 
in the range indicated in table~\ref{table:domains}.

In practice, it is wasteful to consider values of $q^{ }_-$
at which the integrand vanishes. This can be 
avoided by first determining the points where the curve defined
by the zeros of $\mathcal{F}$ crosses the borders of our integration
domain ($q^{ }_- = \pm k/2$, $q^{ }_+ = k/2$ or $\omega/2$). From 
the crossing points, 
the non-trivial $q^{ }_-$ range
can be deduced before starting the integration. 

\vspace*{0.3mm}

Returning to the other quadrants, the plasmon in the T channel is 
heavier than that in the E channel, 
i.e.\ $p^\rmii{T}_0(p) > p^\rmii{E}_0(p)$. 
Therefore, for small energies 
$\omega$, there is the possibility of a 
$1\to 2$ decay 
$\mbox{T}\to\mbox{E}+\varphi$. 
Given that the $2\to 1$ processes 
$\mbox{T}+\mbox{T}\to\varphi$ and 
$\mbox{E}+\mbox{T}\to\varphi$ only take place 
when $\omega > 2 \mE^{ }/\sqrt{3}$, 
there is a gap between the $1\to 2$ and $2\to 1$ channels. 

For treating the $\mbox{T}\to\mbox{E}+\varphi$ process, 
it is simpler to stay with the momenta $p,q$ rather than $q^{ }_\pm$
from \eq\nr{qpm}. If we integrate over $q$, the Jacobians from 
\eqs\nr{poleintT_qp} and \nr{poleintE_qp} are larger by a factor~2, 
compensating for the absence of the factor~2 in \eq\nr{qpm}.

Choosing $\Q$ to be the four-momentum of the decaying T, 
the function whose zero we are searching for, now reads
\be
 \mathcal{G}(p, q) 
 \; \equiv \; 
 q^\rmii{T}_0(q) - 
 p^\rmii{E}_0(p) - \omega
 \;. 
 \la{def_G}
\ee
The integration domain in $(p,q)$
is shown in \fig\ref{fig:domains}. Setting $p=0$, 
the starting point of the curve of zeros satisfies 
$q^\rmii{T}_0(q^{ }_\rmi{min}) = \mE^{ }/\sqrt{3} + \omega$.
Given that $q^\rmii{T}_0(0) = \mE^{ }/\sqrt{3}$ and 
$q^\rmii{T}_0(q)$ grows less rapidly than linearly with $q$, 
the value is necessarily in the domain $q^{ }_\rmi{min} > \omega$.
At $q \ge q^{ }_\rmi{min}$, 
the zeros of \eq\nr{def_G} allow us to solve for $p$. 
The solution crosses into the allowed domain when $p > q - k$ and out
of it when $p > q + k$ (cf.\ \fig\ref{fig:domains}). 
At very large $q$ and $p$, the zero of \eq\nr{def_G} is 
at $p = q - \omega < q - k$ (cf.\ \eqs\nr{p0T} and \nr{p0E}), 
so we are out of the domain to the left. 
This means that in general there are two separate 
ranges in which we are inside the allowed domain.
For $\omega\ll \mE^{ }$, 
we illustrate the contribution of the first range in
\eq\nr{IRlimitppfinal}. Otherwise, 
after having determined the integration boundaries, 
numerical integration is straightforward. 

%
\subsection{Pole-cut contribution}
\la{ss:pc}

For the pole-cut contributions, the integration domains can be 
determined from geometric considerations similar to those 
in \ses\ref{ss:cc} and \ref{ss:pp}, which lead to the ranges 
shown in table~\ref{table:domains}. 
There is one Dirac-$\delta$, so that one of the integrals 
(say, 
over $p^{ }_0$, 
after a renaming $\P \leftrightarrow \Q$ whenever necessary) 
can be carried out, similarly 
to \eqs\nr{poleintT_p0} and \nr{poleintE_p0}. 
In analogy to the discussion in \se\ref{ss:pp}, 
the main challenge is that the integrand vanishes in parts
of the integration domain, because the argument of the 
Dirac-$\delta$ does not cross zero, as we vary~$p^{ }_0$
such that $q^{ }_0 = \omega - p^{ }_0$ is within
the cut domain ($|q^{ }_0| < q$). 
For an efficient evaluation, 
we need to determine the reduced integration domain, 
in which the integrand is non-zero.  

Let us view $q^{ }_- \in (-k/2,+k/2)$ as the outermost
integration variable (cf.\ table~\ref{table:domains}).
As we increase $q^{ }_+$, both $q = q^{ }_+ + q^{ }_-$
and $p = q^{ }_+ - q^{ }_-$ increase, with a fixed
difference, $q - p = 2 q^{ }_-$. Asymptotically, 
$p^\rmii{T/E}_0(p) \approx p$ (cf.\ \eqs\nr{p0T} and \nr{p0E}).  
The upper bound of the $p^{ }_0$ integral (originating from the 
requirement
$q^{ }_0 > -q$) is $\omega + q$ (cf.\ table~\ref{table:domains}).
We find
\be
 p^\rmii{T/E}_0(p)
 \;
 \overset{q^{ }_+\;\to\;\infty}{\approx}
 \;
 p
 \;
 =
 \;
 q - 2 q^{ }_-
 \;
 \overset{q^{ }_-\; >\; -k/2}{<} 
 \; 
 q + k 
 \; 
 \overset{k\; < \; \omega}{<}
 \; 
 q + \omega
 \;. 
\ee
Therefore, for large enough $q^{ }_+$, the $p^{ }_0$ pole does fall 
within the cut domain, irrespective of the values of 
$\omega$, $k$, and $q^{ }_-$, 
and the integrand is indeed non-zero. 

In contrast, the low-$q^{ }_+$ domain does not always yield 
a non-vanishing integrand. To determine what happens, we can 
locate the points in the $(q^{ }_+,p^{ }_0)$ plane in which
the pole, $p^\rmii{T/E}_0(p)$, hits one boundary of the 
integration domain (cf.\ table~\ref{table:domains}), 
\begin{align}
 p^\rmii{T/E}_0(\overbrace{q^{ }_+ - q^{ }_-}^{p})
 &\; 
 \underset{\rmii{boundary}}{
 \overset{\rmii{upper}}{=}} 
 \;  
 \omega + (\overbrace{q^{ }_+ + q^{ }_-}^{q})
 \;,
 & 
 \frac{k}{2}
 \;
 < 
 q^{ }_+ 
 & 
 <
 \; 
 \infty
 \;, \\[3mm]
 p^\rmii{T/E}_0(\overbrace{q^{ }_+ - q^{ }_-}^{p})
 &\; 
 \underset{\rmii{boundary}}{
 \overset{\rmii{lower}}{=}} 
 \;  
 \omega - (\overbrace{q^{ }_+ + q^{ }_-}^{q})
 \;,
 & 
 \frac{k}{2}
 \;
 < 
 q^{ }_+ 
 & 
 <
 \; 
 \frac{\omega}{2}
 \;. 
\end{align}
If a non-trivial solution to either equation is found, 
it determines the lower bound
of the $q^{ }_+$ integration. In the discussion leading to \eq\nr{polecutIR},
we show how this can be found analytically for small $\omega/\mE^{ }$.
If no crossing of the boundaries is found, the $q^{ }_+$
integration starts at $k/2$.

%
\subsection{Hints for numerical evaluation}
\la{ss:num}

Though in principle straightforward, 
the numerical evaluation of the integrals from 
\ses\ref{ss:cc}--\ref{ss:pc} is 
non-trivial, because there are weak singularities close
to the integration boundaries, and because the integrand 
contains terms of the type ``$0/0$'', which need to be resolved
for a proper evaluation. Here we describe a few ingredients 
that facilitate these tasks. 

\vspace*{3mm}

The first point is to realize, concretely from table~\ref{table:domains}, 
that if $\omega,k \sim \mE^{ }$, the variables $p^{ }_0$ and $q^{ }_0$
are generically also $p^{ }_0,q^{ }_0 \sim \mE^{ }$
(this is verified more explicitly in \app\ref{se:IR}, where we also 
show how the situation changes when $k, \omega \ll \mE^{ }$). 
Within the weak-coupling expansion, we can then
assume $p^{ }_0, q^{ }_0 \ll \pi T$, and expand the Bose distributions 
in \eq\nr{rho_K} as
\be 
 1 + \nB^{ }(p^{ }_0) + \nB^{ }(q^{ }_0)
 \; 
 \overset{p^{ }_0\,,\,q^{ }_0 \; \ll \; \pi T}{\approx} 
 \; 
 \frac{T}{p^{ }_0} + \frac{T}{q^{ }_0}
 \; 
 \overset{q^{ }_0 \; = \; \omega - p^{ }_0}{=} 
 \; 
 \frac{\omega T}{p^{ }_0 q^{ }_0}
 \;. 
 \la{classical}
\ee
Physically, these leading terms manifest 
Bose enhancement (many soft quanta), and \eq\nr{classical}
is therefore often referred to as the {\em classical approximation}.

The classical approximation can be used to practical benefit, 
after noting that the thermal widths, 
$\Gamma^\rmii{T}_\P$ from \eq\nr{Gamma_T} 
and 
$\widetilde\Gamma^\rmii{E}_\P$ from \eq\nr{Gamma_E}, 
are proportional to $p^{ }_0$.
In the cut contributions, we can therefore analytically
cancel the apparent poles from \eq\nr{classical},  
which makes the numerical evaluation 
faster and more accurate. 

\vspace*{3mm}

Accounting for the pole contributions requires care as well. 
For a general $p$, the poles need to be located numerically, 
however it may be difficult to do this precisely if $p \ll \mE^{ }$
or $p \gg \mE^{ }$. Therefore, the numerical root solving should be
replaced by asymptotic formulas in these domains, 
namely\hspace*{0.3mm}\footnote{%
 We note that the cut-cut, pole-pole, and pole-cut 
 contributions to $\gamma^{ }_\rmii{HTL}(\omega,k)$ 
 generically differ by many orders of magnitude
 (cf.\ \fig\ref{fig:HTL_channels} on p.~\pageref{fig:HTL_channels}).
 If we have $k < \omega$, and consider
 the large-$\omega$ asymptotics, the pole-pole contribution 
 is the leading term. If we want to subtract the large-$\omega$
 asymptotics from $\gamma^{ }_\rmii{HTL}(\omega,k)$
 (cf.\ \se\ref{se:asy}), the leading term 
 needs to be determined with high relative accuracy. 
 To achieve this, many further terms need to be added
 to \eq\nr{p0T}.
 }
\ba
 p^\rmii{T}_0 (p)
 & 
 = 
 & 
 \left\{
  \begin{array}{ll}
  \displaystyle 
  \sqrt{
  \frac{\mE^2}{3} + \frac{6 p^2_{ }}{5} - \frac{81 p^4_{ }}{175\mE^2}
  + \frac{1266 p^6_{ }}{875 \mE^4}
  - \frac{356913 p^8_{ }}{67375 \mE^6} + \ldots 
  }
  & 
  \mbox{for} \quad p \ll \mE^{ }\,,
  \\[3mm]
  \displaystyle 
  \sqrt{
   p^2_{ } + \frac{\mE^2}{2}
  + \frac{\mE^4 \bigl( 2 - \ln^{ }_\rmii{T} \bigr) }{8p^2_{ }}
  + \frac{\mE^6 \bigl( 5 - 6 \ln^{ }_\rmii{T}
  + \ln^2_\rmii{T} \bigr) }{32p^4_{ }} 
  + \ldots
  }
  & 
  \mbox{for} \quad p \gg \mE^{ }\,,
  \end{array}
 \right.
 \hspace*{4mm}
 \nn \la{p0T} \\[3mm]
 p^\rmii{E}_0 (p) 
 & 
 =
 & 
 \left\{
  \begin{array}{ll}
  \displaystyle 
  \sqrt{
  \frac{\mE^2}{3} + \frac{3 p^2_{ }}{5} + \frac{36 p^4_{ }}{175\mE^2}
  - \frac{48 p^6_{ }}{875 \mE^4} + \frac{432 p^8_{ }}{67375 \mE^6} + \ldots 
  }
  & 
  \mbox{for} \quad p \ll \mE^{ }\,,
  \\[3mm]
  \displaystyle 
  p \, \biggl[ 1 + e^{ }_\rmii{E}
  + \frac{e^2_\rmii{E}}{2}\biggl( \frac{4 p^2_{ }}{\mE^2} + 5 \biggr) 
  + \frac{e^3_\rmii{E}}{4}\biggl( \frac{24 p^4_{ }}{\mE^4}
           + \frac{52 p^2_{ }}{\mE^2}  + 29 \biggr) 
   + \ldots 
  \biggr]
  & 
  \mbox{for} \quad p \gg \mE^{ }\,,
  \end{array}
 \right.
 \nn \la{p0E}
\ea
where we have abbreviated 
\be
 \ln^{ }_\rmii{T} 
 \; \equiv \; 
 \ln\biggl( \frac{8 p^2_{ }}{\mE^2 } \biggr)
 \;, \quad
 e^{ }_\rmii{E}
 \; \equiv \; 2 \exp\biggl[ -2 \biggl( \frac{p^2_{ }}{\mE^2} + 1\biggr) \biggr]
 \;. 
\ee

\vspace*{3mm}

At the end of the computation, we present results for the 
interaction rate, $\Upsilon(\omega,k)$, from \eq\nr{Ups_def}. 
For a numerical evaluation, we should identify a dimensionless
quantity from which unnecessary terms have been factored out. 
Pulling out a number of trivial coefficients, 
originating from $c^{ }_\chi$ (cf.\ \eq\nr{c_chi}), 
the Bose distributions 
(cf.\ \eq\nr{classical}), and the overall dimension, 
we re-parametrize $\Upsilon$ through a dimensionless 
coefficient, $\gamma$, as 
\ba
 \Upsilon(\omega,k) 
 & 
 \equiv
 &  
 \frac{(\Nc^2 - 1)\alphas^2 m_\rmiii{E}^2 T}{(4\pi)^3_{ } f^2_\ax } 
 \, 
 \gamma(\omega,k)
 \;. \la{repara}
\ea
In particular, the result of the HTL computation is expressed as 
\ba
 \gamma_\rmii{HTL}^{ }(\omega,k)
 &
  \overset{\rmii{\nr{repara}}}{
  \underset{\rmii{\nr{Ups_def},\nr{c_chi}}}{=}}
 & 
 \frac{\pi}{4 (\Nc^2 - 1) c_\chi^2 m_\rmiii{E}^2 T}
 \,
 \frac{\rho^\rmiii{HTL}_\chi(\K)}{\omega}
 \;. \label{gamma_def}
\ea
Our numerical results will be shown in terms of $\gamma(\omega,k)$
in \figs\ref{fig:chi_htl} and \ref{fig:chi_assembled}.

%
\section{Analytic extrapolation towards $\omega \gg \mE^{ }$}
\la{se:asy}

An essential part of our investigation is to match the HTL
computation, valid for $\omega\sim \mE^{ }$, and lattice results, 
valid for $\omega \ll \mE^{ }$, 
to existing results from the literature, 
valid for $\omega \ge \pi T$. 
In order to achieve this, we need to work out 
the asymptotics of the HTL result for $\omega \gg \mE^{ }$, 
so as to avoid double counting contributions that are already 
included in the $\omega \ge \pi T$ calculations. The main principles and 
practical form of the matching procedure will be presented 
in the paragraph around \eq\nr{gamma_assembled}.
In the current derivation of the asymptotic $\omega \gg \mE^{ }$ limit,
we adopt the classical approximation from \eq\nr{classical}, 
as this is what classical-statistical lattice simulations do
(cf.\ \se\ref{se:lattice}).

Once we adopt the replacement from \eq\nr{classical}, 
\eq\nr{rho_K} is a function of three dimensionful parameters 
($k$, $\omega$ and $\mE^2$). 
If we keep $k/\omega$ fixed and take $\omega \gg \mE^{ }$, we exit
the domain where HTL resummation has had influence. 
In other words, we recover the most IR sensitive 
terms of the unresummed computation. 
The two leading terms in an expansion in 
$\mE^2/\omega^2_{ }$ can be worked out analytically, the leading
being of $\rmO(\omega^2_{ }/ \mE^2)$ and next-to-leading of 
$\rmO(1)$, up to logarithms. Let us anticipate that there is
an ordering-of-limits issue in the sense that the asymptotics
is different if we take $(\omega - k)/\mE^{ }$ large or keep 
it fixed, when we send $\omega \gg \mE^{ }$.

\vspace*{0.3mm}

For a practical computation, it is helpful to envisage that rather
than keeping $\mE^{ }$ fixed and making $\omega$ large, we keep $\omega$
fixed and make $\mE^{ }$ small. This way, we immediately see that 
the cut contributions are small, since 
$\Gamma^\rmii{T}_\P$ and $\widetilde\Gamma^\rmii{E}_\P$
are proportional to $\mE^2$ 
(cf.\ \eqs\nr{Gamma_T} and \nr{Gamma_E}, respectively). 
Furthermore, as $\widetilde\Pi^\rmii{E}_\P\to 0$, 
the spectral function 
$\rhoE_\P$ from 
\eq\nr{rho_E} vanishes. Referring to these as 
the {\em Born limit}, we thus have  
\ba
 \rhoTB_\P
 &
 \overset{\rmii{\nr{rho_T}}}{
 \underset{m_\rmiii{E}^{ }\; \to \; 0}{\equiv}}  
 &
 \im\biggl[ \frac{1}{-(p^{ }_0 + i 0^+_{ })^2 + p^2_{ }} \biggr]
 \; 
 = 
 \;
 \pi\,\sign(p^{ }_0)\,\delta(\P^2_{ }) 
 \;, \la{rho_T_0}
 \\[3mm]
 \rhoEB_\P
 &
 \overset{\rmii{\nr{rho_E}}}{
 \underset{m_\rmiii{E}^{ }\; \to \; 0}{\equiv}}  
 &
 0
 \;. \la{rho_E_0}
\ea
The corresponding limit of $\rho^\rmii{HTL}_\chi$
is denoted by $\rho^\rmii{Born}_\chi$. 

Computing the pole-pole contribution in the Born limit
(which is not entirely trivial despite the simple outcome), and 
normalizing the result according to \eq\nr{repara}, we find
\be
 \gamma^{ }_\rmii{Born}(\omega,k) 
 \; 
 \overset{\omega\;\ge\; k \vphantom{\big|}}{=} 
 \; 
 \frac{(\omega^2_{ } - k^2_{ })^2_{ }}{2\omega k \hspace*{0.3mm} \mE^2}
 \ln \biggl( \frac{\omega + k }{\omega - k} \biggr)
 \;. \la{gamma_Born}
\ee
The result vanishes on the lightcone, but is finite
and non-vanishing for $k\to 0$, with
$\gamma^{ }_\rmiii{Born}(\omega,0) = \omega^2_{ }/\mE^2$.
A numerical evaluation of
$
 \gamma^{ }_\rmii{Born}(\omega,k)
$
is shown in \fig\ref{fig:chi_htl}(left). 

\vspace*{3mm}

It is considerably more difficult to work out the next term 
in the expansion. We write
\be
 \gamma^{ }_\rmii{HTL}(\omega,k)
 \; \equiv \; 
 \gamma^{ }_\rmii{Born}(\omega,k) 
 +  
 \Delta\gamma(\omega,k)
 \;, \la{Delta_gamma} 
\ee
and denote the next term as 
\be
 \gamma^{ }_\rmi{asy}(\omega,k) 
 \; \equiv \; 
 \lim^{ }_{m^{ }_\rmiii{E}\;\to\;0}
 \Delta\gamma(\omega,k)
 \;, \la{gamma_asy} 
\ee
where it is understood that possible logarithmic dependences
on $\mE^{ }$ are retained. The subscript stands for ``asymptotic'', 
signalling that $\gamma^{ }_\rmi{asy}$ gives the limiting value at
$\omega \gg \mE^{ }$, with the would-be 
higher terms containing a positive
power of $\mE^{ }$ and therefore, for a fixed $k/\omega$,
a negative power of $\omega$.
The term
$
 \gamma^{ }_\rmi{asy}
$
is of next-to-leading order (NLO) from two separate 
perspectives, namely as the 
NLO term in the expansion of $\gamma^{ }_\rmii{HTL}$ 
in $\mE^2/\omega^2_{ }$, and as the IR limit of the  
NLO evaluation of  
$\rho^{ }_\chi$ in the domain $k,\omega \ge \pi T$.
For $k = 0$, these computations yield the remarkably simple result 
$ \gamma^{ }_\rmi{asy} (\omega,0) = 1$~\cite[eq.~(5.15)]{Bulk_wdep}.

\vspace*{3mm}

We now proceed to how the result for
$\gamma^{ }_\rmi{asy}(\omega,k)$ can be extended to $k > 0$.
Writing 
$\rho^\rmii{HTL}_\chi = \rho^\rmii{Born}_\chi + \Delta \rho^\rmii{HTL}_\chi$
in analogy with \eq\nr{Delta_gamma}, and similarly for the individual spectral
functions $\rhoT_\P$ and $\rhoE_\P$; 
making use of the symmetry 
$\Q\leftrightarrow \P$ to factor out $\varrho_\Q^\rmii{T,Born}$ and  
$\widetilde\varrho_\Q^{\hspace*{0.3mm}\rmii{E,Born}}$;  
and inserting for them the forms from \eqs\nr{rho_T_0}
and \nr{rho_E_0}, 
we then find 
\begin{eqnarray}
 \Delta\rho^\rmii{HTL}_\chi(\K)
 & 
 \approx 
 & 
 \frac{ 2 (\Nc^2 - 1)\, c_\chi^2 }{\pi^3_{ }k} \! 
    \int_0^\infty \! {\rm d}p \, p 
    \int_{|p-k|}^{p+k} \! {\rm d}q \, q  
 \int_{-\infty}^{\infty}
 \! {\rm d} p^{ }_0 
 \, \frac{\omega T}{p^{ }_0 q^{ }_0}
 \,\pi\,\sign(q^{ }_0)\,\delta(\Q^2)^{ }_{q^{ }_0 \,=\, \omega - p^{ }_0}
 \hspace*{5mm}
 \nn[3mm]
 & \times & 
 \biggl\{\; 
 2 \, \Delta \rhoE_\P \,\bigl[\, k^2_{ } - (p-q)^2_{ } \,\bigr]
 \bigl[\, (p+q)^2_{ } - k^2_{ } \,\bigr]
 \label{rho_K_NLO}
 \\[3mm]
 &  & \; +\, 2\,\Delta \rhoT_\P \,  
 \biggl[\,
 \biggl( \frac{p_0^2}{p^2_{ }} + \frac{q_0^2}{q^2_{ }} \biggr) 
 \bigl[\,
  \bigl( p^2_{ } + q^2_{ } - k^2_{ } \bigr)^2_{ } 
 + 
  4 p^2_{ } q^2_{ }
 \,\bigr]
 + 8 p^{ }_0 q^{ }_0 
   \bigl( p^2_{ } + q^2_{ } - k^2_{ } \bigr) 
 \,\biggr] \, 
 \;\biggr\}
 \;.  \nonumber 
\end{eqnarray}
We can use \eq\eqref{gamma_def} to obtain $\Delta\gamma$. 
We also integrate the $\delta$-function over $q$, to find
\begin{eqnarray}
\Delta\gamma^\rmii{ }(\omega,k)
 & 
 \approx
 & 
 \frac{ 1  }{4 \pi_{ } m_\rmiii{E}^2\hspace*{0.3mm}k} \! 
    \int_0^\infty \! {\rm d}p \, p 
 \int_{-\infty}^{\infty}
 \! {\rm d} p^{ }_0 
 \, \frac{\theta(p+k-\vert \omega - p^{ }_0\vert)
          \theta(\vert \omega - p^{ }_0\vert-\vert p -k\vert)}
  {p^{ }_0 \vert \omega - p^{ }_0\vert}
 \hspace*{5mm}
 \nn[3mm]
 & \times & 
 \biggl\{\; 
 2 \, \Delta \rhoE_\P
   \, \Bigl[\,-(\K^2-\P^2)^2-4(\omega-p_0)
 \bigl(\,\omega\P^2-p_0\K^2\,\bigr) \,\Bigr]
 \nn[3mm]
 &  & \; +2\,\Delta\rhoT_\P \,  
  \biggl[\,(\K^2-\P^2)^2+\frac{\bigl[\K^2p^{ }_0
  -\P^2(2\omega-p^{ }_0)\bigr]^2}{p^2}
 \,\biggr]
 \;\biggr\}
 \;. \label{oneloopexpmd2p}
\end{eqnarray}
The integration
domain is illustrated in \fig\ref{fig:asy}.
We remark that rewriting the coefficients 
as in \eq\nr{oneloopexpmd2p}
may take some effort to verify, and of course
the representation is not unique. 

\begin{figure}[t]

\hspace*{-0.1cm}
\centerline{%
 \epsfysize=7.3cm\epsfbox{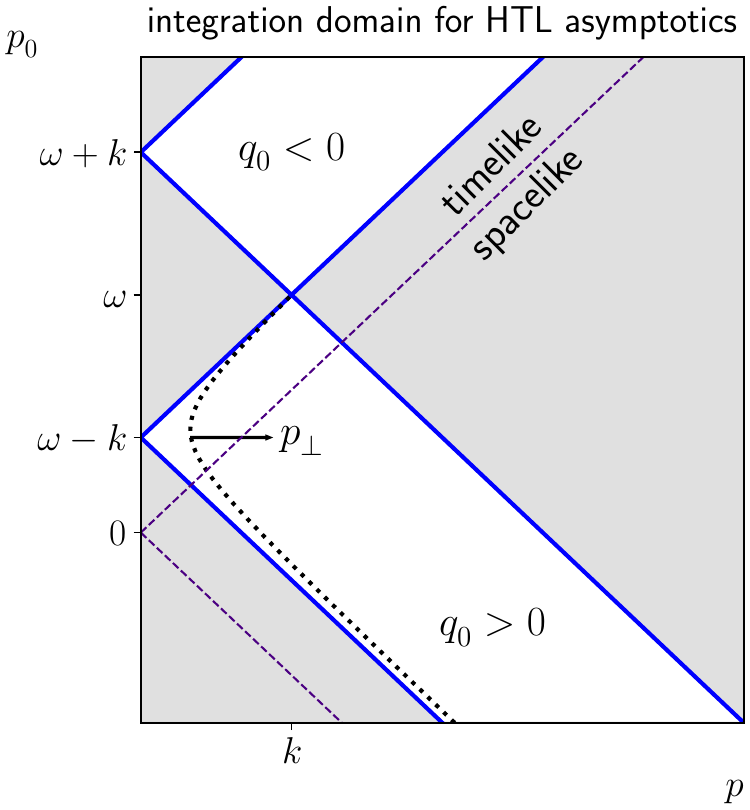}%
}

\vspace*{-4mm}

\caption[a]{\small
  The integration domain for \eqs\nr{oneloopexpmd2p}
  and \nr{oneloopexpmdperp}. The meanings of the domains
  and of the dotted line, 
  indicating a contour of constant $p^{ }_\perp$,
  are explained between \eqs\nr{oneloopexpmd2p}
  and \nr{oneloopexpmdperp}.
  }

\la{fig:asy}
\end{figure}

In order to connect \fig\ref{fig:asy}
with the terminology in \tabl\ref{table:domains} on 
p.~\pageref{table:domains}, 
we note that $\delta(\Q^2_{ })$
in \eq\nr{rho_K_NLO} sets $\Q$ on a pole. For $\P$, 
\fig\ref{fig:asy} shows that we could either have  a pole (``timelike'')
or a cut (``spacelike''). The domain with $q^{ }_0 < 0$ 
corresponds to the third pole-pole process
in \tabl\ref{table:domains}. 
As explained
at the end of \se\ref{ss:pp}, it can only contribute at small $\omega$. 
The large-$\omega$ asymptotics therefore
originates from the pole-pole and pole-cut domains with $q^{ }_0 > 0$.

Focussing thus on $q^{ }_0 > 0$, 
it is helpful to substitute variables as 
\be
 p^2_{ }
 \; \equiv \; 
 p_\perp^2\frac{\omega-p_0^{ }}{k}+(p_0^{ }-2k_-^{ })^2_{ }
 \;, \quad
 {\rm d}p\, p 
 \; = \; 
 {\rm d}p^{ }_\perp\, p^{ }_\perp \, \frac{\omega-p_0^{ }}{k}
 \;, \quad
 k^{ }_\pm 
 \; \equiv \; 
 \frac{\omega\pm k}{2}
 \;. \la{new_vars}
\ee 
The geometrical meaning of a constant-$p^{ }_\perp$
contour is illustrated in \fig\ref{fig:asy}; for $k\approx \omega$, 
it represents transverse momentum with respect to lightlike propagation.
With these variables, and after some algebra, 
the $q^{ }_0 > 0$ part
of \eq\nr{oneloopexpmd2p} can be expressed as 
\begin{eqnarray}
\Delta\gamma^\rmii{ }(\omega,k)
 & 
 \approx
 & 
\frac{1}{m_\rmiii{E}^2\, k^2} \! 
 \int_0^{2k} \! {\rm d}p_\perp \, p_\perp
 \int_{-\infty}^{\omega}
 \! \frac{ {\rm d} p^{ }_0 }{ 2 \pi_{ } } 
 \, \frac{1}{p^{ }_0 }
 \biggl\{\; 
  \Delta \rhoE_\P 
  \,\frac{(\omega-p^{ }_0 )^2 \, p_\perp^2
  \, (\, 4k^2 -p_\perp^2 \,)}{k^2}
 \nn[3mm]
 &  & \; +\Delta\rhoT_\P \,  
 \biggl[\,(\K^2-\P^2)^2
 +\frac{\bigl[\K^2p^{ }_0-\P^2(2\omega-p^{ }_0)\bigr]^2}{p^2}
 \,\biggr] \, 
 \;\biggr\}
 \;. 
 \label{oneloopexpmdperp}
\end{eqnarray}

We can now carry out the integral over $p^{ }_0$ with the residue theorem. 
Since the integral is convergent and we are interested in the large-$\omega$
limit, the upper end of the $p^{ }_0$-integral 
can be extended to infinity, with 
an error suppressed by $\sim \mE^2/\omega^2_{ }$.
%
%
We express the spectral functions as half-differences of retarded 
and advanced functions,
\begin{align}
  2\Delta \rhoT_\P 
 \;& = \; 
 \biggl[ 
   \frac{i}{\P^2_{ } - \Pi^\rmii{T}_\P} - \frac{i}{\P^2_{ }}
 \biggr]^{ }_{p_0^{ }\,+\,i 0^+_{ }}
 -
 \biggl[ 
 \frac{i}{\P^2_{ } - \Pi^\rmii{T}_\P}  - \frac{i}{\P^2_{ }} 
 \biggr]^{ }_{p_0^{ }\,-\,i 0^+_{ }}
 \;, \la{rho_expl_T} \\[2mm] 
 2\Delta \rhoE_\P 
 \; & = \;
 \biggl[  
 \frac{i}{p^2_{ } + \widetilde\Pi^\rmii{E}_\P} - \frac{i}{p^2_{ }}
 \biggr]^{ }_{p_0^{ }\,+\,i0^+_{ }}
 -
 \biggl[ 
 \frac{i}{p^2_{ } + \widetilde\Pi^\rmii{E}_\P}  - \frac{i}{p^2_{ }}
 \biggr]^{ }_{p_0^{ }\,-\,i 0^+_{ }}
 \;.\la{rho_expl_E}
\end{align}
After the change of variables in \eq\nr{new_vars}, 
the subtractions in \eq\nr{rho_expl_E} yield non-trivial functions. 
The integration contours are illustrated 
in \fig\ref{fig:contours}, and amount algebraically to
\be
 \int_{-\infty}^{\infty} \! \frac{{\rm d}p^{ }_0}{2\pi} \,
 \frac{i}{p^{ }_0} \, 
 \bigl[
  \phi^{ }_\rmii{R}(p^{ }_0 + i 0^+_{ }) 
       - 
  \phi^{ }_\rmii{R}(p^{ }_0 - i 0^+_{ }) 
 \bigr]
  = 
  - 
 \hspace*{-3mm}
 \overbrace{
 \phi^{ }_\rmii{R}(0)
 }^{\rm zero~mode}
 \hspace*{-3mm}
 \,-  
 \overbrace{
 \sum
 \mbox{Res}\biggl[ \frac{\phi^{ }_\rmiii{R}(p^{ }_0)}{p^{ }_0} \biggr]
 }^{\rm other~poles}
  +  
 \overbrace{
 \lim_{|p^{ }_0|\to\infty}\phi^{ }_\rmii{R}(p^{ }_0)
 }^{\rm arcs} 
 \;. \la{residue}
\ee
The analytic structures appearing in $\phi^{ }_\rmii{R}$, 
related to the physics of causality, were explained 
in ref.~\cite{CaronHuot:2008ni}, who also pointed out
that the contributions of the ``other poles'' cancel between
the T and E channels. The role that the arcs
play was worked out in 
refs.~\cite{Ghiglieri:2013gia,Ghiglieri:2015ala}.
The procedure is frequently referred 
to as ``light-cone sum rules''~\cite{aurenche,db_sum}.

\begin{figure}[t]

\hspace*{-0.1cm}
\centerline{%
 \epsfysize=5.3cm\epsfbox{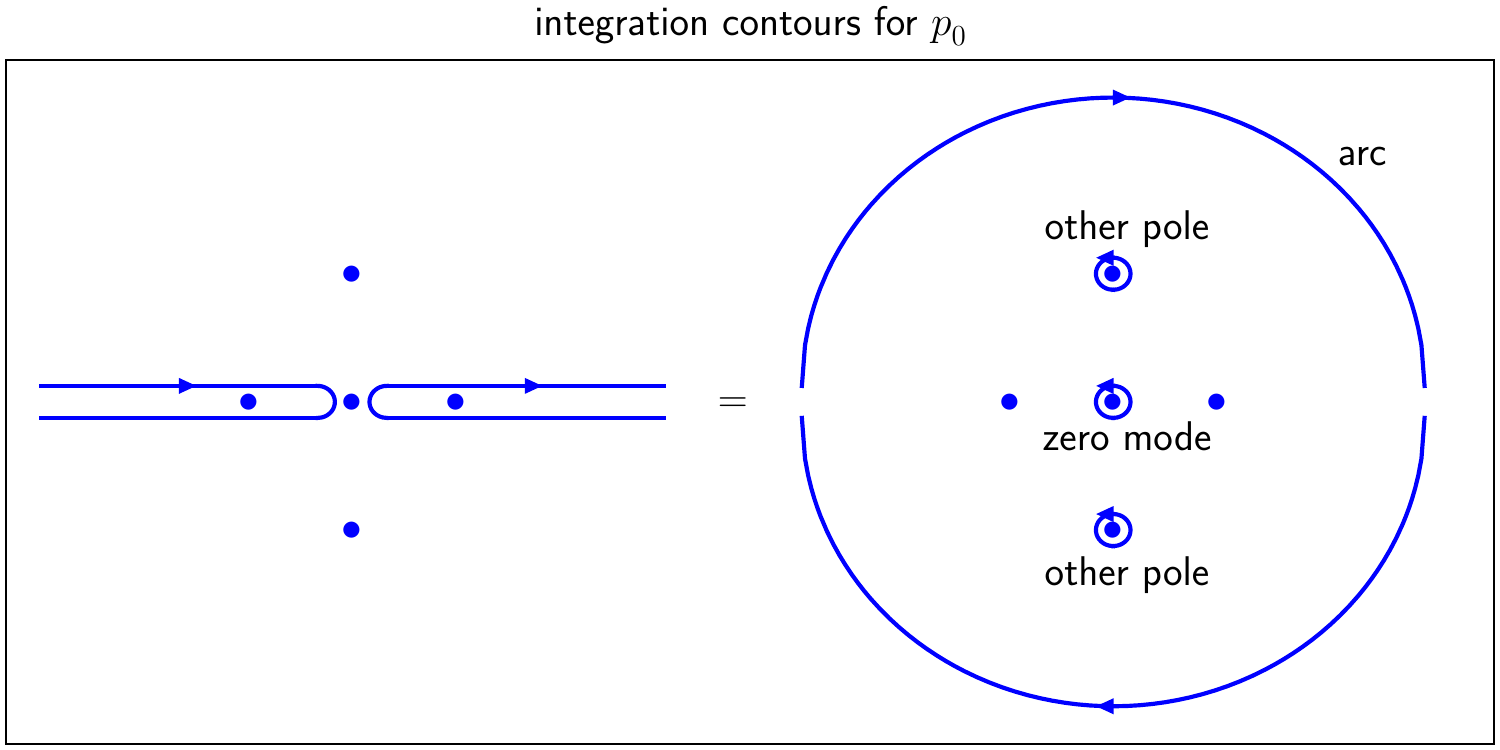}%
}

\caption[a]{\small
  The integration contours corresponding  
  to \eq\nr{residue}, leading to \eq\nr{oneloopexpmdperpeucl}.
  }

\la{fig:contours}
\end{figure}

Collecting together the contributions 
of the poles and arcs, and noting
that a few terms drop out, either because of their 
antisymmetry in $p^{ }_0$, or because of the mentioned
cancellation of the poles 
originating from $p^2_{ } = 0$ 
between the E and T parts, we obtain
\begin{eqnarray}
 \Delta\gamma^\rmii{ }(\omega,k)
 & 
 \approx
 & 
 \frac{1}{2 m_\rmiii{E}^2 k^2} \! 
 \int_0^{2k} \! {\rm d}p_\perp \, p_\perp
 \label{oneloopexpmdperpeucl} \\[3mm]
 & \times & 
 \Biggl\{\; 
 \frac{p_\perp^2\,\bigl(\, 4k^2 -p_\perp^2\,\bigr)}{k^2} 
 \bigg[\overbrace{\frac{-\omega^2}{\frac{\omega}{k} p_\perp^2+4k_-^2+\mE^2}}
  ^{\text{E zero mode}}
 +\underbrace{1}_\text{E arc}
 -\bigg(\overbrace{\frac{-\omega^2}{\frac{\omega}{k} p_\perp^2+4k_-^2}}
  ^{\text{E zero mode Born}}
 +\underbrace{1}_\text{E arc Born}\bigg)\bigg]
 \nn[3mm]
 &  &+
  \overbrace{\frac{-\omega^2\,p_\perp^2\,\bigl(\, 4k^2 -p_\perp^2\,\bigr)}
  {k^2\bigl(\frac{\omega}{k} p_\perp^2+4k_-^2\bigr)}
 +8 \omega^2}^{\text{T zero mode}}
 \,\underbrace{-4p_\perp^2 +\frac{p_\perp^4}{k^2}
- 2 \left(\frac{\omega}{k}  p_\perp^2 + 4k_-^2\right)  - 
 4\K^2+ \mE^2}_\text{T arc}\nn[1mm]
 & &- \bigg[ \overbrace{\frac{-\omega^2\,p_\perp^2
  \,\bigl(\, 4k^2 -p_\perp^2\,\bigr)}
  {k^2\bigl(\frac{\omega}{k} p_\perp^2+4k_-^2\bigr)}
 +8 \omega^2}^{\text{T zero mode Born}}
 \,\underbrace{-4p_\perp^2 +\frac{p_\perp^4}{k^2} 
 - 2 \left(\frac{\omega}{k}  p_\perp^2 + 4k_-^2\right)  - 
 4\K^2}_\text{T arc Born}\bigg]
 \, 
 \;\Biggr\}
 \;. \nonumber 
\end{eqnarray}
%
%
Adding everything together the expression
gets greatly simplified, yielding finally
\begin{eqnarray}
 && \hspace*{-1.3cm}
 \Delta\gamma^\rmii{}(\omega,k)
 \; 
 \approx
 \;
\frac{1}{2 k^2} \! 
 \int_0^{2k} \! {\rm d}p_\perp \, p_\perp
 \biggl\{ \; 
 \frac{p_\perp^2\,\bigl(\, 4k^2 -p_\perp^2\,\bigr)}
 {\bigl[ p_\perp^2+\frac{k}{\omega}( 4k_-^2 + \mE^2 )\bigr]\,
 \bigl[ p_\perp^2+ \frac{k}{\omega} ( 4k_-^2) \bigr]}
 +1
 \;\biggr\}
 \nn[3mm]
 & = &  \frac{1}{4 m_\rmiii{E}^2\omega k}
 \bigg\{\,
 \mE^2\bigl(4k_+^2+4k_-^2+\mE^2\bigr)
 \ln\frac{4k_+^2+\mE^2}{4k_-^2+\mE^2}
 +
  (\omega^2_{ } - k^2_{ })^2_{ }
 \ln\frac{k_-^2[4k_+^2+\mE^2]}{k_+^2[4k_-^2+\mE^2]}
 \,\bigg\}\;.\nn
 &&
 \label{oneloopexpmdfinal}
\end{eqnarray}

\begin{figure}[t]

  \begin{center}
    \includegraphics[width=7.5cm]{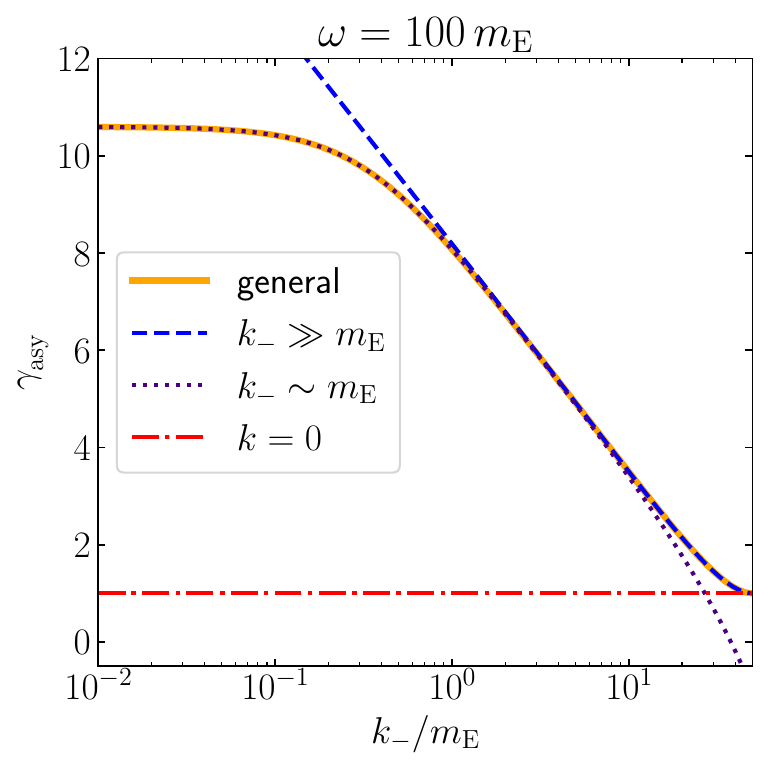}
  \end{center}

  \vspace*{-3mm}

  \caption[a]{\small 
  Comparison of $\gamma_\rmi{asy}$ 
  from \eq\eqref{gammaasyresumhack}, denoted by ``general'',  
  with the first two approximations  
  from \eq\eqref{gamma_asy_cases}, denoted by 
  $k^{ }_- \gg \mE^{ }$ and $k^{ }_- \sim \mE^{ }$, 
  respectively. 
  The line $k=0$ gives $\gamma^{ }_\rmi{asy}(\omega,0) = 1$.
  The variable $k^{ }_-\equiv(\omega-k)/2$ runs from 0 (at $k=\omega$)
  to $\omega/2$ (at $k=0$).}

  \label{fig:g_asy}

\end{figure}

Subsequently, we are interested in the asymptotics, given by 
\eq\nr{gamma_asy}. However, if $k^{ }_- = (\omega - k)/2$ is smaller 
that $\mE^{ }$, the limit cannot be taken literally, because of 
logarithmic singularities in \eq\nr{oneloopexpmdfinal}. Concretely, 
we find 
\be
 \gamma^{ }_\rmi{asy}(\omega,k) 
 \; \approx \; 
 \left\{ 
 \begin{array}{ll} 
 \displaystyle
 \frac{\omega^2+k^2}{\omega k}\ln\frac{\omega+k}{\omega-k}-1
 & 
 \displaystyle
 \;,\quad 
 k^{ }_- \gg \mE^{ } \,
 \\[4mm]
 \displaystyle
   \ln \frac{4 k_+^2}{4 k_-^2+\mE^2}
  -\frac{4 k_-^2}{\mE^2} \ln \left(1+\frac{\mE^2}{4 k_-^2}\right)
 & 
 \displaystyle
 \;,\quad 
 k^{ }_- \sim \mE^{ } \,
 \\[4mm]
 \displaystyle 
 \ln \frac{4 k^2_{ }}{\mE^2}
 & 
 \displaystyle
 \;,\quad 
 k^{ }_- \ll \mE^{ } \,
 \end{array}
 \right. 
 \la{gamma_asy_cases}
\ee 
The first line reproduces 
$\gamma^{ }_\rmi{asy}(\omega,0) = 1$~\cite{Bulk_wdep}.
All cases can be represented simultaneously with 
\begin{equation}
 \gamma_\rmi{asy}^{ }(\omega,k) 
 \; \approx \; 
 \frac{\omega^2+k^2}{2\omega k}
 \ln \frac{(\omega+k)^2+\mE^2}{(\omega-k)^2+\mE^2}
 -\frac{(\omega-k)^2}{\mE^2} 
  \ln \biggl[ 1+\frac{\mE^2}{(\omega-k)^2} \biggr]
   \;,
   \label{gammaasyresumhack}
\end{equation}
which for $\omega\gg \mE$ is in excellent agreement 
with \eq\eqref{oneloopexpmdfinal} for all $k$. The result
from \eq\nr{gammaasyresumhack}, together with its various
limiting values from \eq\nr{gamma_asy_cases}, is illustrated
in \fig\ref{fig:g_asy}.

%
%

\vspace*{3mm} 

In summary, for $\omega\gg\mE$, 
the coefficient $\gamma^{ }_\rmii{HTL}$ from 
\eq\nr{gamma_def} 
is approximated 
by a Born term, given in \eq\eqref{gamma_Born}, 
and its first correction $\gamma_\rmi{asy}$, 
given in \eq\eqref{gammaasyresumhack}. 
If $k < \omega$, the Born term dominates, 
and $\gamma_\rmi{asy}$ 
can be determined without resummations,
as visible on the first line of \eq\eqref{gamma_asy_cases}. 
However, if we go close to lightcone, 
whereby $k^{ }_- = (\omega - k)/2$ becomes 
smaller than $\mE^{ }$, the behaviour changes. 
The Born term becomes smaller than~$\gamma_\rmi{asy}^{ }$, 
which develops a logarithmic enhancement compared with its value
at $k=0$
(cf.\ \fig\ref{fig:g_asy}).

%
\section{Lattice results at $\omega \ll \mE^{ }$}
\la{se:lattice}

The HTL computation presented in 
\se\ref{se:htl} is valid for 
$k \le \omega \sim \mE^{ }
 \sim \sqrt{\alphas^{ }\pi\vphantom{|}}\hspace*{0.3mm} T$, 
which is called the soft domain. 
If we decrease $\omega$ and $k$ to values $\le \alphas^{ }T$, 
which is called the ultrasoft domain, 
non-Abelian plasma fluctuations are  
expected to become non-perturbative, even 
at temperatures where $\alphas^{ } \ll 1$~\cite{linde}.
We therefore need lattice methods. 
The quantity that is often addressed is 
called the {\em strong sphaleron rate} or the 
{\em Chern-Simons diffusion rate},
\ba
 \Gamma^{ }_\rmi{sph}
 & \equiv & 
 \lim_{\omega\to 0^+_{ }}
 \int_{\X} 
 e^{i \omega t }_{ } 
 \, 
 \biggl\langle\,
 \frac{1}{2}
 \bigl\{ \,
 \hat\chi(\X),\hat\chi(0) 
 \,\bigr\} 
 \,\biggr\rangle
 \; = \;
 \lim_{\omega\to 0^+_{ }} \frac{2 T \rho^{ }_\chi(\omega,\vec{0})}{\omega}
 \;. 
 \la{Gamma_sph}
\ea
Combining with \eqs\nr{Ups_def} and \nr{gamma_def}, we find that 
\ba
 \Upsilon_\rmi{sph}
 \; 
 \equiv 
 \;
 \lim_{\omega\to 0^+_{ } } \Upsilon(\omega,0)
 & 
 \underset{\rmii{\nr{Gamma_sph}}}{
 \overset{\rmii{\nr{Ups_def}}}{=}}
 & 
 \frac{T^3_{ }}{2 f_\ax^2}\,
 \times
 \frac{\Gamma^{ }_\rmi{sph}}{T^4_{ } }
 \;,
  \la{Ups_sph}
 \\[3mm]
 \lim_{\omega\to 0^+_{ } } \gamma(\omega,0)
 &
   \overset{\rmii{\nr{c_chi},\nr{gamma_def}}}
 {\underset{\rmii{\nr{Gamma_sph}}}{=}} 
 &  
 \frac{32 \pi^3_{ } T^2_{ }}{(\Nc^2 - 1) \alphas^2 m_\rmiii{E}^2 }
 \,
 \times
 \frac{ \Gamma^{ }_\rmi{sph} }{T^4_{ }}
 \;. \la{Gamma_vs_gamma}
\ea

The strong sphaleron rate has been investigated with two types of 
lattice methods. First, there are attempts at determining
it with full quantum-statistical
4d lattice simulations~\cite{eucl1,eucl2,eucl3} (LQCD). 
It should be kept in mind, however, the 4d lattice simulations
operate in imaginary time, and their analytic continuation to 
Minkowskian signature is not a numerically robust procedure. 
Effectively, any measurement represents an average over the 
true spectral function (cf.\ \fig\ref{fig:kinematics}), 
so that narrow features, such as transport peaks, are missed. If 
transport peaks happen to be present, the lattice measurement
likely yields an {\em underestimate} of the true value. With 
these reservations, the data from ref.~\cite{eucl3} are
shown in table~\ref{table:lattice},
together
with perturbative values for $\alphas^{ }$ and $\mE^{ }$
from ref.~\cite{mE2}, permitting for the use of 
\eq\nr{Gamma_vs_gamma}, through which we compare lattice
and perturbative results.

%
\begin{table}[t]

\vspace*{-3mm}

{\fontsize{9pt}{11pt}\selectfont
$$
\begin{array}{|c|c|c|c|} 
  \hline %
  & & & \\[-3mm]
  T \, / \,\mbox{GeV} &
  \Gamma^{ }_\rmii{sph} \, / \, T^4_{ } \hspace*{2mm} 
  [\mbox{\href{https://arxiv.org/abs/2308.01287}{2308.01287}}] & 
  \alphas^{ } \hspace*{2mm}
  [\mbox{\href{https://arxiv.org/abs/1911.09123}{1911.09123}}] & 
  m^{ }_\rmiii{E} \, / \, T \hspace*{2mm} 
  [\mbox{\href{https://arxiv.org/abs/1911.09123}{1911.09123}}]
 \\[1mm]
  \hline %
  & & & \\[-3mm]
 0.230 & 0.310 \pm          0.080 & 0.359 & 2.79 \\ 
 0.300 & 0.165 \pm          0.049 & 0.311 & 2.55 \\
 0.365 & 0.115 \pm          0.030 & 0.284 & 2.41 \\ 
 0.430 & 0.065 \pm          0.020 & 0.265 & 2.32 \\ 
 0.570 & 0.045 \pm          0.012 & 0.237 & 2.20 
   \\[1mm]
  \hline
\end{array}
$$
}

\vspace*{-3mm}

\caption[a]{\small
 Lattice data for the strong sphaleron rate, 
 from ref.~\cite{eucl3}, 
 and a perturbatively estimated effective
 coupling constant ($\alphas^{ }$)
 and Debye mass ($\mE^{ }$)
 at the same temperatures,
 from ref.~\cite{mE2}. 
 The perturbative values are needed for converting
 lattice results to the same units as our HTL computation
 (cf.\ \eqs\nr{Gamma_vs_gamma} and \nr{clgt}).
 To do this in a numerically smooth way, we have
 given more digits than are accurately known. 
 In the same spirit, 
 we remark that the effects of the charm quark 
 have been included in the perturbative values
 of $\alphas^{ }$ and $\mE^{ }$, but not in the lattice data.
 }

 \la{table:lattice}
\end{table}
%

The second set of lattice measurements only works in the 
regime $\alphas^{ }\ll 1$, and is aimed at resolving the
non-perturbative physics that is still present in this 
domain~\cite{linde}. The great benefit of this 
{\em classical lattice gauge theory} (CLGT) approach, with ``classical'' 
referring to the same Bose enhancement that was mentioned
in \eq\nr{classical}, is that it works directly in Minkowskian
signature. However, there is also a problem, which is that the framework
is non-renormalizable, and the interpretation of the numerical 
result close to the continuum limit requires an analytic 
understanding of lattice artifacts, which is made technically 
challenging by the spatial discretization~\cite{pba}. In any case, 
a measurement of the strong sphaleron rate was 
presented in ref.~\cite{mt}
(with two different formulations), 
and an extension to finite 
$\omega > 0$ in ref.~\cite{clgt}.
In terms of $\Delta\gamma$ from \eq\nr{Delta_gamma}, 
the result of ref.~\cite{clgt} can be expressed as 
\ba
   \Delta \gamma^{ }_\rmii{CLGT}(\omega,0)
 &
 \overset{\omega \;\le\; \mE^{ }}{\simeq}
 & 
 \kappa \, 
 \frac{ (4\pi \alphas^{ } \Nc^{ })^3_{ }\, T^{\hspace*{0.3mm}2}_{ }}{\mE^2 }
 \frac{ 1
   + \Bigl( \frac{\omega  \vphantom{\big|} }
                 {c^{ }_\rmiii{IR} \alphas^2 \Nc^2 T \vphantom{\big|} }
      \Bigr)^2_{ } }
      { 1 
   + \Bigl( \frac{\omega  \vphantom{\big|} }
                 {c^{ }_\rmiii{M} \alphas^{ } \Nc^{ } T  \vphantom{\big|} }
      \Bigr)^2_{ } }
 \;, 
 \la{clgt} 
 \\[3mm]
 && 
 \kappa \; \simeq \; 1.5
 \;, \quad
 c^{ }_\rmii{IR} \; \simeq \; 106 
 \;, \quad
 c^{ }_\rmii{M}  \; \simeq \; 5.1
 \;. \la{kappa}
\ea
Though $\kappa$ depends
logarithmically on $\alphas^{ }$~\cite{db0}, the simulations
in ref.~\cite{clgt} 
could not resolve this, due to the 
large lattice discretization effects mentioned above.
Moreover, the functional form of \eq\nr{clgt} was not derived
theoretically in ref.~\cite{clgt}, 
but rather came out as an 
empirically successful representation of the data.  
In retrospect, the large-$\omega$ limit of \eq\nr{clgt}
corresponds to $\gamma^{ }_\rmi{asy}(\omega,0)$ 
(cf.\ \se\ref{se:asy}). The value gets probably
moderately modified on the spatial lattice, an effect that 
could be studied with the methods of ref.~\cite{pba}, but we have
not done this, adjusting the behaviour rather 
to the continuum expression (cf.\ \se\ref{se:assembled}).

\begin{figure}[t]

\hspace*{-0.1cm}
\centerline{%
 \epsfysize=7.3cm\epsfbox{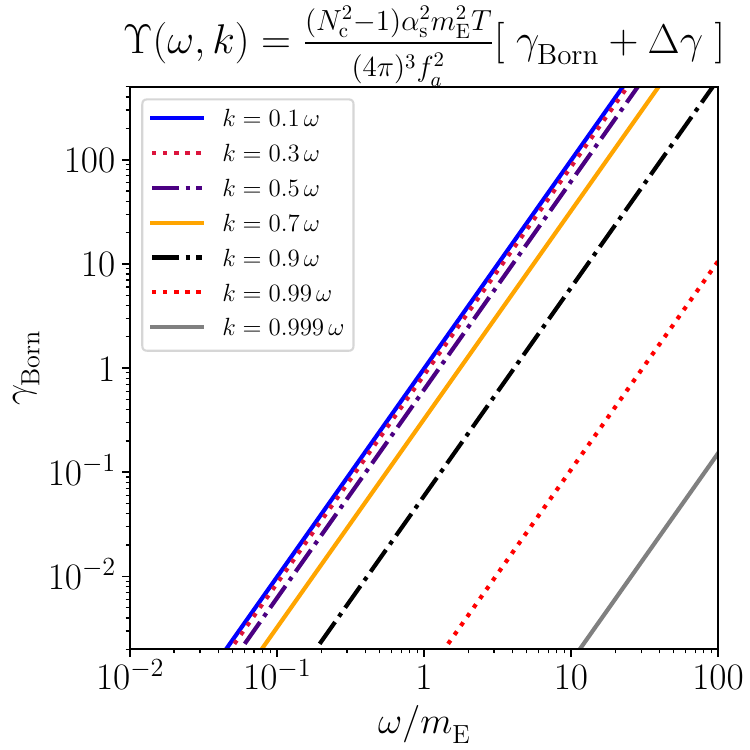}%
 \hspace{0.4cm}%
 \epsfysize=7.3cm\epsfbox{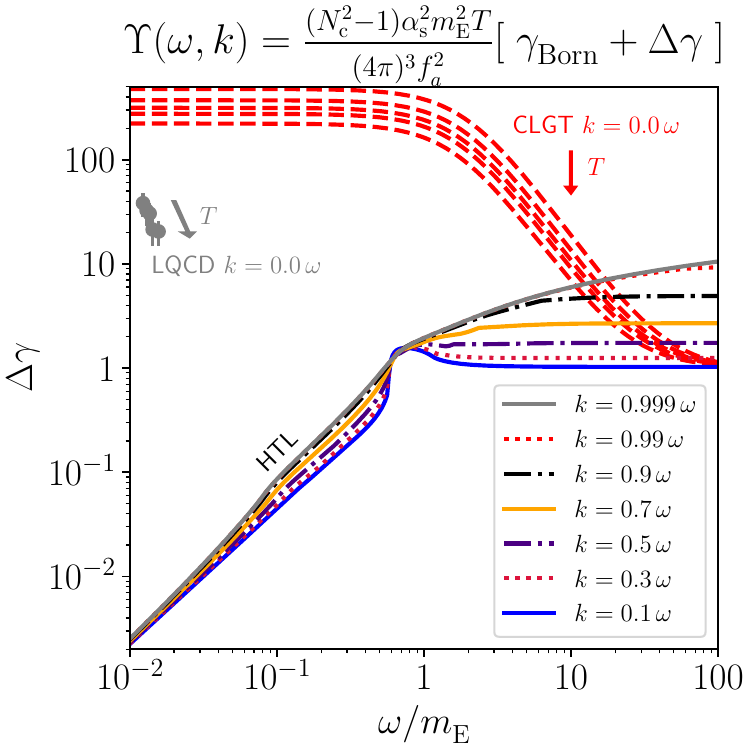}%
}

\caption[a]{\small
  Left: the Born approximation from \eq\nr{gamma_Born}, 
  $\gamma^{ }_\rmii{Born}$. 
  If $k < \omega$, $\gamma^{ }_\rmii{Born}$
  dominates the result at $\omega \gg \mE^{ }$. 
  Right: 
  the remainder beyond $\gamma^{ }_\rmii{Born}$, 
  denoted by $\Delta\gamma$, 
  compared with  
  real-time classical (CLGT)
  and
  full 4d (LQCD) 
  lattice results.
  The CLGT and LQCD sets correspond to  
  a few separate temperatures,
  as shown in table~\ref{table:lattice}.
  The LQCD data points are for $k = \omega = 0$, 
  and have been slightly 
  displaced for better visibility. 
  The approximately 
  constant value of $\Delta\gamma$ at $\omega\gg \mE^{ }$
  amounts to the term $\gamma^{ }_\rmii{asy}$, 
  determined in \se\ref{se:asy}. 
  This plot suggests that at $k < \omega$,  
  non-perturbative phenomena dominate until  
  $\gamma^{ }_\rmii{Born}$ takes over,
  but at $k \approx \omega$, 
  it is $\Delta\gamma$ which takes over.  
  However, lattice data only 
  exist at $k = 0$, and in order to consolidate the picture, 
  it would be valuable to have CLGT 
  data at $k > 0$, verifying in particular the expected
  logarithmic increase of the large-$\omega$ asymptotics with $k$.
  }

\la{fig:chi_htl}
\end{figure}

A comparison of the lattice and perturbative results for $\Delta\gamma$ 
is shown in \fig\ref{fig:chi_htl}(right), with a conversion between the 
two sets based on table~\ref{table:lattice}. The Born result is displayed 
separately in \fig\ref{fig:chi_htl}(left), to 
illustrate the domain in which it dominates. 

\vspace*{3mm}

Despite the uncertainties associated with the lattice results, 
the overall message from \fig\ref{fig:chi_htl} is clear: 
if $\omega \ll \mE^{ }$,
the interaction rate is probably substantially larger than 
the HTL prediction. Perhaps surprisingly,  
this agrees on the qualitative level
with the findings
of ref.~\cite{mainz}, who found that in the same domain, 
loop momenta $p,q \ge \pi T$, beyond those captured by the HTL theory,
play a role, 
and that accounting for them increases the result. 
That said, as we recall 
in \app\ref{se:IR}, loop-wise expanded or naively resummed
perturbation theory is not self-consistent in the domain $\omega \ll \mE^{ }$. 

\section{Update on axion contribution to $\Delta N^{ }_\rmi{eff}$}
\la{se:assembled}

In \ses\ref{se:htl}--\ref{se:lattice} 
and in \app\ref{se:hard}, we have provided information about the axion 
interaction rate, parametrized by $\gamma$ via \eq\nr{repara},  
obtained with different methods and applicable 
in different kinematic domains. 
Here, we show how this 
information can be assembled together. 
Having in mind the most common physical
application, to light QCD axions, we stay on the lightcone in the 
present section, setting $\omega = k$ and denoting
\be
 \gamma(k) \; \equiv \; \gamma(k,k)
 \;. \la{lc}
\ee
We recall that on the lightcone, the Born approximation, 
$\gamma^{ }_\rmi{Born}$ from \eq\nr{gamma_Born}, drops out. 

\vspace*{3mm}

We factorize the determination
of $\gamma$ into parts. By $\gamma^{ }_\rmi{soft}$ we denote 
a computation valid for $\omega \le \mE^{ }$, notably through the 
HTL theory (\ses\ref{se:htl} and \ref{se:asy}) or lattice methods 
(\se\ref{se:lattice}). It would be proper to refer to the 
lattice contribution as arising from {\em ultrasoft} scales, however
the scales are numerically not well separated, and we combine them
into a single {\em soft} term.  
By $\gamma^{ }_\rmi{hard}$ we denote the 
{\em hard} result, valid for $\omega\ge \pi T$ (cf.\ \app\ref{se:hard}). 
Let us summarize the ingredients that are at our disposal, 
starting from $\gamma^{ }_\US$ and $\gamma^{ }_\IR$.

%
\paragraph{(i) Lattice for the ultrasoft domain.}

We have discussed two types of lattice determinations in 
\se\ref{se:lattice}, classical-statistical (CLGT) and 
quantum-statistical (LQCD). 
Even though both employed $k=0$, 
we assume in the following that the 
extrapolation to $\omega\to 0$
is the same along the axis $k=\omega$. The reasoning is that  
non-perturbative dynamics gives a finite correlation length 
to thermal fluctuations~\cite{linde}, and furthermore $\chi$
is not related to conserved currents, whereby it should not couple
to hydrodynamic modes. We thus expect
$G^\rmii{R}_\chi$ to have no 
singularity at $k,\omega = 0$, an assertion also backed 
by the AdS/CFT correspondence~\cite{cs3}.

We have argued that, due to its inherent
nature of taking an average over the true spectral function, LQCD may
underestimate the sphaleron rate, $\Gamma^{ }_\rmi{sph}$. 
In contrast, CLGT only works at weak coupling, $\alphas^{ }\ll 1$.
Its extrapolation to large $\alphas^{ }$ likely yields an overestimate, 
because the result comes with a high power of~$\alphas^{ }$. 
To be conservative, 
we therefore multiply the CLGT result with a fudge factor, 
$\phi^{ }_\kappa \approx 0.1$, chosen so that the CLGT result
agrees with LQCD at low temperatures 
(cf.\ \fig\ref{fig:chi_htl}(right)). Otherwise we keep 
the CLGT parametrization, so that  
\be
 \lim_{k\to 0} \gamma^{ }_\US(k)
 \; \overset{\rmii{\nr{clgt}}}{\equiv} \; 
 \phi^{ }_\kappa \,\kappa\,
 \frac{ (4\pi \alphas^{ } \Nc^{ })^3_{ }\, T^{\hspace*{0.3mm}2}_{ }}{\mE^2 }
 \;, \quad
 \phi^{ }_\kappa \; \approx \; 0.1 
 \;, \quad
 \kappa 
 \; \overset{\rmii{\nr{kappa}}}{\simeq} \;
 1.5
 \;. \la{gamma_IR_small}
\ee
This implies that the decrease of 
$
  \gamma^{ }_\US(0)
$
with increasing temperature 
is determined by perturbative running,
as it is visible in \tabl\ref{table:lattice}
on p.~\pageref{table:lattice}
($\alphas^3$ decreases faster than $\mE^2/T^2_{ }$). 

%
\paragraph{(ii) HTL asymptotics for the soft domain.}

In the context of \fig\ref{fig:chi_htl}(right), we have argued that 
even though the HTL result for $\Delta\gamma$
is overtaken by non-perturbative effects at 
$\omega \ll \mE^{ }$, it comes to dominate 
in the parametric domain 
$\mE^{ } \ll \omega \ll \pi T$. This domain is narrow
in full QCD, but it is broad in CLGT, from which the scale $\pi T$
is absent. The crossover to HTL asymptotics 
is clearly visible in the CLGT
lattice data in  \fig\ref{fig:chi_htl}(right), albeit for $k=0$. The 
asymptotic value of the HTL result, 
$ \gamma^{ }_\rmi{asy} $, 
was determined in \eq\nr{gamma_asy_cases}. 
We therefore assert 
that $\gamma^{ }_\rmi{soft}$ asymptotes to 
$ \gamma^{ }_\rmi{asy} $, 
\be
 \lim_{k\to k^{ }_\rmii{max}} \gamma^{ }_\IR(k)
 \; \equiv \; 
 \gamma^{ }_\rmi{asy}(k^{ }_\rmii{max})
 \;, \la{gamma_IR_large}
\ee
where 
$
 k^{ }_\rmi{max}  
$
needs to be larger than 
the ultrasoft scales visible in \eq\nr{clgt}. 
In practice, we take 
$
 k^{ }_\rmi{max} \simeq 5 \max\{
 c^{ }_\rmiii{IR} \alphas^2 \Nc^2 T, 
 c^{ }_\rmiii{M} \alphas^{ } \Nc^{ } T 
 \}
$. 

%
\paragraph{(iii) Interpolation for the soft domain.}

In order to interpolate between \eqs\nr{gamma_IR_small} and 
\nr{gamma_IR_large}, we adopt the form suggested by CLGT data, 
from \eq\nr{clgt}, but now rotated to the lightcone. 
In order to incorporate the modified limiting values, 
we multiply all fit coefficients with fudge factors, so that
\be
 \gamma^{ }_\IR(k)
 \;
 \overset{\rmii{\nr{clgt}}}{\simeq}
 \; 
 \left\{ 
 \begin{array}{ll}
 \displaystyle
 \phi^{ }_\kappa \kappa \, 
 \frac{ (4\pi \alphas^{ } \Nc^{ })^3_{ }\, T^{\hspace*{0.3mm}2}_{ }}{\mE^2 }
 \frac{ 1
   + \Bigl( \frac{k  \vphantom{\big|} }
                 {\phi^{ }_\rmiii{IR}
                  c^{ }_\rmiii{IR} \alphas^2 \Nc^2 T \vphantom{\big|} }
      \Bigr)^2_{ } }
      { 1 
   + \Bigl( \frac{k  \vphantom{\big|} }
                 {\phi^{ }_\rmiii{M}
                  c^{ }_\rmiii{M} \alphas^{ } \Nc^{ } T  \vphantom{\big|} }
      \Bigr)^2_{ } }
 & 
 \displaystyle 
 , \quad k \; < \; k^{ }_\rmi{max} \;, 
 \\[3mm]
 \displaystyle 
 \gamma^{ }_\rmi{asy}(k)
 & 
 \displaystyle 
 , \quad k \; \ge \; k^{ }_\rmi{max} \;. 
 \end{array}
 \right.
 \la{gamma_IR}
\ee
In order to satisfy \eq\nr{gamma_IR_large}, the fudge factors need to 
be related by 
\be
 \frac{ \phi^{ }_\rmii{IR} }{ \phi^{ }_\rmii{M} } 
 \; \approx \; 
 \frac{4\pi c^{ }_\rmiii{M}}{c^{ }_\rmiii{IR}}
 \biggl[\, 
   \frac{\phi^{ }_\kappa \kappa}
        {\gamma^\rmii{ }_\rmii{asy}(k^{ }_\rmii{max})}
   \frac{4\pi \alphas^{ } \Nc^{ } T^2}{m_\rmiii{E}^2}
 \,\biggr]^{1/2}_{ }
 \;. \la{phi_IR}
\ee
We vary $\phi^{ }_\rmii{M}$ within the 
range 0.01...10.0, thereby 
allowing for the ultrasoft domain to be 
narrower or broader 
than suggested by the $k=0$ CLGT data. 
The function $\gamma^{ }_\rmi{soft}(k)$ 
that follows from this procedure is illustrated
in \fig\ref{fig:chi_assembled}(left), where it is also 
compared with the full HTL result from \se\ref{se:htl}.

%
\paragraph{(iv) Combination of the soft and hard domains.}

\begin{figure}[t]

\hspace*{-0.1cm}
\centerline{%
 \epsfysize=7.3cm\epsfbox{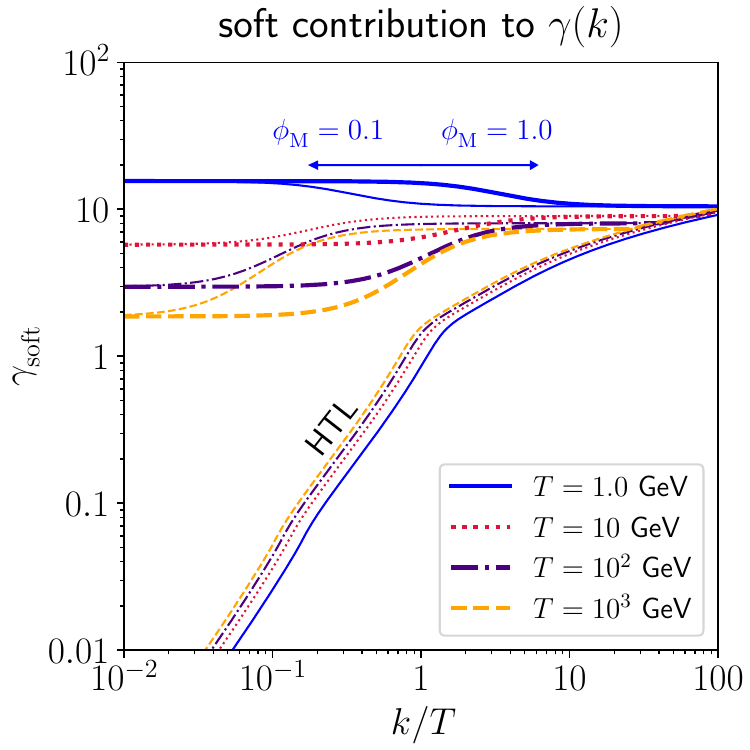}%
 \hspace{0.4cm}%
 \epsfysize=7.3cm\epsfbox{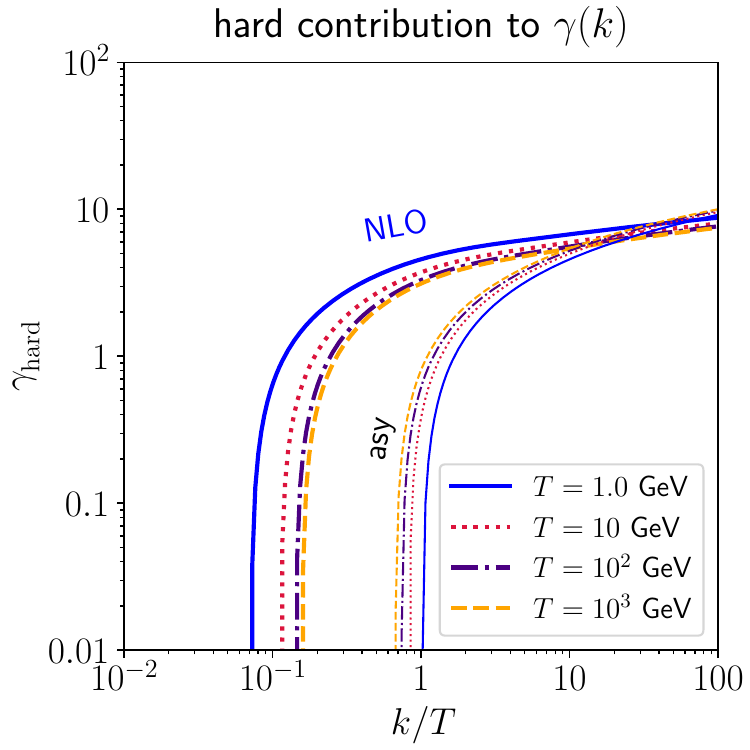}%
}

\vspace*{-4mm}

\caption[a]{\small
  Left: the soft contribution to $\gamma$,
  as described by \eq\nr{gamma_IR}. For comparison 
  we also show the purely perturbative HTL result 
  from \se\ref{se:htl}.
  Right: the NLO hard contribution to $\gamma$, 
  as described by \eq\nr{gamma_UV}. 
  With ``asy'' we show the subtraction 
  needed for \nr{gamma_assembled}.  
  For $k \gg \pi T$, the NLO result 
  for $\gamma_\rmi{hard}$ can be larger or smaller than 
  $\gamma_\rmi{asy}$ (cf.\ \fig\ref{fig:hardrate} on 
  p.~\pageref{fig:hardrate}), 
  whereas its extrapolation 
  to $k \ll \pi T$ yields a positive contribution
  on top of $\gamma_\rmi{asy}$
  (cf.\ \eq\nr{NLO_hard}).
  While the latter does {\em not} amount to a theoretically
  consistent computation in this domain, it nevertheless 
  appears to ``anticipate'' the emergence of
  the positive ultrasoft contribution, though it displays 
  a different parametric dependence. We note
  that in numerical evaluations of $\gamma_\rmi{hard}$,
  we adopt the variable-$\Nf$ scheme from ref.~\cite{bg}.
  }

\la{fig:chi_assembled}
\end{figure}

The NLO result for the domain of hard momenta~\cite{bg} 
is explained in \app\ref{se:hard}
(cf.\ \eqs\nr{LO_hard} and \nr{NLO_hard}), and is now denoted by
\be 
 \gamma^{ }_\UV(k) 
 \; \equiv \; 
 \gamma^\rmii{strict}_\rmii{hard,NLO}(k) 
 \;. \la{gamma_UV}
\ee
As our overall interpolation, we then assemble
\eqs\nr{gamma_IR} and \nr{gamma_UV} together, as 
\be
  \gamma^{ }_\rmi{full}(k)
  \; \equiv  \;
  \gamma^{ }_\IR(k)
 -\gamma^{ }_\rmi{asy}(k) 
 +\gamma^{ }_\UV(k)
 \;. \la{gamma_assembled}
\ee
The function $\gamma_\rmi{asy}(k) = \ln(4 k^2_{ }/\mE^2)$ 
is present both on the soft (cf.\ \eq\nr{gamma_asy_cases}) 
and 
the hard (cf.\ \eq\nr{LO_hard}) side, and 
needs to be subtracted to avoid double counting.
In an effective theory language, the difference 
$\gamma_\rmi{hard} - \gamma_\rmi{asy}$ is a Wilson or 
matching coefficient, accounting for the hard contribution, 
which needs to be added to the soft contribution.  
We remark that \eq\nr{gamma_assembled} is a generalization 
of the procedure introduced in ref.~\cite{Bulk_wdep}, 
but there the functional form of the subtraction was non-singular,
$
 (\gamma^{ }_\rmii{Born} + \gamma^{ }_\rmi{asy})(\omega,0) =
 \omega^2_{ } / \mE^2 + 1
$. 
The functions $\gamma_\rmi{hard}$ and $\gamma_\rmi{asy}$
are illustrated
in \fig\ref{fig:chi_assembled}(right), 
and the assembled result 
$\gamma_\rmi{full}$ in 
\ref{fig:chi_Y}(left). 

%
\begin{figure}[t]

\hspace*{-0.1cm}
\centerline{%
 \epsfysize=7.3cm\epsfbox{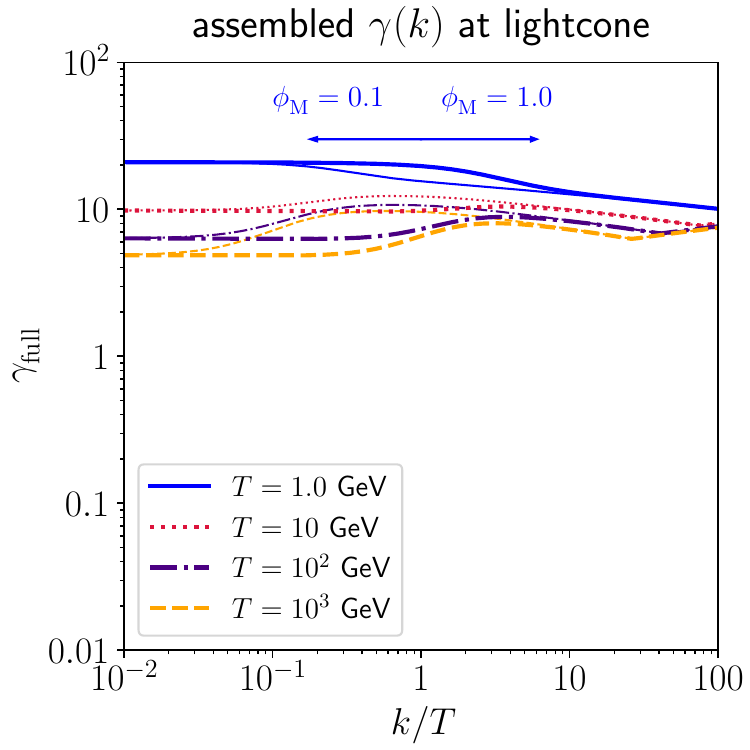}%
 \hspace{0.4cm}%
 \epsfysize=7.3cm\epsfbox{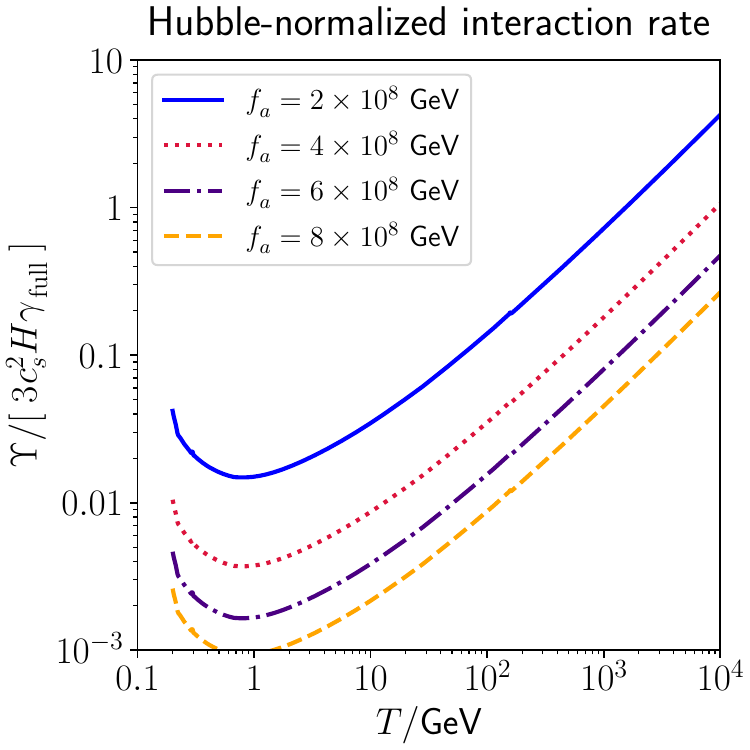}%
}

\caption[a]{\small
  Left: 
  the full interaction rate,
  as described by
  \eq\nr{gamma_assembled}. 
  For $\alphas^{ }$
  we use the running coupling from
  \tabl\ref{table:lattice} on p.~\pageref{table:lattice}. 
  Right: the $k$-independent coefficient 
  $\Upsilon(k) / [ 3\hspace*{0.3mm}c_s^2 H \gamma^{ }_\rmii{full}(k) ]$
  from \eq\nr{def_Y}, for typical values of $f^{ }_\ax$.
  The cosmologically relevant Hubble-normalized rate,
  $\Upsilon(k) / ( 3\hspace*{0.3mm}c_s^2 H )$, 
  is given by the product
  of the left and right panels (cf.\ \eq\nr{kinetic2}), 
  and exceeds unity at $T = 10^4_{ }$~GeV.
  }

 \la{fig:chi_Y}
\end{figure}
%

%
\paragraph{(v) Rewriting of the kinetic equation.}

Having constructed $\gamma^{ }_\rmi{full}(k)$
for all momenta, we can insert it 
into \eq\nr{repara}, to obtain $\Upsilon(k)$; 
and then to \eq\nr{kinetic}, to integrate for $f^{ }_\varphi(t,k)$. 
Moving along a trajectory of comoving momentum, 
$
 k(t) = k(t^{ }_\now)\, a(t^{ }_\now) / a(t)
$, 
the left-hand side turns into an ordinary time derivative. 
It is convenient to replace time through
temperature as the integration variable, so we define
\be
 x \; \equiv \; \ln\biggl( \frac{T^{ }_\iini }{T} \biggr)
 \;, \quad
 \frac{{\rm d}}{{\rm d}t}
 \; = \; 
 3 \hspace*{0.3mm} c_s^2 H \, 
 \frac{{\rm d}}{{\rm d}x}
 \;, \la{def_x}
\ee 
where $T^{ }_\ini$ is an initial (high) temperature, chosen
so that $\Upsilon \gg 3 \hspace*{0.3mm} c_s^2 H$; 
$c_s^2$ is the speed of sound squared;  
and we made use of the identity   
$
 {{\rm d}T}/{{\rm d}t}
  = 
 - 3 \hspace*{0.3mm} c_s^2 T  H
$,
derived from Friedmann equations and 
thermodynamic identities. Thereby \eq\nr{kinetic} becomes
\be
 \frac{{\rm d}f^{ }_\varphi(t^{ }_x,k^{ }_x)}{{\rm d}x}
 \; \approx \;
 \frac{\Upsilon}{3 \hspace*{0.3mm}c_s^2 H}
 \, \bigl(
  \nB^{ } - f^{ }_\varphi 
 \bigr) 
 \,\biggr|^{ }_{k \;=\; k^{ }_x\,,\; T \; = \; T^{ }_x}
 \;, \la{kinetic2}
\ee
where $k^{ }_x \equiv k^{ }_\now \hspace*{0.3mm} a^{ }_\now/a^{ }_x$ 
denotes a redshifting momentum, 
and $T^{ }_x$ the temperature for a given~$x$,
\be
 \frac{k^{ }_x}{T^{ }_x}
 \; = \; 
 \frac{k^{ }_\inow}{T^{ }_\inow}
 \biggl( 
         \frac{h^{ }_{*,x}}
              {h^{ }_{*,\inow}}
 \biggr)^{1/3}_{ } 
 \;, \quad
 T^{ }_x
 \; 
 \overset{\rmii{\nr{def_x}}}{\equiv}
 \; 
 T^{ }_\ini \, e^{-x}_{ }
 \;. \la{a}
\ee
The function $h^{ }_*$ parametrizes the overall expansion of 
the universe through an effective (non-equilibrium) 
entropy density, with the low-temperature limit
$h^{ }_{*,\now} \approx 3.930$~\cite{mea}.

As \eq\nr{kinetic2} shows, 
the nature of the dynamics depends on the ratio
$
 \Upsilon / (3\hspace*{0.3mm}c_s^2 H)
$.
If 
$
 \Upsilon \gg 3\hspace*{0.3mm}c_s^2 H
$, 
axions are in equilibrium, 
and the solution is given by 
$f^{ }_\varphi(t^{ }_\ini,k) \approx \nB^{ }(k)$.
If 
$
 \Upsilon \ll 3\hspace*{0.3mm}c_s^2 H
$, 
the right-hand side of \eq\nr{kinetic2} vanishes, 
and axions free-stream. 
Making use of \eq\nr{repara}, together with
the Hubble rate,  
$
 H =  
 \sqrt{{8\pi e}/({3 \mpl^2})}
$, 
where 
$\mpl^{ } = 1.2209 \times 10^{19}_{ }\,\mbox{GeV}$ is the Planck mass, 
we can express the ratio as 
\be
 \frac{\Upsilon(k^{ }_x)}{3 \hspace*{0.3mm}c_s^2 H}
 \; = \;  
 \frac{(\Nc^2 - 1) \hspace*{0.3mm}
       \alphas^2  \hspace*{0.3mm}
       m_\rmiii{E}^2  \hspace*{0.3mm}
       \mpl   \hspace*{0.3mm}
       T}
      {(4\pi)^3_{ } 
      f^2_\ax   \hspace*{0.3mm}
      c_s^2  \hspace*{0.3mm}
      \sqrt{24\pi e}}
 \, \gamma^{ }_\rmi{full}(k^{ }_x)
 \;. \la{def_Y}
\ee
There is dependence on $k^{ }_x$ only through
$\gamma^{ }_\rmi{full}(k^{ }_x)$. The $k^{ }_x$-independent  
coefficient is plotted in \fig\ref{fig:chi_Y}(right), 
after inserting thermodynamic functions from ref.~\cite{eos15},\footnote{%
 The axion contribution has not been included in the thermodynamic
 functions, which yields an 
 uncertainty on the $\sim 1\%$ level. To improve on the precision, 
 a numerically inexpensive possibility would be to solve the 
 equations iteratively, including the axion contribution 
 to the Hubble rate and other required quantities, 
 once they are available from the zeroth-order solution.
 However, this is not necessary at our current resolution.    
 } 
and restricting to a range of $f^{ }_\ax$ close to the phenomenological
lower bound $f^{ }_\ax \gsim 4 \times 10^8_{ }$~GeV~\cite{exp}
(however we note that this bound originates from fermionic pseudoscalar
operators rather than directly \eq\nr{L}, and the relations
between the operators depend on the UV completion of the theory).
We see that in the chosen $f^{ }_\ax$-range, axions are in 
equilibrium at $T \ge 10^4_{ }$~GeV, but start falling out of it
at lower temperatures, until perhaps around the QCD crossover, 
where stronger interactions tend to pull them closer to 
equilibrium again. 

%
\begin{figure}[t]

\hspace*{-0.1cm}
\centerline{%
 \epsfxsize=7.3cm\epsfbox{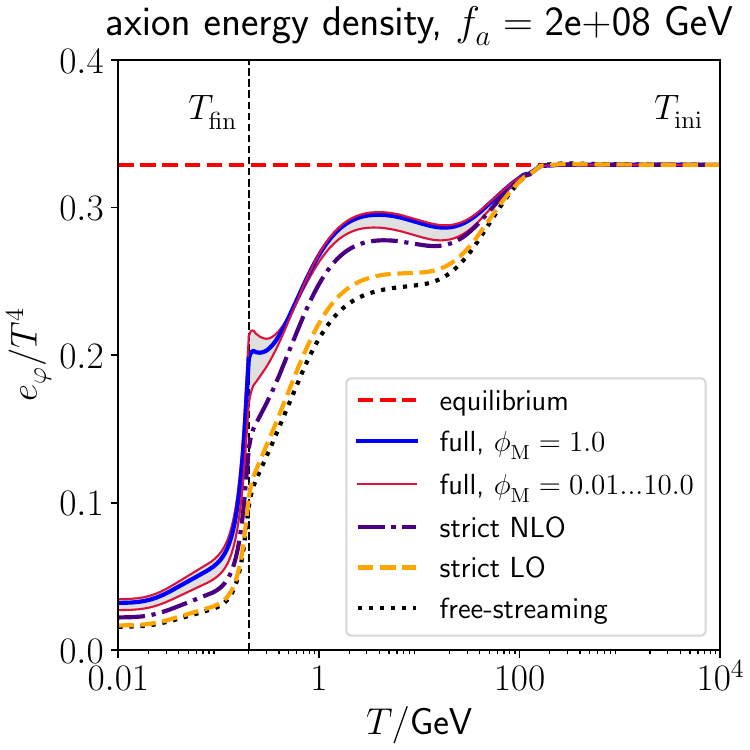}%
 \hspace{0.4cm}%
 \epsfxsize=7.3cm\epsfbox{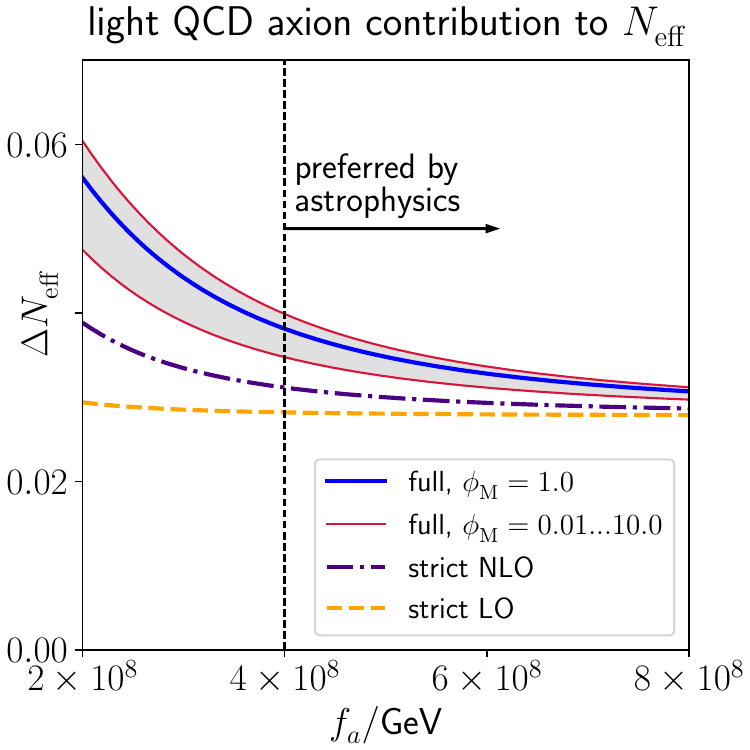}%
}

\caption[a]{\small
  Left: 
  the evolution of $e^{ }_\varphi/T^4_{ }$, obtained 
  from \eqs\nr{kinetic2}, \nr{e_varphi} and, 
  for $T < T_\rmii{fin}$, \nr{Neff}.
  An upper bound
  is given by the equilibrium value,
  $\pi^2_{ }/30 \approx 0.329$
  (dashed red line), 
  whereas a lower bound is given by the free-streaming value,
  $\pi^2_{ }/30\, (h_{*,T}/h_{*,\iini})^{4/3}_{ }$ 
  (black dotted line).
  The larger the interaction rate,
  the more the system attempts to equilibrate. 
  Right: 
  the axion contribution to $\Delta N^{ }_\rmii{eff}$, 
  as given by \eq\nr{Neff}. The strict LO and NLO results
  correspond to those obtained in ref.~\cite{bg}
  (albeit now evolved down to $T^{ }_\ifin = 200$~MeV rather 
  than 300~MeV), and the 
  astrophysical bound originates from ref.~\cite{exp}. The ``full'' result refers to \eq\nr{gamma_assembled}, and 
  $\phi^{ }_\rmii{M}$ is defined in \eq\nr{phi_IR}.
  }

 \la{fig:Neff}
\end{figure}
%

%
\paragraph{(vi) Determination of $\Delta N^{ }_\rmi{eff}$~.}

After having fixed the parameters, we integrate \eq\nr{kinetic2}
from an initial temperature, $T_\ini \equiv 10^4_{ }$~GeV, 
where $f_\varphi(t_\ini,k) \approx \nB^{ }(k)$ $\forall k$
as argued above, 
until a final temperature, $T_\fin \equiv 200$~MeV, 
below which we lack the data to handle the QCD crossover 
reliably. 
At $T_\fin$, the axion energy density reads
\be
 e^{ }_{\varphi,\fin}
 \; = \;  
 \int_0^\infty 
 \! 
 \frac{ 
 {\rm d} k^{ }_{ } k^3_{ }
 }{2\pi^2_{ }}
 \, 
 f^{ }_\varphi(t^{ }_\fin,k^{ }_{ })
 \;. \la{e_varphi}
\ee
Subsequently, assuming that axions are effectively 
massless down to the recombination temperature, 
i.e.\ $m^{ }_\ax \ll 0.3~$eV, 
we free-stream $e^{ }_\varphi$ with 
$
 (a^{ }_\fin/a^{ }_\now)^4_{ }
 = 
 (T^{ }_\now/T^{ }_\fin)^4_{ }
 (h^{ }_{*,\now}/h^{ }_{*,\fin})^{4/3}_{ }
$, 
and extract $\Delta N^{ }_\rmi{eff}$ by comparing
the low-temperature limit with the photon energy density, 
\be
 \Delta N^{ }_\rmi{eff} 
 \; \equiv \; 
 \frac{8}{7}\,
 \biggl( \frac{11}{4}\biggr)^{4/3}_{ }\,
 \frac{e^{ }_{\varphi,\inow}/T^4_\inow}{2 \pi^2 / 30}
 \;, \quad 
 \frac{e^{ }_{\varphi,\inow}}{T^4_\inow}
 \; = \; 
 \frac{e^{ }_{\varphi,\ifin}}{T^4_\ifin}
 \biggl( 
         \frac{h^{ }_{*,\inow}}
              {h^{ }_{*,\ifin}}
 \biggr)^{4/3}_{ }
 \;. \la{Neff}
\ee
The evolution of $e^{ }_\varphi/T^4_{ }$ is 
shown in \fig\ref{fig:Neff}(left), and the final
$\Delta N^{ }_\rmi{eff}$ is illustrated in 
\fig\ref{fig:Neff}(right), where $\Delta N^{ }_\rmi{eff}$ is also 
compared with previous strict 
LO and NLO estimates from ref.~\cite{bg}.
 (In the latter ones, 
 the negative small-$k$ tails of 
 $\gamma$ have been artificially cut off.)

We observe from \fig\ref{fig:Neff} 
how the positive ultrasoft contribution to the axion interaction
rate pulls the axion energy density closer to the equilibrium
value by up to $\sim 50$\%, continuing the trend initiated by the 
NLO hard contribution~\cite{bg}. The shapes of the curves 
suggest that there is a further boost from the QCD crossover, 
but currently there is no data to handle this domain reliably. 
On the other hand, for $f^{ }_\ax \gg 4\times 10^8_{ }$\hspace*{0.3mm}GeV, 
the interaction rate has only a small effect. In this case
the solution is close to that obtained from a free-streaming 
approximation, corresponding to axion freeze-out before the
electroweak crossover, where $h^{ }_*$ is to a good approximation
temperature-independent. 

\section{Conclusions and outlook}
\la{se:concl}

The technical core of this paper 
(cf.\ \ses\ref{se:htl} and \ref{se:asy})
has been to compute 
the axion interaction rate, $\Upsilon$, 
in the domain of soft momenta and energies, 
$
 0 \le k \le \omega \sim \mE^{ } 
 \sim \sqrt{\alphas^{ }\pi\vphantom{|}} T
$, 
where $T \ge 200$~MeV so that 
$\alphas \le 0.4$, 
and $\mE^{ }$ denotes the Debye mass that screens 
colour-electric interactions in a thermal environment. 
We have demonstrated how the regimes $k=0$ 
and $k=\omega$ can be interpolated into each other
(cf.\ \fig\ref{fig:chi_htl} on p.~\pageref{fig:chi_htl}), 
which physically corresponds to accounting for
the effects of a finite axion mass 
(cf.\ \fig\ref{fig:kinematics} on p.~\pageref{fig:kinematics}).
An analytic approximation to our result, valid at $\omega > \mE^{ }$, 
is given by \eqs\nr{gamma_Born} and \nr{gammaasyresumhack}. 

In previous determinations 
of $\Upsilon$ at $k =\omega \ge \pi T$, there has been 
a long-standing issue that extrapolating the result to $k = \omega \ll \pi T$
tends to turn it negative (cf.\ ref.~\cite{bg} for an overview).
In our systematic HTL computation at $k = \omega \sim \mE^{ }$, 
which agrees with ref.~\cite{mainz} 
up to that 
our subdominant pole-pole contribution 
appears to be twice as large, 
the result stays positive, and the issue
of negativity has been resolved
(cf.\ \fig\ref{fig:chi_assembled}(left)). 

However, despite its positivity, 
the HTL result is not reliable when $\omega \ll \mE^{ }$. 
In this domain, physics becomes sensitive also 
to ultrasoft ($p\sim g^2_{ }T/\pi$) and 
hard loop momenta ($p\sim \pi T$), and to higher-order
scatterings, which require
further resummations (cf.\ \app\ref{se:IR}). 

In the non-Abelian case, 
where ultrasoft momenta are non-perturbative, 
the domain $\omega \ll \mE^{ }$ needs
be tackled with lattice methods. 
As can be observed from \fig\ref{fig:chi_htl}(right), 
the true value of $\Upsilon$ is probably substantially larger than 
the HTL one in this domain. We have drawn this conclusion
both from classical-statistical (CLGT) and 
quantum-statistical (LQCD) simulations, 
despite their different systematic uncertainties.  
However, for the moment lattice data only exists for $k=0$, 
while our perturbative results suggest 
a noticeable logarithmic dependence on~$k$. 
In order to confirm this
dependence on the non-perturbative level,  
it would be valuable to extend the CLGT 
simulations from ref.~\cite{clgt} to $k > 0$. 

Ultimately, the problem should be studied with
full LQCD simulations. As has been discussed in the 
context of photon production from a QCD plasma
(cf.,\ e.g.,\ refs.~\cite{photon1,photon2,photon3}), 
where the physical interest lies on the lightcone
like for light axions, 
both the timelike and the spacelike frequency domains 
influence the measurement, because the simulations
are carried out in imaginary time
(cf.\ \fig\ref{fig:kinematics} on p.~\pageref{fig:kinematics}). 
Due to the difficulties associated
with analytic continuation, perturbative information of the type that
we have derived, complemented by its extension to the spacelike domain,  
can facilitate such investigations~\cite{photon1,photon2}.

Apart from offering perturbative support to 
future lattice studies, we have also made use of existing data, 
in order to update the light-axion contribution to $\Delta N^{ }_\rmi{eff}$.
Specifically, we have assembled together 
a phenomenological estimate for $\Upsilon$ that encompasses 
all available information (cf.\ \se\ref{se:assembled}), 
from lattice investigations 
(both CLGT and LQCD, cf.\ \se\ref{se:lattice}); 
from our HTL computation that determines the asymptotic value 
for the soft domain ($\omega\ge \mE^{ }$, 
cf.\ \se\ref{se:asy}); and from an 
NLO computation valid for the hard domain 
($\omega\ge \pi T$, cf.\ \app\ref{se:hard}).

Inserting this information into a kinetic equation 
(cf.\ \eq\nr{kinetic2}) and integrating over a cosmological history, 
we find 
$\Delta N^{ }_\rmi{eff} \approx 0.04 $ for 
$f^{ }_a \approx 4\times 10^8_{ }$\,GeV,
clearly above the previous NLO determination in ref.~\cite{bg}
(cf.\ \fig\ref{fig:Neff}). 
The increase is due to the large value of $\Upsilon$ in the 
ultrasoft domain, originating from the lattice data. 
We note that our treatment of the ultrasoft domain
has been rather conservative, with a fudge factor 
$\phi^{ }_\kappa = 0.1$ to reduce the amplitude
of the strong sphaleron rate, 
and $\phi^{ }_\rmii{M} = 0.01 ... 10.0$ to probe uncertainties
associated with the width of the ultrasoft domain. 
Therefore, 
the new result for $\Delta N^{ }_\rmi{eff}$ is arguably closer
to the true physical value 
than the previous one~\cite{bg}, and represents 
our current best estimate for the integrated light QCD 
axion contribution to $\Delta N^{ }_\rmi{eff}$ from $T \ge 200$~MeV
(of course, the domain $T < 200$~MeV has also been included, but
only in a free-streaming approximation).

%
\section*{Acknowledgements}

We thank Mathias Becker for helpful clarifications
concerning ref.~\cite{mainz}.
K.B.\ and G.S.S.S.\ thank the University of Bern for hospitality 
during initial stages of this work. 
J.G.\ is partly funded by the Agence
Nationale de la Recherche under grant ANR-22-CE31-0018 (AUTOTHERM).
M.L.\ thanks Dietrich B\"odeker for sharing his views on the 
difference between the $k = 0$ and $k = \omega$ approaches to the
sphaleron rate in an informal seminar in Bern in February 2025.
G.S.S.S.\ acknowledges the support of Funda\c{c}\~ao de Amparo \`a Pesquisa do
Estado de S\~ao Paulo (FAPESP) under the grant numbers 2022/15419-6 and
2023/17722-0.

\newpage

%
\appendix
\renewcommand{\thesection}{\Alph{section}} 
\renewcommand{\thesubsection}{\Alph{section}.\arabic{subsection}}
\renewcommand{\theequation}{\Alph{section}.\arabic{equation}}

%
\section{Role of light chiral fermions in the strong sphaleron rate}
\la{se:fermions}

It has long been thought that light chiral fermions play
a major role for sphaleron-related processes, 
however the arguments were not rigorous, and in fact 
they have been revised recently~\cite{dz}. Here we summarize the 
current understanding, in particular why the sphaleron rate
from \eq\nr{Gamma_sph} plays a role for axion dynamics
in cosmology at $k = 0$, even if strongly interacting 
light chiral fermions ($u$, $d$, $s$, $c$ and $b$ quarks) are present
in the Standard Model. 

As a starting point, we write down the Euler-Lagrange equation
following from \eq\nr{L}, 
\be
 \partial^{\mu}_{ }\partial^{ }_{\mu} \varphi
 + \partial^{ }_\varphi V^{\vphantom{|}}_0(\varphi) 
 + \frac{1}{f^{ }_\ax} \, \langle \chi \rangle^{ }_\varphi
 \; 
 \overset{\rmii{\nr{L}}}{=} 
 \; 0
 \;, \la{E-L}
\ee
as well as the chiral anomaly equation for $\Nf^{ }$
degenerate vector-like fermions of mass $m$~\cite{ad,bj},
\be
 \partial^{ }_\mu 
 \langle \bar\psi \gamma^\mu_{ } \gamma^{ }_5 \psi \rangle 
 - 2 \Nf^{ }\langle\chi\rangle^{ }_\varphi
 - 2 i m \langle\bar\psi \gamma^{ }_5 \psi \rangle
 \; 
 = 
 \; 
 0
 \;. \la{abj}
\ee 

We now need to expand 
$
 \langle\chi\rangle^{ }_\varphi
$
in the background of a non-vanishing~$\varphi$. Going to 
first order in $\varphi$, 
this corresponds to {\em linear-response theory}.
If we take $\varphi$ spatially constant, like in \eq\nr{eom_varphi}, 
and slowly varying, the answer can be expressed in terms of 
the retarded correlator from \eq\nr{G_R} and its imaginary
part from \eq\nr{rho}, 
\be
 \langle 
 \chi
 \rangle^{ }_{\bar\varphi}
 \; 
 = 
 \; 
 \langle 
 \chi
 \rangle^{ }_{ 0 }
 + 
 \frac{1}{f^{ }_\ax}
 \biggl[\, 
  - \, \bar\varphi(t) \, G^\rmii{R}_\chi(0,\vec{0})  
  + \dot{\bar\varphi}(t) 
  \lim_{\omega\to 0} \frac{\rho^{ }_\chi(\omega,\vec{0})}{\omega}
 \,\biggr]
 + 
 \rmO\biggl( 
 \frac{\ddot{\bar\varphi}}{f^{ }_\ax}, 
 \frac{\bar\varphi^2_{ }}{f^2_\ax}
 \biggr)
 \;. \la{lin_resp_1}
\ee

The second ingredient is to realize that if the axial
charge evolves slowly compared with typical plasma processes, 
then in general $\langle \chi \rangle^{\vphantom{|}}_0 \neq 0$. 
To see this, we consider \eq\nr{abj} with 
$\bar\varphi = m = 0$, 
and identify the ensemble 
average of the first term with $\dot{n}^{ }_\A$, 
where $n^{ }_\A$ is the axial charge density, 
$
 n^{ }_\A \equiv n^{ }_\rmii{R} - n^{ }_\rmii{L}
$.
Starting from the Landau theory of hydrodynamic fluctuations, 
it can be shown that the {\em non-equilibrium} evolution rate, 
$\dot{n}^{ }_\A$, is proportional to the
2-point {\em equilibrium} correlator of the $\dot{n}^{ }_\A$ operator, 
times the axial chemical potential, $\mu^{ }_\A$~\cite{kubo}.
The equilibrium 2-point correlator yields $\Gamma_\rmi{sph}$, 
whereas $\mu^{ }_\A$ can be related to $n^{ }_\A$, 
through a susceptibility. 
Physical arguments leading to the same result were known
much earlier (cf., e.g., ref.~\cite{KhSh}). 

In practice, 
the expressions following from the above logic
are a bit intransparent, given that 
the susceptibility contains
numerical factors, originating from integrals 
over the Fermi distribution 
and from the dimension of the fermion representation. We capture 
these by a coefficient $c^{ }_\psi\sim 1$, and then write down
the 2-point correlator as 
a linear-response relation analogous 
to the second term of \eq\nr{lin_resp_1}, 
\be
 \langle \chi \rangle^{\vphantom{|}}_0 
 \; 
 = 
 \; 
 - n^{ }_\A \, \frac{c^{ }_\psi \Gamma^{ }_\rmii{sph} }{T^3_{ }}
 + \rmO\bigl( \dot{\mu}^{ }_\A,\mu^2_\A \bigr) 
 \;. \la{lin_resp_2}
\ee 

We can now insert \eqs\nr{lin_resp_1} and \nr{lin_resp_2} 
into \eqs\nr{E-L} and \nr{abj}. Making use of 
$\Upsilon_\rmi{sph}$ from \eq\nr{Ups_sph};
defining 
$\chi^{ }_\rmi{topo} \equiv - G^\rmii{R}_\chi(0,\vec{0})$; 
and going to an expanding background with the Hubble rate $H$, 
\eq\nr{E-L} becomes
\be
 \ddot{\bar\varphi}
 + 3 H \dot{\bar\varphi}
 + \partial^{ }_\varphi V^{\vphantom{|}}_0 (\bar\varphi) 
 +
 \overbrace{
 \biggl[\, 
 - \frac{c^{ }_\psi \Gamma^{ }_\rmii{sph}}{f^{ }_\ax T^3_{ }}
   \,  n^{ }_\A
 + \frac{\chi^{ }_\rmii{topo}\bar\varphi}{f^2_\ax}
 + \Upsilon_\rmi{sph} \dot{\bar\varphi}
 \,\biggr]
 }^{ \approx\;\langle\chi\rangle^{ }_\varphi / f^{ }_\ax
 \;{\rm from}\; \nr{lin_resp_1},\nr{lin_resp_2} }
 \; 
 \underset{\rmii{ }}{
 \overset{\rmii{\nr{E-L}}}{\approx}} 
 \; 
 0
 \;. \la{E-L_rev}
\ee  

As for \eq\nr{abj}, the last term implies the presence 
of quantum-mechanical oscillations between the chiral states.
If we go to temperatures above the electroweak scale, the explicit
mass term is absent, however chirality-changing processes can 
be induced by $2\leftrightarrow 2$ and $1+n\leftrightarrow 2+n$
scatterings. In this regime we expect that the system can be 
described by a classical version of \eq\nr{abj}~\cite{bs}, 
\be
 \dot{n}^{ }_\A + 3 H n^{ }_\A 
 - 2 \Nf^{ } f^{ }_\ax 
 \overbrace{
 \biggl[\, 
 - \frac{c^{ }_\psi \Gamma^{ }_\rmii{sph}}{f^{ }_\ax T^3_{ }}
   \,  n^{ }_\A
 + \frac{\chi^{ }_\rmii{topo}\bar\varphi}{f^2_\ax}
 + \Upsilon_\rmi{sph} \dot{\bar\varphi}
 \,\biggr]
 }^{ \approx\;\langle\chi\rangle^{ }_\varphi / f^{ }_\ax
 \;{\rm from}\; \nr{lin_resp_1},\nr{lin_resp_2} }
 + 
 \Upsilon^{ }_\rmi{yuk}\, n^{ }_\A  
 \; 
 \underset{\rmii{ }}{
 \overset{\rmii{\nr{abj}}}{\approx}} 
 \; 
 0
 \;, \la{abj_rev}
\ee
where the rate induced by Yukawa couplings ($h$) is of order 
$
 \Upsilon_\rmi{yuk} \sim h^2_{ } g^2_{ }T/(128\pi)
$~\cite{bs}.

Let us consider the solution of \eq\nr{abj_rev} 
in a {\em stationary} regime, presumably reached after 
initial transients have died off. We find
\be
 \biggl(
  3 H + \Upsilon^{ }_\rmi{yuk} 
  + \frac{2 c^{ }_\psi \Nf^{ } \Gamma^{ }_\rmii{sph}}{T^3_{ }} 
 \biggr)
 \, n^{ }_\A 
 \; 
 \underset{\dot{n}^{ }_\A \; \approx \; 0}{
 \overset{\rmii{\nr{abj_rev}}}{\approx}}
 \; 
 2 \Nf^{ } f^{ }_\ax 
 \biggl(\, 
 \frac{\chi^{ }_\rmii{topo}\bar\varphi}{f^2_\ax}
 + \Upsilon_\rmi{sph} \dot{\bar\varphi}
 \,\biggr)
 \;. \la{stationary}
\ee 
Solving for $n^{ }_\A$ and inserting into \eq\nr{E-L_rev}, yields
\be
 \ddot{\bar\varphi}
 + 3 H \dot{\bar\varphi}
 + \partial^{ }_\varphi V^{\vphantom{|}}_0 (\bar\varphi) 
 + 
 \biggl(\, 
 \frac{\chi^{ }_\rmii{topo}\bar\varphi}{f^2_\ax}
 + \Upsilon_\rmi{sph} \dot{\bar\varphi}
 \,\biggr)
 \frac{
     3 H + \Upsilon_\rmii{yuk} 
      }
      {
     3 H + \Upsilon_\rmii{yuk} 
    + \frac{2 c^{ }_\psi \Nf^{ } \Gamma^{ }_\rmii{sph}}
           {T^3_{ } \vphantom{\big | }}
      \vphantom{\bigg |}
      } 
 \; 
 \underset{\rmii{\nr{stationary}}}{
 \overset{\rmii{\nr{E-L_rev}}}{\approx}}
 \; 
 0
 \;. \la{E-L_final}
\ee

One implication of \eq\nr{E-L_final} is that if the rate of explicit 
chirality violation is in equilibrium, $\Upsilon_\rmi{yuk} \gg 3 H$,
and if it is also large compared with the sphaleron rate, then the 
friction felt by the axion condensate is given by $\Upsilon_\rmi{sph}$.
Physically,  
chirality violation is a fast process
in this situation, and there would have been 
no need to introduce $n^{ }_\A$ and $\mu^{ }_\A$ in the first place. 

On the other hand, suppose that there are $\Nf^{ }$ light fermions, 
whose chirality is violated only 
very slowly, $\Upsilon_\rmi{yuk} \ll 3 H$.
In ref.~\cite{mms}, the correction factor appearing in \eq\nr{E-L_final}
was estimated, but the Hubble rate, $H$, was omitted. Then the correction
factor appears to go to zero 
proportionally to $\Upsilon_\rmi{yuk}$. This has 
led to a wide-spread belief that an axion 
condensate does not feel sphaleron friction in the presence of 
light fermions. 

The Hubble rate was included in the estimates in ref.~\cite{dz}. 
In its presence, the correction factor 
in \eq\nr{E-L_final} is finite but non-zero
for $\Upsilon_\rmi{yuk} \to 0$, 
and needs to be incorporated in its full form. This plays 
an important role in the recently proposed Standard Model embedding of 
warm inflation with the help
of a QCD axion~\cite{sm1,sm2,sm3}.

\newpage

%
\section{Why the HTL calculation fails at $\omega \ll \mE^{ }$}
\label{se:IR}

The purpose of this appendix is to demonstrate why 
the HTL computation is leading-order consistent in the regime
$\omega\sim \mE^{ }$, and yet fails if we
take $\omega \ll \mE^{ }$.
We indicate three separate reasons for 
the failure at $\omega \ll \mE^{ }$: 
the computation becomes sensitive to ultrasoft internal momenta, 
$p\sim g^2_{ }T/\pi$, which are non-perturbative in 
a non-Abelian theory;
to hard internal momenta, 
$p\sim \pi T$, where the HTL approximation ceases to be kinematically
self-consistent; and to higher-order scattering processes, not contained
in the leading-order HTL computation. Parts of this appendix
follow ref.~\cite[appendix~B]{mainz}. 

\vspace*{0.3mm}

%
\begin{figure}[t]

  \begin{center}
    \includegraphics[width=7.5cm]{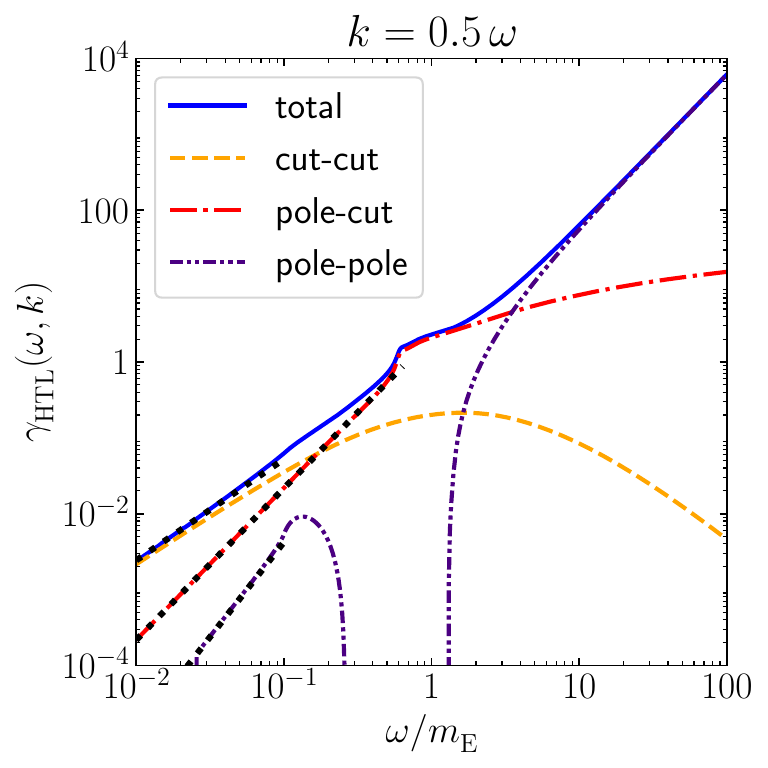}
    \includegraphics[width=7.5cm]{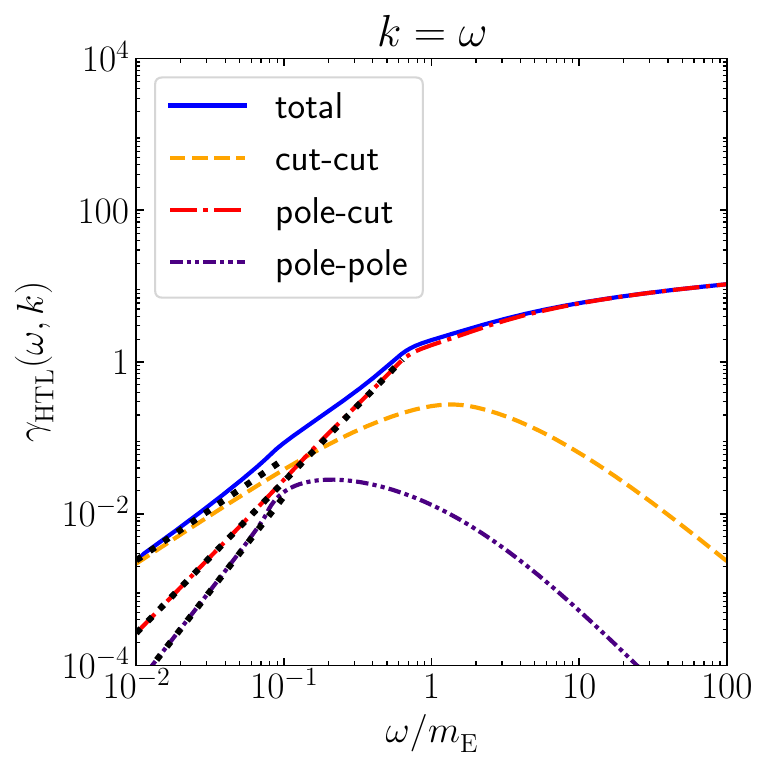}
  \end{center}

  \vspace*{-4mm}

  \caption[a]{\small 
  Different contributions to the coefficient $\gamma^{ }_\rmii{HTL}$, 
  defined in \eq\nr{gamma_def}. The left plot is for $k = 0.5\,\omega$, 
  the right plot for $k=\omega$.
  The dotted black lines represent 
  the respective analytically computable IR limits, 
  given in \eqs\eqref{IRlimit}, \eqref{polecutIR} and \eqref{IRlimitppfinal}.
  }

  \label{fig:HTL_channels}
\end{figure}
%

To get started, 
we plot in \fig\ref{fig:HTL_channels}  our result 
for $\gamma_\rmii{HTL}$ from \eq\eqref{gamma_def} for two 
different values of $k/\omega$. 
While the behaviour
is very different at $\omega/\mE^{ } \gg 1$, 
the qualitative behaviour is the same at $\omega/\mE \ll 1$, 
with a hierarchy cut-cut $\gg$ pole-cut $\gg$ pole-pole. 

We then start by inspecting 
the cut-cut contribution
at $\omega\ll \mE^{ }$. 
In this limit, 
the $q$ integration in \eq\eqref{rho_K} can be 
approximated as 
$1/k\int_{\vert p-k\vert}^{p+k}\mathrm{d}q\,f(q)\approx 2 f(p)$.
Simplifying the coefficients by restricting
to the domain $k\le \omega \ll p \approx q$, we find
\ba
 \gamma_\rmii{HTL,soft}^\rmii{cut-cut}(\omega,k)
 & \approx& 
 \frac{ 4 }{\pi^2_{ }\mE^2} \! 
    \int_0^\infty \! {\rm d}p \, p^4_{ }  
 \int_{-p}^{p}
 \! {\rm d} p^{ }_0 
 \, \frac{\theta(p^2-(\omega-p^{ }_0)^2)}{p^{ }_0\,(\omega-p^{ }_0)}
 \nn[2mm]
 & & \hspace*{2cm} \times \,  
 \bigl[\, 
   k^2_{ } \, (\rhoT_\P \rhoE_Q + \rhoT_\Q \rhoE_\P )
   + 2 \omega^2_{ }\rhoT_\P \, \rhoT_\Q
 \,\bigr]^{q^{ }_0 \; = \; \omega - p^{ }_0}_{q\;=\;p}
 \;.  \label{cutcutlim} 
\ea
As we will verify {\it a posteriori}, the integrals are
dominated by $|p^{ }_0| \ll p$, and then \eqs\nr{rho_T}--\nr{Gamma_E}
imply
\ba
 \rhoT_\P
 & \overset{|p^{ }_0|\;\ll\; p}{\approx} &
 \im\Biggl[\, 
  \frac{1}{p^2_{ } - \frac{i \pi \,\miE^2\, p^{ }_0}{4 p}}
  \,\Biggr]
 \; = \; 
 \frac{ \frac{\pi \,\miE^2 \, p^{ }_0}{4 p}}
      { p^4_{ } + \frac{\pi^2_{ }\,\miE^4 \,p_0^2}{16 p^2_{ }} }
 \;, \la{rhoT_ir_appro} \\[3mm]
 \rhoE_\P
 & \overset{|p^{ }_0|\;\ll\; p}{\approx} &
 \im\Biggl[\, 
  \frac{-1}{p^2_{ } + \mE^2 + \frac{i \pi \,\miE^2\, p^{ }_0}{2 p}}
  \,\Biggr]
 \; \overset{p\;\ll\;\miE}{\approx} \; 
 \frac{ \frac{\pi \,\miE^2 \, p^{ }_0}{2 p}}
      { \mE^4 + \frac{\pi^2_{ }\,\miE^4 \,p_0^2}{4 p^2_{ }} }
 \;. \la{rhoE_ir_appro}
\ea
{}From \eq\nr{rhoT_ir_appro} we see that 
the $p^{ }_0$-integrand gets a contribution from 
$|p^{ }_0|\sim 4 p^3_{ }/(\pi \mE^2)$, 
implying $|p^{ }_0|\ll p$ if $p \ll \mE^{ }$.
{}From \eq\nr{rhoE_ir_appro}
we see that $\rhoE_\P$ cannot grow as much as $\rhoT_\P$ at small
$\P$, as it is regularized by $\mE^2$, so it can be omitted. 
Then \eq\nr{cutcutlim} becomes
\ba
 \gamma_\rmii{HTL,soft}^\rmii{cut-cut}(\omega,k)
 & \approx& 
 \frac{ \mE^2 \omega^2_{ } }{2} \! 
    \int_0^\infty \! {\rm d}p \, p^2_{ }  
 \int_{-p}^{p}
 \! {\rm d} p^{ }_0 
 \, 
 \frac{1}{
 \Bigl[ 
 p^4_{ } + \frac{\pi^2_{ }\,\miE^4 \,p_0^2}{16 p^2_{ }}
 \Bigr]
 \Bigl[
 p^4_{ } + \frac{\pi^2_{ }\,\miE^4 \,(\omega - p_0^{ })^2}{16 p^2_{ }}
 \Bigr]
 }
 \nn[3mm]
 & \approx & 
 \omega^2_{ }
    \int_0^\infty \! {\rm d}p \, p \, 
  \frac{1}{p^4_{ } +  \frac{\pi^2_{ }\,\miE^4\, \omega^2_{ }}{64 p^2_{ }} }  
 \; = \; 
 \frac{4\pi^{1/3}_{ }}{3\sqrt{3}}
 \biggl( \frac{\omega}{\mE^{ }} \biggr)^{4/3}_{ }
 \;, 
 \label{IRlimit}
\ea
where in the second step we extended the $p^{ }_0$-limits to infinity.
As shown in \fig\ref{fig:HTL_channels}, 
\eq\eqref{IRlimit} is in good agreement 
with our numerical results at $\omega \ll \mE^{ }$, 
and it is also in reasonable agreement with the estimate  
for the coefficient $\frac{4 \pi^{1/3} }{3 \sqrt{3}}$
that was given in ref.~\cite{mainz}.

However, we see from \eq\nr{IRlimit} that the final $p$-integral
is dominated by \linebreak
$p \sim (\pi\, \mE^2\, \omega)^{1/3}_{ }/2$.
If we reduce $\omega$ to the domain $\omega \sim 8 g^4_{ }T/\pi^4_{ }$, 
we are sensitive to momenta $p \sim g^2_{ }T/\pi$. This is the regime
in which non-perturbative effects dominate the dynamics~\cite{linde}. 
Therefore the HTL evaluation of the cut-cut contribution ceases to be 
reliable when $\omega \ll \mE^{ }$.

\vspace*{3mm}

In the case of the pole-cut contribution, 
exploiting the $\P\leftrightarrow\Q$ symmetry, 
we can consider 
the pole in $p_0^{ }$, cut in $q_0^{ }$ contribution
to \eq\nr{rho_K}. 
The integrals over the pole can be carried out with 
\eqs\nr{poleintT_p0} and \nr{poleintE_p0}.
We anticipate that the relevant domain
is $p\sim p^{ }_0 \gg \mE^{ }$ for $\omega \ll \mE^{ }$, 
and then the pole locations are given by the latter rows
of \eqs\nr{p0T} and \nr{p0E} for the T and E channel, respectively. 
Given that $p^\rmii{E}_0(p)$ is exponentially close to $p$ 
according to \eq\nr{p0E}, and that the Jacobian is 
proportional to $\P^2_{ }$ according to \eq\nr{poleintE_p0}, 
the contribution from the E pole can be omitted. Like for 
the cut-cut contribution, a small $k$ implies that $p\approx q$, 
rendering the coefficient of the E cut small, whereby it can 
be omitted as well. 
Therefore we focus on the T--T channel, for which 
$[p^\rmii{T}_0(p)]^2_{ } \approx p^2_{ } + \mE^2/2$
according to \eq\nr{p0T}, and 
$
 \int \! {\rm d}p^{ }_0 \,\phi(p^{ }_0)\, \rhoT_\P
 \approx
 \pi\, \phi(p^\rmii{T}_0(p)) / [2 p^\rmii{T}_0(p)]
$
according to \eq\nr{poleintT_p0}.

We now turn to the integration domain, 
given in \tabl\ref{table:domains} on p.~\pageref{table:domains}. 
As discussed in \se\ref{ss:pc}, the pole always enters the 
cut domain, so the full $q^{ }_-$ domain is allowed. We just 
need to estimate the lower boundary of the $q^{ }_+$ integration. 
This originates from where the pole crosses the upper  
$p^{ }_0$ boundary, implying 
$p^{ }_0 \approx p + \mE^2/(4p) = \omega + q$. 
Together with $p\approx q \approx q^{ }_+$,  
this creates the domain
\begin{equation}
  \label{asymdomains}
  q^{ }_- \; \in \; 
  \biggl( -\frac{k}{2} , \frac{k}{2} \biggr)
  \;, \quad
  q^{ }_+ \; > \; \frac{\mE^2}{4(\omega + 2 q^{ }_-)} 
  \;.
\end{equation}
This justifies our previous assumption that 
$ p\approx q \approx q^{ }_+ \gg \mE^{ }$ for $\omega \ll \mE^{ }$.

It remains to estimate $\rhoT_\Q$ and its coefficient, as they 
appear when \eq\nr{rho_K} is combined with \eq\nr{gamma_def}, 
\be
 \gamma^\rmii{pole-cut}_\rmii{HTL,soft}
 \; \approx \; 
 \frac{2}{\pi^2_{ }\mE^2 k}
 \int_{-k/2}^{k/2} \! {\rm d} q^{ }_- 
 \int_{\frac{\miE^2}{4(\omega + 2 q^{ }_-)} }^\infty \! {\rm d}q^{ }_+\, 
 \frac{\pi p \hspace*{0.3mm} q }
      {2 p_0^2 q^{ }_0}
 \biggl\{
 \biggl( \frac{p_0^2}{p^2_{ }} + \frac{q_0^2}{q^2_{ }} \biggr) 
 [\, ... \,] 
  + 8 p^{ }_0 q^{ }_0 
 (\, ... \,) 
 \biggr\}
 \rhoT_\Q
 \;. 
\ee 
To be systematic, we can assign power counting to the 
various variables. Counting $\mE^{ }\sim 1$; 
setting $k\to \epsilon k$, 
$q^{ }_0 \to \epsilon \omega - p^{ }_0$, 
$p^{ }_0 \to p + \mE^2/(4 p) $,
$p \to q^{ }_+/\epsilon - \epsilon q^{ }_-$, 
and 
$q \to q^{ }_+/\epsilon + \epsilon q^{ }_-$; 
and expanding in $\epsilon$, 
we find 
$
 \{ ... \} 
 \approx 8 q_+^2 (\omega + 2 q^{ }_-)^2_{ }
$.
For the $\Q$ virtuality, the same expansion yields
$
 q_0^2 - q^2_{ } 
 \approx 
 \mE^2/2 - 2 q^{ }_+ (\omega + 2 q^{ }_-)
 < 0
$.
Consequently, 
\be
 \rhoT_\Q
 \; \overset{q^{ }_0\;\sim\; q}{\approx} \; 
 \im\Biggl[\, 
  \frac{1}{-q_0^2 + q^2_{ } + \frac{\miE^2}{2}
  - \frac{i \pi \,\miE^2\, q^{ }_0}{4 q}
  \bigl( 1 - \frac{q_0^2}{q^2_{ }} \bigr)}
  \,\Biggr]
 \; \approx \; 
 \frac{\frac{\pi\,\miE^2}{4 q_+^2}
  \bigl[ 2 q^{ }_+ (\omega + 2 q^{ }_-) - \frac{\miE^2}{2} \bigr] }
  {4 q_+^2 (\omega + 2 q^{ }_-)^2}
 \;. \la{rhoT_lc}
\ee
All in all, this yields 
\be
 \gamma^\rmii{pole-cut}_\rmii{HTL,soft}
 \; \approx \; 
 \frac{2}{k}
 \int_{-k/2}^{k/2} \! {\rm d} q^{ }_- 
 \int_{\frac{\miE^2}{4(\omega + 2 q^{ }_-)} }^\infty \! {\rm d}q^{ }_+\, 
 \frac{ 2 q^{ }_+ (\omega + 2 q^{ }_-) - \frac{\miE^2}{2} }
  {4 q_+^3 }
 \; = \; 
 \frac{2(k^2_{ } + 3 \omega^2_{ })}{3 \mE^2}
 \;.  \label{polecutIR}
\ee
As shown in \fig\ref{fig:HTL_channels}, \eq\nr{polecutIR} is in 
good agreement with our numerical results for $\omega \ll \mE^{ }$.

However, there are at least two issues which indicate that 
\eq\nr{polecutIR} is not physically reliable for $\omega \ll \mE^{ }$.
First, when $\omega \ll \mE^2/(\pi T) \sim g^2 T/\pi$, we have that,
following \eq\nr{asymdomains}, 
$q^{ }_+\equiv(q+p)/2\gsim \mE^{2}/\omega\gg \pi T$ 
and $ q^{ }_-\equiv(q-p)/2\sim k \ll \pi T$.
This in turn implies that 
$p\gg \pi T$, and the transverse pole dispersion 
relation implies that $p^{ }_0\approx p\gg \pi T$. In this regime, 
$\nB^{ }(p^{ }_0)\approx \exp(-p^{ }_0/T)$. However, the onset of 
this exponential suppression~\cite{mainz} is entirely missed by
our HTL calculation, where we approximate $\nB^{ }(p^{ }_0)$ by
its classical limit, \eq\eqref{classical}.
The other is that the width implied by the HTL 
propagator, \eq\nr{rhoT_lc}, 
is unreliable when $q^{ }_0 \approx q \approx q^{ }_+ \gg \mE^{ }$. 
Recalling $1 - q_0^2/q^2 \sim \mE^2/q^2$, 
the width that we have used is of magnitude
$\sim \pi \mE^4 q^{ }_0 / q^3_{ } $, however 
gluons with a hard momentum have a much larger width
$
 \sim (g^2_{ }\Nc^{ } T q^{ }_0/\pi) \ln(1/g)
$. 
The presence of a logarithm reflects the fact 
that trying to incorporate
this physics with a naive perturbative computation in a non-Abelian
theory yields a logarithmic divergence~\cite[appendix~B]{bg}.
Its proper treatment likely requires 
Landau--Pomeranchuk--Migdal (LPM) resummation~\cite{lpm1,amy1,bb1}. 

\vspace*{3mm}

Finally, let us investigate how the pole-pole contribution
behaves at $\omega \ll \mE^{ }$. As discussed in the paragraphs
around \eq\nr{def_G}, it 
originates from the $\mbox{T}\to\mbox{E}+\varphi$ process.
Using the $\P\leftrightarrow\Q$ symmetry in 
\eqs\nr{rho_K} and \nr{gamma_def}, and adopting 
the relevant integration domain from \tabl\ref{table:domains}
on p.~\pageref{table:domains}, 
we thus have
\begin{eqnarray}
   \gamma_\rmii{HTL,soft}^\rmii{pole-pole}(\omega,k)
 & 
 =
 & 
 \frac{ 1}{\pi^2_{ }\mE^2 k} \! 
    \int_0^\infty \! {\rm d}p \, p 
    \int_{|p-k|}^{p+k} \! {\rm d}q \, q  
 \int_{-\infty}^{-p}
 \! {\rm d} p^{ }_0 
 \frac{
 1
 }{p_0^{ } q^{ }_0 }
 \hspace*{5mm}
 \nn[3mm]
 & \times & 
 \bigl[\, k^2_{ } - (p-q)^2_{ } \,\bigr]
 \bigl[\, (p+q)^2_{ } - k^2_{ } \,\bigr]
 \, 
  \rhoT_\Q \, \rhoE_\P 
 \; \Bigl|^{ }_{q^{ }_0 \;=\; \omega - p^{ }_0}
 \la{IRlimitpp}
 \;. 
\end{eqnarray}
The integral over $p^{ }_0$ can be carried out with \eq\nr{poleintE_p0}, 
and the integral over $q$ with \eq\nr{poleintT_qp}, 
multiplying with a factor~2 in the latter case. 

In order for the decay to produce a low-energy axion, the spatial momenta
of the plasmons need to be small, with $p,q \ll \mE^{ }$. The dispersion
relations then follow from the first lines of \eqs\nr{p0T} and \nr{p0E}. 
The phase-space constraint is given by the zeros of \eq\nr{def_G}, yielding 
\be
 \frac{3\sqrt{3} q^2_{ }}{5\mE^{ }}
 \; \approx \; 
 \omega 
 \; + \; 
 \frac{3\sqrt{3} p^2_{ }}{10\mE^{ }}
 \;. \la{pp_constraint}
\ee
The $p$-range is given by when the solution
of \eq\nr{pp_constraint} crosses the boundaries
$q = p+k$ and $q = p-k$ from \eq\nr{IRlimitpp}, 
which for $k^2_{ } \ll \mE^{ }\hspace*{0.3mm}\omega$ yields
the range
\be
 p \; = \; \bar{p} + \Delta p
 \;, \quad
 \bar{p} 
 \; \equiv \; \sqrt{\frac{10 \mE^{ }\omega }{3\sqrt{3}}}
 \; \ll \; \mE^{ }
 \;, \quad
 \Delta p \; \in \; (-2k,2k)
 \;. \la{def_pc}
\ee
Substituting back in \eq\nr{pp_constraint}, we find that 
$q \approx \bar{p} + \Delta p/2$,  
whereby $p + q \approx 2 \bar{p}$
and $p - q \approx \Delta p / 2$.
The factors from \eq\nr{IRlimitpp} that require special care are 
$
 k^2_{ } - (p-q)^2_{ } \approx (4 k^2_{ } - \Delta p^2_{ })/4
$, 
as well as the Jacobian following from $\rhoT_\Q$, 
\be
  \frac{\pi\, \sign(q^{ }_0) \,  q (q_0^2 - q^2_{ })}
         { \mE^2\, q_0^2 - 3 (q_0^2 - q^2_{ })^2_{ } }
 \bigg|^{ }_{ q_0^2\;\approx\; 
 \frac{\miE^2\vphantom{\big | }}{3}
 + \frac{6 q^2_{ } \vphantom{\big | }}{5} }
 \; 
 \overset{q\;\approx\; \bar{p}}{\approx} 
 \; 
 \frac{5 \pi}{12 \bar{p}}
 \;. 
\ee
All in all this yields 
\ba
 \gamma_\rmii{HTL,soft}^\rmii{pole-pole}(\omega,k)
 & \approx & 
 \frac{1}{\pi^2_{ }\mE^2 k}
 \int_{-2k}^{2k} \! {\rm d}\Delta p \, 
 \frac{\bar{p}^2_{ }}{p^{ }_0 q^{ }_0}
 \times \, \frac{4 k^2_{ }- \Delta p^2_{ }}{4}
 \times \, 4 \bar{p}^2_{ } 
 \times \, \frac{5\pi}{12 \bar{p}} 
 \times \, \frac{\pi p^{ }_0}{2 \bar{p}^{\hspace*{0.3mm} 2}_{ }} 
 \nn[3mm]
 & \approx & 
 \frac{2^{5/2}5^{3/2} }
 {3^{9/4}}\frac{\sqrt{\omega}\, k^2}{\mE^{5/2}}
 \la{IRlimitppfinal}
 \;. 
\ea
As \fig\ref{fig:HTL_channels} shows, 
this describes well the 
$\omega \ll \mE^{ }$ asymptotics of the pole-pole contribution. 

Yet, there are once again reasons to doubt that 
\eq\nr{IRlimitppfinal} is physically accurate if $\omega \ll \mE$.
First of all, we see from \eq\nr{def_pc} that if 
$\omega$ is sufficient small, we become sensitive to 
non-perturbative momenta, $p\sim g^2_{ }T/\pi$. Furthermore, 
ref.~\cite{mainz} finds that, upon including 
a 1-loop width for the plasmon self-energy, 
the T--T channel opens up and becomes 
dominant over the T--E one.
However, the propagator in the T--T case
is like in \eq\nr{rhoT_lc}, only with $\Q^2_{ } > 0$, 
and the same reservations as expressed
below \eq\nr{polecutIR} apply. 

%
\begin{figure}[t]
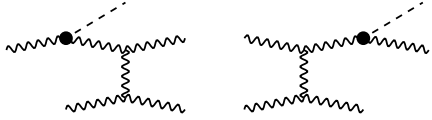


\vspace*{-1.0cm}%

\begin{eqnarray*}
 & &
 \hspace*{0cm}
 \GraphggCtwo\qquad
 \GraphggC
\end{eqnarray*}

\vspace*{-1.2cm}%

\caption[a]{\small 
 Two diagrams for the $2\leftrightarrow3$ scattering contribution 
 to soft axion production. Gluons are denoted by curly lines, 
 the axion by a dashed line, and the axion--gauge vertex
 by a black blob.
 The $t$-channel rung is soft and 
 leads to a logarithmic divergence if the axion energy
 is $\omega \ll \mE^{ }$. 
 We note that higher-order $1+n \to 2+n$ processes, 
 with $n \ge 2$, which are not shown, 
 are expected to contribute at the same order. 
} 
\label{fig:1nto2n}
\end{figure}
%

In fact, 
we expect that the T--T pole-pole and pole-cut processes
may be viewed as parts of a single phenomenon,  
``soft axion bremsstrahlung''.
Specifically, the pole-cut contribution corresponds 
to the square of diagrams such as on the left 
in \fig\ref{fig:1nto2n}. There,
the virtual gluon attached to the right-hand side
of the axion--gauge vertex 
must be space-like. Conversely,
the pole-pole contribution corresponds to the square 
of the diagram on the right, where the virtual gluon attaching 
to the left-hand side of the axion--gauge vertex is time-like. 
There are similar diagrams where the external particles 
on the bottom half of each diagram 
are quarks or antiquarks.
For a full description, higher-order iterations of such 
diagrams, as well as interference terms between them, 
need to be included, which probably amounts to 
an LPM-type resummation~\cite{lpm1,amy1,bb1}. 

\vspace*{3mm}

To summarize this appendix, we have shown that the $\omega\ll \mE$
limits of the cut-cut, pole-cut, and pole-pole contributions shown
in \fig\ref{fig:HTL_channels} can be understood analytically. The
computations demonstrate explicitly that as long as 
$\omega \sim \mE^{ }$, the loop momenta that contribute are of soft
magnitude, $p\sim \mE^{ }$, and the HTL computation is self-consistent. 
However, when $\omega \ll \mE$, some processes become sensitive
to ultrasoft loop momenta $p \sim g^2_{ }T/\pi$, others to 
hard loop momenta $p \sim \pi T$. Both cases fall out of the 
purview of the HTL theory, which only describes the propagation and 
self-interactions of soft modes. Moreover, additional processes,
not contained in the HTL computation, start to play 
a role (cf.\ \fig\ref{fig:1nto2n}). In the body of the paper, 
we have therefore taken over the description at $\omega \ll \mE^{ }$
from non-perturbative lattice simulations. 

\newpage

%
\section{What is known about $k \approx \omega \ge \pi T$}
\la{se:hard}

In \se\ref{se:asy}, we have extrapolated 
the HTL result, $\rho^\rmii{HTL}_\chi$, 
to the domain $\omega \gg \mE^{ }$. For $k = 0$, it
is known that this UV limit of the HTL result matches the IR limit 
of an unresummed NLO computation
at $\omega \sim \pi T$~\cite{Bulk_wdep}. 
The situation is technically 
more complicated for $k> 0$, 
where no NLO result is known for $\rho^{ }_\chi$
(even though the general techniques are available and have
been applied to other spectral functions~\cite{dilepton,gj}). 
The complications increase further when approaching $k \approx \omega$, 
where resummations become necessary. 
In the present appendix, we explain the structures
of the leading-order (LO) and 
NLO results for $k = \omega \ge \pi T$, by summarizing 
and extending the work presented in ref.~\cite{bg}.
We denote $\gamma(k) \equiv \gamma(k,k)$.

%
\hiddenappsubsection[se:hardLO]{Leading order (LO)}

\noindent
The LO result at $k \ge \pi T$
has been discussed in several
schemes differing at relative order~$g$, arising
from different ways of extrapolating the HTL-resummed
expression to $k \sim \mE^{ } \ll \pi T$.
At \emph{strict} leading order, we anticipate that the
result has the form  
\begin{equation}
  \gamma^{ \rmii{strict}}_\rmii{LO,hard}(k) 
   \;=\;
  \ln\biggl( \frac{4k^2}{\mE^2} \biggr)
  + f_\rmii{LO}^{ }\left(\frac kT\right)\;.
  \label{LO_hard}
\end{equation}
The logarithm corresponds to the last row 
of \eq\nr{gamma_asy_cases}, and $f^{ }_\rmii{LO}$ will be 
given in \eq\nr{fLO}.

To pave the way for presenting $f_\rmii{LO}^{ }$, 
we start by rederiving
the logarithm. 
For this we can take 
the $\omega=k$ limit of \eq\eqref{rho_K_NLO}
as a starting point, i.e.
\begin{eqnarray}
 && \hspace*{-1.1cm}
 \rho^{\rmii{HTL}}_{\chi,\rmii{hard}}(k)
 \;
 \approx
 \;
 \frac{ 4 (\Nc^2 - 1)\, c_\chi^2 }{\pi^2_{ }k} \! 
    \int_0^\infty \! {\rm d}p \, p 
    \int_{|p-k|}^{p+k} \! {\rm d}q \, q  
 \int_{-\infty}^{\infty}
 \! {\rm d} p^{ }_0 
 \, \bigr[1 + \nB^{ }(p^{ }_0) + \nB^{ }(q^{ }_0)\bigl]
 \sign(q^{ }_0)\delta(\Q^2)
 \hspace*{5mm}
 \nn[3mm]
 & \times & 
 \biggl\{\; 
  \rhoE_\P \,\bigl[\, k^2_{ } - (p-q)^2_{ } \,\bigr]
 \bigl[\, (p+q)^2_{ } - k^2_{ } \,\bigr]
 \label{rho_K_lc}
 \\[3mm]
 &  & \; +\rhoT_\P \,  
 \biggl[\,
 \biggl( \frac{p_0^2}{p^2_{ }} + \frac{q_0^2}{q^2_{ }} \biggr) 
 \bigl[\,
  \bigl( p^2_{ } + q^2_{ } - k^2_{ } \bigr)^2_{ } 
 + 
  4 p^2_{ } q^2_{ }
 \,\bigr]
 + 8 p^{ }_0 q^{ }_0 
   \bigl( p^2_{ } + q^2_{ } - k^2_{ } \bigr) 
 \,\biggr] \, 
 \;\biggr\}^{ }_{q^{ }_0 \;=\; \omega - p^{ }_0}
 \;. \nonumber 
\end{eqnarray}
Compared with \eq\eqref{rho_K_NLO}, 
we have dropped the subtraction ($\Delta$), given that there
is no Born term at $\omega = k$, and we have
undone the classical limit for the statistical functions. 
Upon integrating 
over $q$ and re-expressing the result 
according to \eq\nr{gamma_def}, we find 
\begin{eqnarray}
\gamma^\rmii{HTL}_\rmii{LO,hard}(k)
 & 
 \approx 
 & 
 \frac{ 1  }{2\pi_{ } m_\rmiii{E}^2\,k^2T} \! 
 \int_{-\infty}^{k}
 \! {\rm d} p^{ }_0 
 \int_{\vert p_0^{ }\vert}^{2k-p_0^{ }}
 \! {\rm d}p \, p \,\bigl(\, p^2-p_0^2 \,\bigr)\,
 \bigr[1 + \nB^{ }(p^{ }_0) + \nB^{ }(k-p^{ }_0)\bigl]
 \hspace*{5mm}
 \nn[3mm]
 & \times & 
 \biggl\{\; 
 \, \rhoE_\P\,\bigl[\,(2k-p_0^{ })^2-p^2 \,\bigr]
\; +\rhoT_\P \,  
\frac{ p^2-p_0^2 }{p^2}\,\bigl[\, (2k-p_0^{ })^2+p^2 \,\bigr] \, 
 \;\biggr\}
 \;.\label{gamma_htl_lc} 
\end{eqnarray}
As discussed after \eq\eqref{oneloopexpmd2p}
and illustrated in \fig\ref{fig:asy}, 
we restrict to $q^{ }_0 > 0$ here.

HTL resummation is only needed for $p^{ }_0,p\sim \mE^{ }$. 
Approximating $p_0^{ },p \ll k$, we find
\begin{eqnarray}
\gamma^\rmii{HTL}_\rmii{LO,hard}(k)
 & 
 \approx
 & 
 \frac{ 2  }{\pi_{ } m_\rmiii{E}^2\,T} \! 
 \int_{-\infty}^{k}
 \! {\rm d} p^{ }_0 
 \int_{\vert p_0^{ }\vert}^{2k-p_0^{ }}
 \! {\rm d}p \, p \,\bigl(\, p^2-p_0^2 \,\bigr)\,
 \frac{T}{p_0^{ }}
 \biggl\{\; 
 \, \rhoE_\P 
 \; +\rhoT_\P \,  
 \frac{ p^2-p_0^2 }{p^2}
 \;\biggr\}\nn
 & =& \ln\left(1+\frac{4k^2}{\mE^2}\right)
 \;+\;\mathcal{O}\biggl(\frac{\mE^2}{k^2}\biggr)
 \;\approx\; \ln\biggl( \frac{4k^2}{\mE^2} \biggr)
 \;,\label{gamma_htl_lc_LO}  
\end{eqnarray}
thus recovering the logarithm in \eq\eqref{LO_hard}. 
The $p^{ }_0$-integral was carried out  
with lightcone sum rules, 
as discussed around \eq\nr{residue}; 
the subleading term of order $\mE^2/k^2$ arises from 
stretching the $p_0^{ }$ integration to infinity.

The remainder in \eq\nr{LO_hard}, 
$f_\rmii{LO}$, is coupling-independent, and can be determined
without resummation, provided that the part of the integrand 
leading to \eq\nr{gamma_htl_lc_LO} is subtracted. 
In the language of ref.~\cite[appendix~A]{bg}, $f_\rmii{LO}^{ }$ reads
\begin{align}
 f_\rmii{LO}\left(\frac kT\right)
 \;=\; &
 \frac{ 1 }{  8\pi^2_{ } \hat{m}_\rmii{E}^2 k^2_{ }T^3_{ }}
  \biggl\{ \int^k_{-\infty} \! {\rm d}p^{ }_0
  \int^{2k - p^{ }_0}_{|p^{ }_0|} \! {\rm d}p
 \bigg[
      \Nc^{ }\,{\rm I}\hspace*{0.3mm}{ }^t_{+++}(-1,0)
      +2T^{ }_\F\Nf^{ }\,{\rm I}\hspace*{0.3mm}{ }^t_{-+-}(1,0) \nn[2mm]
 & \hspace*{5.0cm} 
  -\,\frac{12\pi^2_{ } \hat{m}_\rmii{E}^2 k^2_{ }T^3_{ }
  (p^2_{ }-p_0^2)}{p^4_{ }}\bigg]\nn[2mm]
 &+ \int_k^\infty \! {\rm d}p^{ }_0
 \int_{|2k - p^{ }_0|}^{p^{ }_0} \! {\rm d}p
 \bigl[
 \Nc^{ }\, {\rm I}\hspace*{0.3mm}{ }^s_{+++}(-1,0)
 +T^{ }_\F \Nf^{ }\, {\rm I}\hspace*{0.3mm}{ }^s_{--+}(2,0)
 \bigr] \biggr\}\;,
 \label{fLO}
\end{align}
where $T^{ }_\F=1/2$, and 
$\hat{m}_\rmii{E}^2\equiv(\Nc^{ }+T^{ }_\F\Nf^{ })/3$ is the 
leading-order Debye mass in units of $g^2T^2$. 
The functions ${\rm I}\hspace*{0.3mm}{}^t_{ }$ 
(for $t$-channel) and 
${\rm I}\hspace*{0.3mm}{}^s_{ }$ 
(for $s$-channel) 
are defined in \eqs(A.9) and (A.13) 
of ref.~\cite{bg}. The second line subtracts the IR-sensitive
exchange that led to \eq\nr{gamma_htl_lc_LO}.

Even if the integrals in \eq\nr{fLO} are not
analytically solvable, their limiting values 
at $k\gg \pi T$ and $k\ll \pi T$ can be worked out.  
Using techniques from~ref.\cite{pa_qhat}, we find
\ba
  && \hspace*{-1.3cm}
  f_\rmii{LO}\left(\frac kT\gg \pi \right)
  \;=\; 
  \biggl[ \ln\biggl(\frac k T\biggr)-\frac12+\gammaE^{ }\biggr]
  \biggl[ \frac{ (2 \Nc^{ }+ 3  T^{ }_\F\Nf^{ } )\, \zeta(3) }
  {\pi^2_{ } \hat{m}_\rmii{E}^2} - 2 \biggr]
   +8\,\frac{ c^{ }_\rmii{B} \Nc^{ }+ c^{ }_\rmii{F} 
   T^{ }_\F \Nf^{ } }{\pi^2
   \hat{m}_\rmii{E}^2} \nn[3mm]
 && \; - \, \frac{ (2 \Nc^{ }+3 T^{ }_\F \Nf^{ } )\, \zeta'(3) }
   {\pi^2_{ } \hat{m}_\rmii{E}^2}
   - \frac{ [ 10 \Nc^{ }+(1+12\ln2 ) T^{ }_\F \Nf^{ }
            ]\, \zeta(3) }
   {12 \pi^2_{ } \hat{m}_\rmii{E}^2 } 
   +\mathcal{O}\biggl(\frac{\pi T}{ k } \biggr)
   \;,
   \label{farUV_full}
\ea
where the coefficients $c^{ }_\rmii{B}$ and $c^{ }_\rmii{F}$
are given by 
\ba
  c^{ }_\rmii{B}
  & = &
  -\frac12\sum_{n=1}^\infty \frac{\ln(n!)}{n^3}
  = \frac{1}{4} \sum_{m=1}^\infty  \ln (m)\,\psi ^{(2)}(m)
 \approx -0.292085
  \,,\nn[-3mm] \la{cB_cF} \\[1mm]
  c^{ }_\rmii{F}
  & = &
  \sum_{n=1}^\infty (-1)^n\frac{\ln(n!)}{n^3}
   = \frac{1}{8} \sum_{m=1}^\infty (-1)^m_{ } \ln (m)
   \biggl[ \zeta \biggl( 3,\frac{m}{2} \biggr)
   -\zeta\biggl( 3,\frac{m+1}{2}\biggr)\biggr]
   \approx 0.0484891
  \;. \nonumber
\ea
We have used $\ln(n!)=\sum_{m=1}^n\ln(m)$ 
to get to the $m$-series representations, 
which converge faster than the $n$-series ones.
%
%
%
Eq.~\eqref{farUV_full} remains $\mathcal{O}(10\%)$ 
accurate down to $k\approx 2 T$.
In the opposite limit $k\ll \pi T$, 
it can be shown from \eq\eqref{fLO} that
\begin{equation}
  \label{farIR}
  f^{ }_\rmii{LO}\left(\frac kT\ll \pi\right)
  \; = \; -\frac{1}{18\hat{m}_\rmii{E}^2}
   \biggl(\frac{17}{2}\Nc^{ }+ T^{ }_\F\Nf^{ }\biggr)
   \frac kT + \mathcal{O}\biggl(\frac{k}{\pi T}\biggr)^2_{ }
 \;,
\end{equation}
which remains $\mathcal{O}(20\%)$ accurate up to $k\approx 2 T$. 
Combining \eqs\nr{LO_hard}, \nr{farUV_full}, 
and \nr{farIR}, 
\begin{equation}
  \gamma^{ \rmii{strict}}_\rmii{LO,hard}(k) 
  \; \approx \;  
  \left\{ 
  \begin{array}{ll}
   \displaystyle
   \ln\biggl(\frac{4T^2_{ }}{\mE^2}\biggr)
  +\frac{ (2\Nc^{ }+3T^{ }_\F\Nf^{ } )\, \zeta(3)}
   {\pi^2_{ }\hat{m}_\rmii{E}^2}\ln\biggl( \frac{k}{T} \biggr)
  & \\[4mm]
  \displaystyle 
  \hspace*{1cm}
  -\frac{0.992309\,\Nc^{ }+0.054922^{ }\,T^{ }_\F \Nf^{ }}
        {\Nc^{ }+T^{ }_\F\Nf^{ }}
  +\mathcal{O}\biggl(\frac{\pi T}{ k }\biggr)
  & 
  \displaystyle 
  \;, \quad k \gg \pi T \;,
  \\[4mm]
  \displaystyle 
  \ln\biggl( \frac{4k^2_{ }}{\mE^2} \biggr)
  -\frac{1}{18\hat{m}_\rmii{E}^2}
   \biggl(\frac{17}{2}\Nc^{ }+ T^{ }_\F\Nf^{ }\biggr)
   \frac kT +\mathcal{O}\biggl(\frac{k}{\pi T}\biggr)^2_{ }
  & 
  \displaystyle 
  \;, \quad k \ll \pi T \;.
  \end{array} 
  \right. 
  \label{LO_hard_IR}
\end{equation}
Hence, in the IR domain $\mE\ll k\ll \pi T$, 
the hard LO result is dominated by the logarithm, 
and agrees with the UV limit of the HTL result 
on the last row of \eq\eqref{gamma_asy_cases}. 

However, if the last row of \eq\nr{LO_hard_IR}
is employed at $k < \mE/2$, it is negative. 
At LO, schemes were presented in ref.~\cite{bg} which moderate
or eliminate the negativity.
In the present paper, the issue is solved by other ingredients, 
namely by a systematic HTL computation in the soft domain, 
which remains positive 
(cf.\ \fig\ref{fig:HTL_channels} on p.~\pageref{fig:HTL_channels});
by making use of non-perturbative results 
in the ultrasoft domain, which are even larger
(cf.\ \fig\ref{fig:chi_assembled} on p.~\pageref{fig:chi_assembled}); 
and by including a large positive NLO correction 
in the hard domain, to which we now turn.

%
\hiddenappsubsection[se:hardNLO]{Next-to-leading order (NLO)}

\noindent
Apart from soft effects (cf.\ \ses\ref{se:htl} and \ref{se:lattice}), 
another increase of $\gamma$ is given by hard NLO effects. 
Specifically, in thermal field theory, NLO corrections
can be large, of $\mathcal{O}(g)$, arising from soft momenta, 
$p^{ }_0,p \sim \mE^{ } \sim gT$. This was  
clarified in the context of jet quenching (the physics of 
damping rather than particle production, but still 
characterized by an interaction rate)~\cite{CaronHuot:2008ni}. 
In the following,
we argue that ref.~\cite{CaronHuot:2008ni} offers more than an analogue,
in that NLO corrections can be quantitatively
taken over from there, 
once the correspondence is properly identified. 

Let us start by inspecting 
the LO soft-gluon contribution that led 
to \eq\eqref{gamma_htl_lc_LO}.
One might wonder if the steps could hide  
corrections of $\mathcal{O}(g)$. We have checked, following similar 
calculations in refs.~\cite{Ghiglieri:2013gia,Ghiglieri:2015ala}, 
that this is not the case, i.e.\  
that all potential $\mathcal{O}(g)$ effects 
in \eqs\eqref{rho_K_lc} and \eqref{gamma_htl_lc} vanish. 
Consider for illustration the NLO term in the expansion 
of $1 + \nB(p^{ }_0) + \nB(k - p^{ }_0)$. Its
inclusion amounts to replacing the classical $T/p_0^{ }$  
on the first line of \eq\eqref{gamma_htl_lc_LO}
with a $p_0$-independent $1/2+\nB(k)$ term, which leads 
to an odd, vanishing $p^{ }_0$-integral. Similarly, 
all other terms vanish, e.g.\ those coming from the NLO terms 
in the $p_0,p\sim \mE\ll k$ expansion, or the inclusion of 
a thermal mass for the $\Q$ propagator, 
i.e.\ 
$
 \delta(\Q^2_{ })\to\delta(\Q^2_{ }-\mE^2/2) 
 \overset{q^{ }_0\; =\; k - p^{ }_0}{\approx}
 \delta(q^2_{ }-k^2_{ })+(2 k p_0^{ }-p_0^2+\mE^2/2)
 \delta'(q^2_{ }-k^2_{ })
$.

\begin{figure}[t]

 \vspace*{-2mm}

  \begin{center}
    \includegraphics[width=7.5cm]{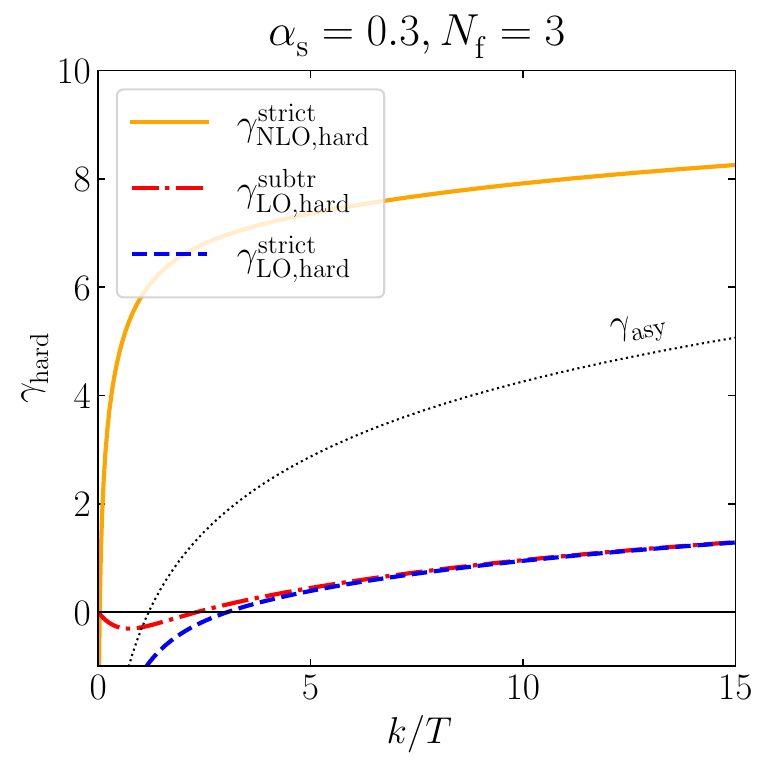}
    \includegraphics[width=7.5cm]{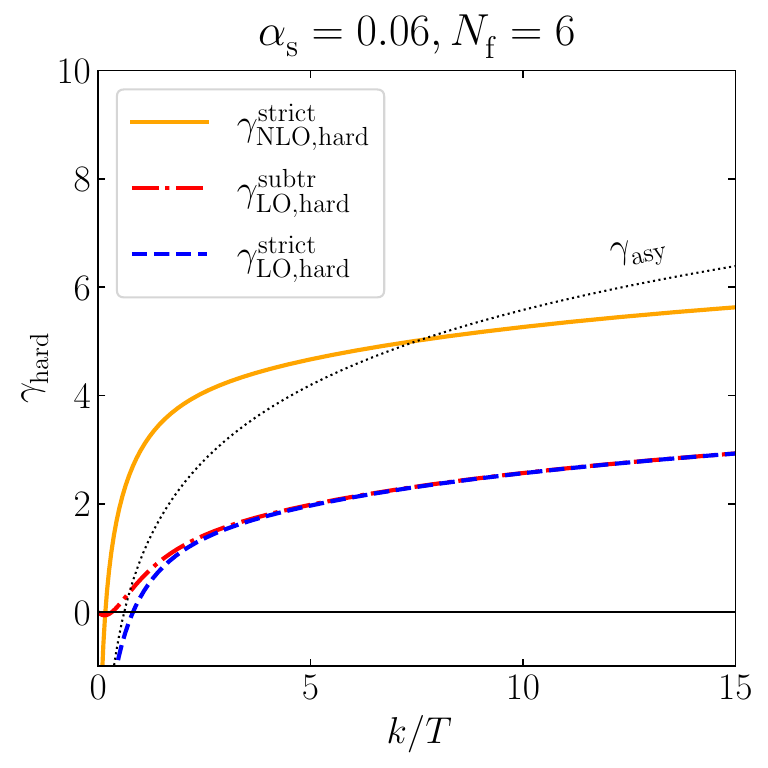}
  \end{center}

 \vspace*{-7mm}

  \caption[a]{\small The LO and NLO hard interaction rates from 
  \eqs\eqref{LO_hard} and \eqref{NLO_hard}, respectively. 
  The ``subtr'' version refers to a recipe suggested in 
  ref.~\cite{bg}, which makes the hard LO rate somewhat better 
  behaved when extrapolated to $k \ll \pi T$.
  For comparison we also show 
  $\gamma^{ }_\rmi{asy} \equiv \ln (4 k^2_{ }/ \mE^2)$.
 }
 
  \label{fig:hardrate}
\end{figure}

Having discounted the possibility of 
hidden $\mathcal{O}(g)$ effects 
in the LO contribution, 
we now turn to genuine NLO effects.
As observed in ref.~\cite{bg}, the LO soft-gluon 
contributions to the hard-axion 
rate and to the so-called 
transverse momentum broadening coefficient, $\hat{q}$, 
are proportional to each other. 
Furthermore, as shown in ref.~\cite[footnote~12]{bg},
the linear dependence on a cutoff scale, $q^*_{ }$
(which corresponds to $p_\perp^*$ in our notation), which 
separates soft and hard transverse gluon momenta, 
and needs to cancel from 
physical observables, appears in an identical way. 
The resulting NLO contribution 
is given in \eq(A.29) of ref.~\cite{bg}, 

\vspace*{-3mm}

\begin{equation}
  \gamma^{ \rmii{strict}}_\rmii{NLO,hard}(k) 
  \;=\;
  \gamma^{ \rmii{strict}}_\rmii{LO,hard}(k) 
  \;+\; 
  \overbrace{
  \frac{2\Nc}{\hat{m}_\rmii{E}^2}  
  \frac{\mE}{T}\frac{10+3 \pi ^2-4 \ln2}{16 \pi }
   }^{\rmO(g)}
  \;.
  \label{NLO_hard}
\end{equation}
Numerical examples of \eqs\eqref{LO_hard} and \eqref{NLO_hard} 
are plotted in \fig\ref{fig:hardrate}.

%
%

We now go beyond ref.~\cite{bg} and explain why 
this is the \emph{only} NLO correction to the hard-axion
interaction rate. The argument relies on the 
structure of the axion--gauge vertex. The physics that we have
to watch out for is that of ``collinear emissions''.
If present, it requires 
Landau--Pomeranchuk--Migdal (LPM) resummation~\cite{lpm1,amy1,bb1},
in fact already at LO. 
This is generically the case for dimension-four vertices, such as 
the coupling of photons to a QCD plasma, or right-handed neutrinos
to an electroweak plasma 
(cf.\ ref.~\cite{lpm} for a review). 
Instead, if the vertex is higher-dimensional and contains more derivatives, 
collinear emission is suppressed. This has been shown
for hard axions~\cite{salvio,bg}, and also hard
gravitons~\cite{gw}, for which collinear $1+n\leftrightarrow 2+n$
processes amount to corrections of $\mathcal{O}(g^2_{ })$.

If we look at the NLO rate
for photons or right-handed neutrinos, 
there are further phase-space regions apart from 
collinear emission that contribute,
such as those associated with  so-called ``semi-collinear'' 
processes~\cite{Ghiglieri:2013gia}. Let us now argue that these 
further regions are also suppressed in the axion case. For this, 
we recall the origin of the semi-collinear region for photons. 
Consider the $s$-channel $2\to 2$
process $q+g\to q+\gamma$, which yields a rate
\begin{equation}
    \label{photonsemis}
    \gamma_\gamma(k)\big |^{s-{\rm channel}}_{k \;\ge\; \pi T}\propto 
    \int_k^{\infty} \! {\rm d} p^{ }_0 
    \int_{|2k - p^{ }_0|}^{p^{ }_0} \!\! {\rm d}p 
     \int_{p^{ }_-}^{p^{ }_+} 
     \! {\rm d}r 
     \int_{-\pi}^{\pi} \! \frac{{\rm d}\varphi}{2\pi}
     \,\biggl( \frac{-u}{s} \biggr)\,
    \frac{\nF(p_0^{ }-r)\,\nB(r)\,[1 - \nF(p^{ }_0-k)]}{\nB(k)}
  \;,
\end{equation}
where $p^{ }_\pm \equiv (p^{ }_0 \pm p)/2$, 
$r$ is the energy of the incoming gluon, 
and the variables have been chosen such that $s=\P^2$. 
The variable $t$ is given in \eq\eqref{tschannel}, with $\varphi$ 
being the azimuthal angle between $\mathbf{r}$ and $\mathbf{k}$.
The semi-collinear region emerges when the incoming gluon is soft, 
$r \sim gT$. The full computation requires 
all crossings and the $t$-channel contribution. 
Furthermore, HTL resummation
and the semi-collinear limit of the collinear contribution 
need to be considered. Here we only illustrate how 
this region emerges and show its parametric size.

To do so, let us see when $r$ can become soft. 
This requires $p^{ }_-$ to be soft. Noting that the bounds 
on $p^{ }_0$ and $p$ imply $p^{ }_+>k$ 
and $0<p^{ }_-<k$, this is allowed. 
In this regime one has 
$s=(p_0^{ }-p)(p_0^{ }+p)\sim r\, k \sim g T^2_{ }$: 
this is the semi-collinear scaling, where
the virtual quark carries a hard (thermal) component 
but has a virtuality $s$ that lies between 
$g^2_{ }T^2_{ }$ and~$T^2_{ }$.
Let us now look at $t$:
\begin{align}
    \label{tschannel}
    t \;=\; &
   \frac{s}{2 p^2_{ }}\bigg[\,
   (2k - p^{ }_0)(2r - p^{ }_0) - p^2_{ } 
  + \cos(\varphi)\, 
   \sqrt{p^2_{ } - (2 k - p^{ }_0 )^2_{ }}
   \sqrt{p^2_{ } - (2 r - p^{ }_0 )^2_{ }}
  \,\bigg]\\[2mm]
   \label{tschannelsemi}
   \;\approx\; & \frac{(p^{ }_0-p)(p^{ }_0 + p)}{2 p^2_{ }}\big[
   p^{ }_0 (p^{ }_0 - 2 k ) - p^2_{ }\big]
   \;\approx\;
   -2k(p^{ }_0-p)
   \;\sim\; gT^2_{ }
 \;,
\end{align}
where in the second line we expanded for 
$r\sim p^{ }_0-p \sim gT \ll p^{ }_0 + p \sim k$. 
This implies that $s$ and $t$, and subsequently also $u$, are small, 
however the matrix element squared itself is of order unity, 
$-u/s \approx ({p}^{ }_+-k)/{p}^{ }_+$.
Inserting $\nB^{ }(r) \approx T/r$
and $\nF^{ }(p^{ }_0)[1-\nF^{ }(p^{ }_0-k)]/\nB^{ }(k) = 
\nF^{ }(p^{ }_0 - k) - \nF^{ }(p^{ }_0)$,  
\eq\eqref{photonsemis} then behaves as
\begin{equation}
    \label{photonsemisfinal}
   \gamma_\gamma(k)\big |^{s-\text{channel}}_{k \;\ge\; \pi T}\propto 2T 
   \int_{0}^{p^*_{ }}
    \!\! {\rm d}{p}_-^{  } \int_{{p}_-^{ }}^{r^*} 
 \frac{{\rm d}r}{r}  
 \int_k^{\infty} \! {\rm d} {p}^{ }_+ 
 \frac{{p}^{ }_+ - k}{{p}^{ }_+}
 \,\big[\nF({p}^{ }_+-k) - \nF({p}^{ }_+)\big]\;.
\end{equation}
The $p^*$ and $r^*$ regulators have been included for illustration only. 
The point 
is that the phase-space region with ${p}^{ }_- \sim r \sim gT$ 
and ${p}^{ }_+ > k \ge \pi T$ is only suppressed
by a factor of $g$: 
this is because $\int {\rm d}r/r\sim 1$
for all scalings of $r$, so the suppression 
comes from $\int \! \mathrm{d}{p}^{ }_- \sim gT$.

In the case of a dimension-five operator, 
the counting changes, given that 
the matrix element squared must have mass dimension two
compared with \eq\nr{photonsemisfinal}. 
As we have shown, $s\sim t\sim u\sim  gT^2$, 
so all combinations $s^at^b$ with $a+b=1$
scale like $gT^2$. The extra suppression factor 
brings this region to the size $\rmO(g^2_{ })$, 
which is of NNLO in the hard domain.

\newpage

\small{
%

}

\end{document}
